\documentclass[RNAAS,twocolumn]{aastex631}
\usepackage{graphicx,url,amssymb,amsmath,color,units,wasysym,epsfig,epstopdf,enumerate}

 \usepackage[hang,flushmargin]{footmisc}
\usepackage{newtxtext,newtxmath}
\usepackage[T1]{fontenc}
\usepackage{ae,aecompl}
\usepackage[normalem]{ulem}

\renewcommand{\sc}[1]{\textsc{#1}}

\newcommand{\new}[1]{#1}

\newcommand{\SPA}{School of Physics and Astronomy, Monash University, Clayton VIC 3800, Australia}
\newcommand{\OzGravMonash}{OzGrav: The ARC Centre of Excellence for Gravitational Wave Discovery, Clayton VIC 3800, Australia}
\newcommand{\DAMTP}{Department of Applied Mathematics and Theoretical Physics, Cambridge CB3 0WA, United Kingdom}
\newcommand{\Curtin}{International Centre for Radio Astronomy Research – Curtin University, GPO Box U1987, Perth, WA 6845, Australia}
\newcommand{\KavliCambridge}{Kavli Institute for Cosmology Cambridge, Madingley Road Cambridge CB3 0HA, United Kingdom}

\begin{document}
\title{Rapid population synthesis of black-hole high-mass X-ray binaries: implications for binary stellar evolution}

\author{Isobel Romero-Shaw}
    \email{Corresponding author's email address: ir346@cam.ac.uk}
\affiliation{\DAMTP}
\affiliation{\KavliCambridge}
\affiliation{\SPA}
\affiliation{\OzGravMonash}

\author{Ryosuke Hirai}
\affiliation{\SPA}
\affiliation{\OzGravMonash}

\author{Arash Bahramian}
\affiliation{\Curtin}

\author{Reinhold Willcox}
\affiliation{\SPA}
\affiliation{\OzGravMonash}

\author{Ilya Mandel}
\affiliation{\SPA}
\affiliation{\OzGravMonash}

\begin{abstract}
We conduct binary population synthesis to investigate the formation of wind-fed high-mass X-ray binaries containing black holes (BH-HMXBs).
We evolve multiple populations of high-mass binary stars and consider BH-HMXB formation rates, masses, spins and separations.  We find that systems similar to Cygnus X-1 likely form after stable Case A mass transfer (MT) from the main sequence progenitors of black holes, provided such MT is characterised by low accretion efficiency, $\beta \lesssim 0.1$, with modest orbital angular momentum losses from the non-accreted material. 
Additionally, \new{efficient} BH-HMXB formation relies on a new simple treatment for Case A MT that allows donors to retain larger core masses compared to traditional rapid population-synthesis \new{assumptions}.
At solar metallicity, our Preferred model yields $\mathcal{O}(1)$ observable BH-HMXBs in the Galaxy today, consistent with observations.
In this simulation, $8\%$ of BH-HMXBs go on to merge as binary black holes or neutron star-black hole binaries within a Hubble time\new{; however, none of the merging binaries have BH-HMXB progenitors with properties similar to Cygnus X-1.
With our preferred settings for core mass growth, mass transfer efficiency and angular momentum loss, accounting for an evolving metallicity, and integrating over the metallicity-specific star formation history of the Universe, we find that BH-HMXBs may have contributed $\approx2$--$5$ BBH merger signals to detections reported in the third gravitational-wave transient catalogue of the LIGO-Virgo-KAGRA Collaboration.}
We also suggest MT efficiency \new{should be} higher during stable Case B MT than during Case A MT.
\end{abstract}

\section{1. Introduction}
\label{sec:intro}

\begin{figure*}
    \centering
    \includegraphics[width=0.85\textwidth]{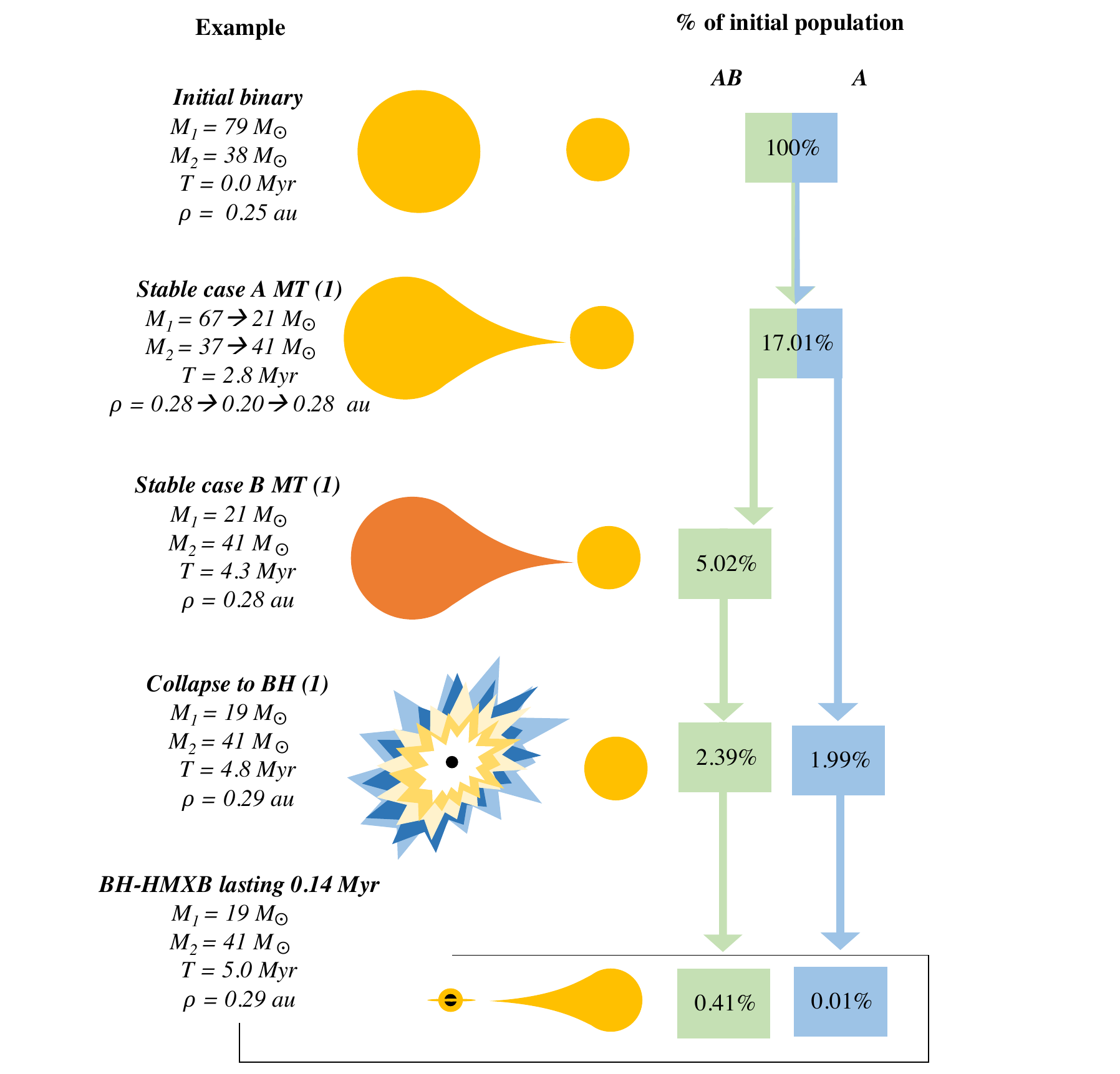}
    \caption{Typical evolutionary stages of the BH-HMXBs produced through pathways involving dynamically stable Case A MT in our simulations. Values provided on this plot are taken from our Preferred simulation, which has MT efficiency $\beta=0.1$, specific angular momentum loss $f_\gamma=0.1$, Wolf-Rayet wind mass loss rate $f_\mathrm{WR}=0.1$, metallicity $Z=Z_\odot$, a minimum zero-age main sequence mass of $10$~M$_\odot$, and increased core mass at the end of Case A MT. The primary and secondary masses $M_1$ and $M_2$, time $T$, and orbital semi-latus rectum distance $\rho$ of an example evolving through stable Case AB MT are provided on the far left. Like the majority of the BH-HMXBs in our simulations, this binary undergoes a major episode of MT as a main sequence star and a minor episode of MT as a Hertzsprung gap star. The first stages tidally lock the primary, spinning it up, and strip the majority of its envelope. This leaves little material to be further stripped during the short early Case B MT phase, and prevents the primary from expanding further and losing angular momentum from the core. In boxes on the right, we provide the percentage of the binaries in our Preferred simulation that proceed through each evolutionary stage for each pathway. As a guide, for stable Case A MT only, this can be read as follows: $17.01\%$ of all binaries undergo stable Case A MT from the primary; $1.99\%$ of all binaries see the primary collapse into a BH with no further MT; and $0.01\%$ of all binaries then proceed to fit our BH-HMXB selection criteria.}
    \label{fig:evolution-example}
\end{figure*}

High-mass X-ray binaries (HMXBs) contain one compact object---a neutron star or a black hole---and one massive star.
The compact object accretes matter from its companion, emitting X-ray radiation in the process.
HMXBs are distinguished from low-mass X-ray binaries by the mass of the companion star, which also dictates the mode of mass transfer (MT): while LMXBs accrete from their $\lesssim 2$~M$_\odot$ companion when the companion overflows its Roche lobe, MT in HMXBs is driven by stellar winds from the early-type companion. HMXBs containing black holes (BH-HMXBs) are of particular interest as potential progenitors of merging binary black holes (BBHs), which are now routinely observed via their gravitational-wave emission \citep{LVK:2022:GWTC-3}.

Of the 94 confirmed HMXBs \citep{Chaty:2022:HMXBs, Fortin:2023:HMXBCatalogue}, only three are known to contain black holes.
These are Cygnus X-1, a $40.6^{+7.7}_{-7.1}$~M$_\odot$ main sequence star with a $21.2 \pm 2.2$~M$_\odot$ BH companion \citep{WebsterMurdin:1972:CygX1Disco, Bolton:1972:CygX1Disco, MillerJones:2021:CygX1}; LMC X-1, a $31.8 \pm 5.5$~M$_\odot$ main sequence star with a $10.91 \pm 1.54$~M$_\odot$ BH \citep{Mark:1969:LMCX1Disco, Wojdowski:1998:LMCX1Disco, Orosz:2009:LMCX1, Orosz:2014:LMCX1}; and M33 X-7, which has a main sequence star of $38^{+22}_{-10}$~M$\odot$ and a BH of $11.4^{+3.3}_{-1.7}$M$_\odot$ \citep{Long:1981:M33X7Disco, Pietsch:2006:M33X7, Orosz:2007:M33X7, Ramachandran:2022:M33X7}.\footnote{~In addition to these known BH-HMXBs, there are some tentative claims, including SS 433 \citep{Seifina:2010:SS433}, Cygnus X-3 \citep{Zdziarski:2018:CygnusX3}, and MWC 656 (\citet{Grudzinska:2015:MWC656}, refuted by \citet{Rivinius:2022:MWC656Refute}), HD96670 \citep{GomezGrindlay:2021:HD96670}, and HR 6819 (\citet{Rivinius:2020:HR6819}, refuted by \citet{Frost:2022:HR6819Refute}).}
Similarities between the observed HMXBs include the near-maximal Roche lobe volume filling factor of the main sequence donor (between $0.899$ and $0.997$) and the apparently near-maximal spin of the BH accretor, with dimensionless spin magnitude measurements $a_* = cJ/GM^2$ ranging from $0.84^{+0.05}_{-0.05}$ to $>0.9985$, where $c$ is the speed of light, $J$ is the angular momentum of the BH, $G$ is the gravitational constant, and $M$ is the BH mass \citep[][]{Gou:2011:CygnusX1Spin, Gou:2014:CygnusX1Spin, Duro:2016:CygnusX1Spin, Qin:2019:HMXBspin, MillerJones:2021:CygX1, Zhao:2021:CygnusX1Spin}.\footnote{~The Roche lobe filling factor is the cubed root of the volume filling fraction; see \citet{Neijssel:2021:WRmassloss} for the definition of this parameter.}
Note that the spins of the BHs have been contested \citep[see, e.g.,][for a discussion of the spin measurement methods and possible systematic biases]{MillerMiller:2015,Reynolds:2020}; for example,  \citet{Kawano:2017:CygnusX1Spin} argue that observations of the soft-state X-rays are better explained by BHs with a lower spin.

Although the BH in a HMXB is accreting from its companion, this is not likely to cause the BH to spin up. In order to drive the BH to high spins, the accreted matter would have to double the mass of the BH, which is difficult to achieve through Eddington-limited accretion from short-lived massive companions \citep{KingKolb:1999,Podsiadlowski:2002:XRayBinaryEvolution}. 
\new{Additionally, natal spins of BHs that formed from single stars, which do not experience spin-up through mass transfer or tides, are expected to be small \citep[e.g.][]{FullerMa:2019:BHNatalSpin, Gottlieb:2023:Collapsars}}.
Thus, the high spin is more likely to have been inherited from a BH progenitor \new{that was spun about by binary interactions, particularly tides} \citep[see][for a brief review of possible spin origins]{MandelFragos:2020:GW190412}. 

Stars in tight binaries can be significantly spun up through tidal locking. If this rapid spin can be retained during the collapse into a BH, the BH can have high spins compatible with the observed values. However, as the progenitor star evolves, a significant fraction of the angular momentum can be transported to the outer layers through strong core-envelope coupling and removed from the star via binary interactions or winds. In order to prevent this, the binary interaction needs to happen early on during its evolution before most of the angular momentum is transported out of the core. Hence it has been proposed that Cygnus X-1 likely evolved via Case A MT, meaning that the BH progenitor transferred mass to its companion whilst on the main sequence \citep{Qin:2019:HMXBspin,Neijssel:2021:WRmassloss}.\footnote{~\citet{Qin:2019:HMXBspin} conclude that chemically homogeneous evolution could produce highly-spinning BH accretors with stellar companions, but with orbital periods larger than observed.}  The formation of a BH-HMXB via Case A MT is illustrated in Figure \ref{fig:evolution-example}. In these models, it is assumed that the MT is relatively non-conservative, in order to maintain a tight orbit and therefore a rapid spin. It has been argued that if the mass transferred in the early stages of the MT can spin up the secondary close to critical rotation, the subsequent material could not be retained due to the centrifugal barrier \citep{Packet:1981:SpinUp,Qin:2019:HMXBspin}; however, see \citet{PophamNarayan:1991,Vinciguerra:2020:MTEfficiency}.

The observability of wind-fed X-ray binaries is determined by the properties of the wind emitted from the donor and the binary orbit. Traditionally, the X-ray luminosity from these systems has been estimated from the mass accretion rate using a simple Bondi-Hoyle-Lyttleton accretion model \citep[]{HoyleLyttleton:939:accretion, BondiHoyle:1944:accretion}. However, the mass accretion rate alone is not sufficient to predict whether the BH will emit X-rays.  Efficient X-ray emission requires the presence of an accretion disk, which in turn relies on having more angular momentum in the accreting material than at the innermost stable circular orbit around the BH \citep[e.g.][]{IllarionovSunyaev:1975:XRayAccretion}. This can only be achieved when the donor star occupies a significant fraction of its Roche lobe, where the companion is located deep within the acceleration zone of the wind \citep[]{Karino:2019,Sen:2021:PopSynthXray,HiraiMandel:2021:AccretionDiskXRays}.

\begin{figure}
    \centering
    \includegraphics[width=0.47\textwidth]{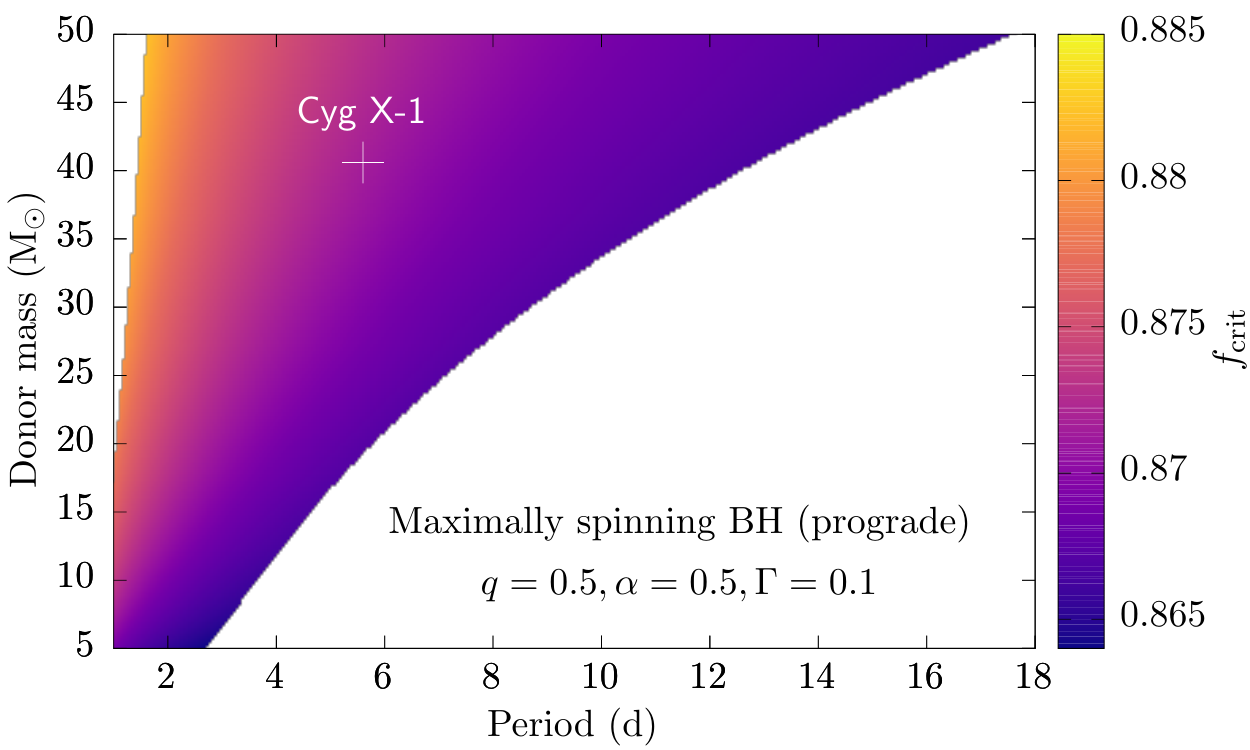}
    \caption{The critical Roche lobe filling factor to create an accretion disk (colour scale) as a function of orbital period (horizontal axis) and donor mass (vertical axis) as predicted by the model of \citet{HiraiMandel:2021:AccretionDiskXRays}. The systems studied here have mass ratio $q\equiv M_\mathrm{BH}/M_\mathrm{d}=0.5$, the wind force multiplier parameter $\alpha=0.5$, and Eddington factor $\Gamma=0.1$. We assume the BH is maximally spinning prograde to the orbit. The white areas are regions of parameter space where either the donor star will overflow its Roche lobe at zero-age main sequence, or it will never reach the required $f_\mathrm{crit}$ during its main sequence phase. The white cross marks the binary parameters of Cygnus X-1 as reference.}
    \label{fig:ryo-prescription}
\end{figure}

\citet{HiraiMandel:2021:AccretionDiskXRays} recently investigated how the tidal force from the BH companion on a nearly Roche-lobe-filling wind donor will significantly affect the wind morphology due to gravity darkening and changes in the wind acceleration process.  The resulting highly asymmetric wind velocity distribution ensures that the accreted matter possesses large amounts of angular momentum, aiding disk formation. They predict a steep transition between observable and dormant BH systems due to the strong separation dependence of tidal forces, with a threshold for disk formation at $f_\mathrm{crit}\gtrsim80~\%$ Roche lobe filling. Above this, the average specific angular momentum of the accretion flow exceeds that of the innermost stable circular orbit around the BH. Otherwise, the accretion will occur in a radiatively inefficient manner.
The exact bound weakly depends on the properties of the binary, such as the mass ratio and wind acceleration parameters.
In Figure~\ref{fig:ryo-prescription}, we plot this critical Roche lobe filling factor $f_\mathrm{crit}$ as a function of the donor mass $M_\mathrm{d}$ and orbital period for a given set of binary parameters. We see that $f_\mathrm{crit}$ is confined to a narrow range between $86.5~\%\lesssim f_\mathrm{crit}\lesssim88.5~\%$ over the entire parameter range. Note that with the classical assumption of a spherical wind \citep[e.g.,][]{IllarionovSunyaev:1975:XRayAccretion}, accretion disk formation is impossible for most of this parameter space.

In this paper, we conduct a binary population synthesis study to investigate the formation rates of wind-fed BH-HMXBs. We vary key parameters that influence the masses, separations, and rates of BH-HMXBs forming through Case A MT, and compare the resulting BH-HMXB populations to the Galactic binary Cygnus X-1. We identify BH-HMXBs as binaries that contain a BH and a main sequence star with $M_\mathrm{d} \geq 10$~M$_\odot$, with the main sequence star having a Roche lobe filling fraction $f\geq f_\mathrm{crit}$, where $80~\%\lesssim f_\mathrm{crit}\lesssim90~\%$. We compute expected yields of BH-HMXBs in our Preferred simulation, and discuss selection effects due to unfortunate positioning of sources in the Galaxy. We discuss alternative formation pathways for BH-HMXBs in our simulations, and argue against them. We also inspect the future evolution of BH-HMXBs \new{in our Preferred simulation}, finding that a non-negligible fraction ($\approx8\%$) go on to merge as compact binaries potentially detectable with ground-based gravitational-wave detectors like Advanced LIGO and Virgo \citep{Aasi13, AdvancedVirgo}.

The component masses of merging BBHs observed via their gravitational-wave emission are similar, although not entirely consistent with BHs in BH-HMXBs \citep[taking the relevant selection effects into account;][]{Fishbach:2022:ApplesOranges, Liotine:2022:SelectionEffects}. However, their spins are not: most merging BBHs contain BHs that have negligible spins \citep[e.g.,][]{LVK:2022:GWTC-3}.
While there are significant uncertainties in the spin measurements for BHs in both HMXBs \citep[e.g.,][]{MillerMiller:2015:Spins, Kawano:2017:CygnusX1Spin, Reynolds:2020} and BBHs \citep[e.g.,][]{LVK:2021:GWTC3-pop, Biscoveanu:2021:NewSpin, Tong:2022:PopSpinGWTC3} that could reduce this discrepancy, we focus on formation channels that can plausibly give rise to rapidly-spinning BHs in HMXBs in this work.

\new{Several recent studies used rapid population synthesis to explore the observed population of BH-HMXBs and their relation to the merging BBH population detected with gravitational waves.}
\citet{Liotine:2022:SelectionEffects} used the COSMIC population synthesis code \citep{Breivik:2020:COSMIC} to investigate observational selection effects affecting the populations of  BHs observed in HMXBs and in merging BBHs.  \citet{Misra:2022:XrayLuminosityPopsynth} studied the influence of physical processes and assumptions on the X-ray luminosity function of all HMXBs (including those containing neutron stars) \new{with} the POSYDON population synthesis code \citep{Fragos:2022:POSYDON}.  \new{\citet{GallegosGarcia:2022:HMXBs} used MESA simulations in combination with COSMIC to study the eventual fates of BH-HMXBs formed via case A MT, comparing the yields of merging BBHs from the BH-HMXB channel.}

In addition to using different population synthesis \new{tools}, our work differs from \new{these three papers} in several \new{important} ways, \new{driven by the different aims of these projects}, \new{which makes it difficult to quantitatively compare the results}.   \new{One important difference is the treatment of BH-HMXB observability:} \citet{Liotine:2022:SelectionEffects} treat any BH with a $\geq 5$~M$_\odot$ stellar companion as a BH-HMXB, and estimate its luminosity assuming spherically-symmetric wind accretion; \new{similarly, \citet{GallegosGarcia:2022:HMXBs} designate any binary comprising a BH and a hydrogen-rich star as an HMXB}.  In contrast, we use the prescription of \citet{HiraiMandel:2021:AccretionDiskXRays} to determine the existence and duration of an observable BH-HMXB phase in a binary containing one BH and one $\geq 10$~M$_\odot$ main sequence star, \new{an approach also taken by \citet{Misra:2022:XrayLuminosityPopsynth}.}  Secondly, while \citet{Liotine:2022:SelectionEffects} use the default behaviour of \citet{Hurley:2002:BSE} to determine the core mass at the end of Case A MT, we allow a more massive core to be retained \citep[e.g.][see next section]{Heger:2000:RotatingMassiveStars}.  Thirdly, \new{different assumptions are made regarding stable mass transfer efficiency; in particular,} we investigate a range of lower MT efficiencies that may be \new{consistent with} nearly maximally spun up accretors \citep{Qin:2019:HMXBspin}.  \new{\citet{GallegosGarcia:2022:HMXBs} and this work focuses on BH-HMXBs formed via case A MT, assuming that this is the channel most likely to lead to highly-spinning BH primaries}, while \citet{Liotine:2022:SelectionEffects} do not distinguish between formation channels in their work.  \new{These studies make different choices of metallicity in the simulations: we consider four metallicities in this study, \citet{Misra:2022:XrayLuminosityPopsynth} explore just one, while \citet{GallegosGarcia:2022:HMXBs} consider five metallicities, but focus on low metallicities given their interest in the formation of merging BBHs.  Other differences include the treatment of angular momentum carried away by non-conservative mass transfer, supernova prescriptions, wind-driven mass loss rates, mass ranges explored, et cetera.}

\new{Despite these differences, clear common trends emerge.   \citet{GallegosGarcia:2022:HMXBs}, \citet{Neijssel:2021:WRmassloss} and previous work (e.g., \citealt{Belczynski:2011:CygX1FutureNSBH}) all conclude that only small fractions of BH-HMXBs will yield merging BBHs, which matches our conclusions.  For example, \citet{GallegosGarcia:2022:HMXBs} find that at most 12\% of BH-HMXBs form merging BBHs. Conversely, the overwhelming majority of merging BBHs observed as gravitational-wave sources (e.g., 93\% according to \citealt{GallegosGarcia:2022:HMXBs}) did not form through the BH-HMXB channel, which is consistent with the observed differences in BH masses and spins between the BH-HMXB and BBH populations \citep[e.g.,][]{Fishbach:2022:ApplesOranges}.}

\new{In this paper, we concentrate on using observations of BH-HMXBs to constrain key uncertain parameters governing binary evolution, particularly mass transfer physics.  The} paper is structured as follows.
In Section \ref{sec:method}, we review the rapid binary population synthesis performed in this work.
In Section \ref{sec:results}, we compare the properties of the simulated BH-HMXB populations produced through Case A MT to observations.
This motivates our Preferred settings, which yield $\approx1$ observable BH-HMXBs per Milky Way-like galaxy today when accounting for selection effects.
In Section \ref{sec:mergers}, we discuss the formation of merging compact-object binaries from BH-HMXBs.
In Section \ref{sec:pathways}, we consider alternative BH-HMXB formation pathways within our Preferred simulation, and detail our reasons for regarding their predictions with scepticism.
In Section \ref{sec:discussion}, we discuss our results and speculate on their implications for binary star evolution.

\section{2. Method: Population synthesis of BH-HMXB candidates}
\label{sec:method}

We perform rapid binary population synthesis using COMPAS \citep{TeamCOMPAS:2021:COMPAS,COMPAS:2021} v02.31.03.
Rapid population synthesis codes like COMPAS are useful for quickly assessing the impact of modelling assumptions on the outcomes of binary evolution; see, e.g., \citet{Broekgaarden:2022,Belczynski:2022:UncertainMassive,StevensonClarke:2022:COMPASConstraints}.
\new{In comparison to 1D stellar evolution codes (e.g., STARS \citep{Eggleton:1972:STARS, Eggleton:2011:STARS}, BPASS \citep{Eldridge:2009:BPASS, Stanway:2018:BPASS}, and MESA \citep{Paxton:2011:MESA,Jermyn:2023:MESA}), the speed of \textit{rapid} population synthesis relies on simplified prescriptions for stellar evolution and binary evolution.}
However, \new{the use of such prescriptions causes rapid population synthesis codes to have intrinsic limitations with respect to their detailed-evolution counterparts}: simplified recipes\new{, while computationally more efficient, may neglect or oversimplify elements of stellar evolution}. 
\new{An example of this is stellar evolution at high masses:} like many other \new{rapid} population synthesis codes \citep[e.g. COSMIC and StarTrack;][]{Breivik:2020:COSMIC, Belczynski:2008:StarTrack}, COMPAS relies on the single stellar evolution (SSE) prescriptions of \citet{Hurley:2000:BSE}. \new{These} are calibrated only up to zero-age main sequence (ZAMS) masses of $50$~M$_\odot$, but extrapolated to higher masses.

For initial conditions, we assume a log-uniform initial distribution of orbital separations between $0.01$~au and $1000$~au \citep{Abt:1983:loguniform, Sana:2012:MassRatio}, the \citet{Kroupa} initial mass function for the primary masses, and a uniform distribution of initial mass ratios.
To improve our sampling efficiency for high-mass systems, we increase the minimum secondary mass to $M_2=10$~M$_\odot$. 
Since we enforce that $M_1 \geq M_2$, the ZAMS mass range sampled for both components spans $10$~M$_\odot$ to $150$~M$_\odot$.
We use the supernova kick and remnant mass prescriptions of \citet{MandelMuller:2020:kicks}\new{, with their default values}.

Our main focus is to explore the Case A MT channel, where the MT occurs when the primary is still on the main sequence. In the original \citet{Hurley:2002:BSE} formalism for Case A MT, the core is not modelled during the main sequence and discretely jumps to a value based on the stellar mass at the terminal-age main sequence (TAMS). This treatment may lead to significantly underestimated core masses for donors that lose a significant amount of mass during late Case A MT. Here, we implement a new recipe for Case A MT similar to the model in \citet{Neijssel:2021:WRmassloss}.  When a donor engages in Case A MT, we determine \new{what} its \new{helium} core mass at TAMS \new{$M_{c, \text{TAMS}}$ would be} according to the \citet{Hurley:2000:BSE} prescription \new{if it evolved to this stage with no further mass loss}.  We then set a minimum core mass \new{$M_{c, \text{min}}$} equal to \new{$M_{c, \text{TAMS}}$}, multiplied by the fraction of the donor's main sequence lifetime at the onset of MT\new{, $\tau_\text{MT}$, such that $M_{c, \text{min}}=\tau_\text{MT}M_{c, \text{TAMS}}$}. The minimum core mass \new{$M_{c, \text{min}}$} is used as a lower limit for the core mass of the donor when its evolution reaches TAMS. In particular, if the total mass at TAMS \new{$M_{*, \text{TAMS}} \leq M_{c, \text{min}}$,} the entire star is considered to become a naked helium star. This treatment allows us to qualitatively account for nuclear fusion prior to the MT phase, leading to larger core masses at TAMS and the subsequent formation of higher-mass black holes \new{(see Section~\ref{sec:core-growth}).}

\begin{figure}
    \vspace{-40pt}
    \centering \includegraphics[width=0.4\textwidth]{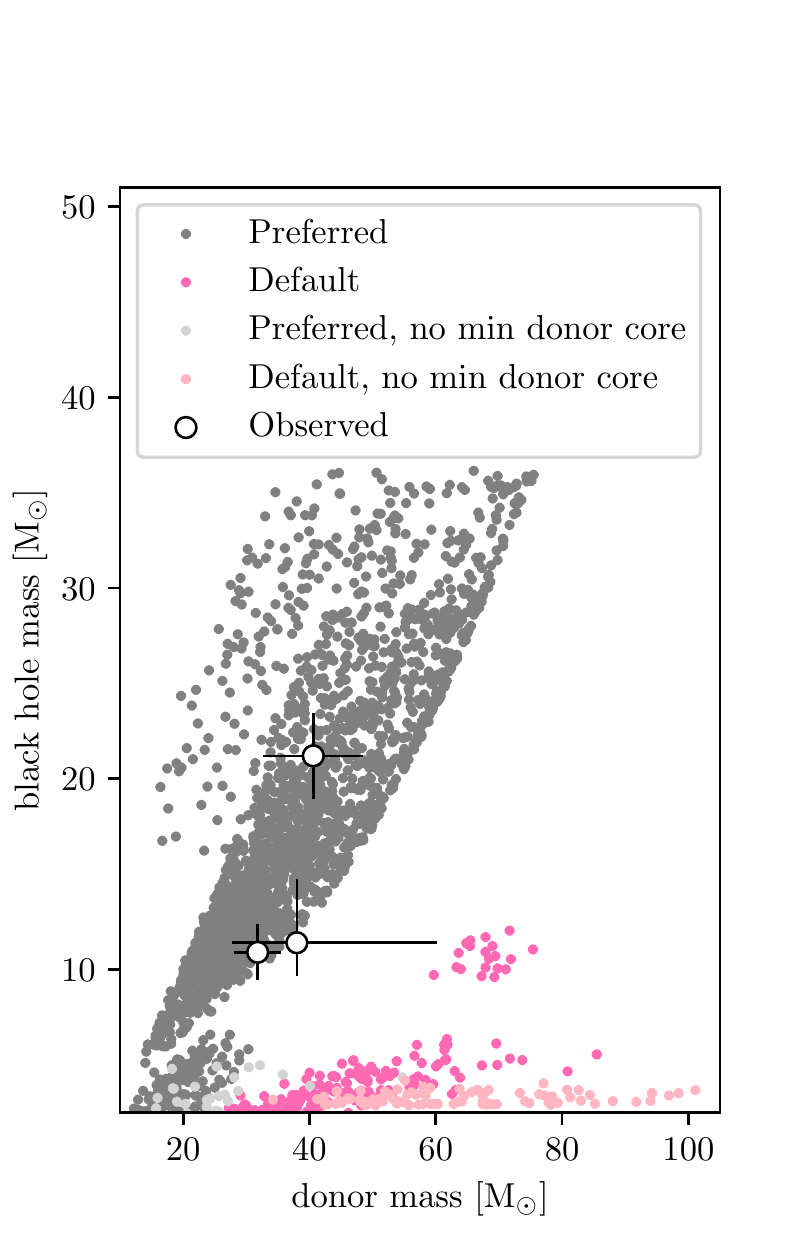}
    \caption{Comparison of BH-HMXB candidates produced through stable Case A and AB MT using \new{four different set-ups.  The first two set-ups assume the new prescription for core mass growth following Case A MT: with our Preferred settings (MT efficiency $\beta=0.1$, specific angular momentum loss $f_\gamma=0.1$, Wolf-Rayet wind mass loss scaling parameter $f_\mathrm{WR}=0.1$, metallicity $Z=Z_\odot$; dark grey points) and with the COMPAS \new{v02.31.03} Default settings ($\beta=\beta_\mathrm{th}$, $f_\gamma = 0$, $f_\mathrm{WR}=1.0$, $Z=Z_\odot$; dark pink points).  The other two populations are evolved  without our new prescription for core mass growth following Case A MT: using Preferred settings (light grey points); and using the Default settings (light pink points).}  The masses of three observed BH-HMXBs (from lightest to heaviest median donor and BH mass: LMC X-1, M33 X-7, Cygnus X-1) are indicated with black-encircled white points and black 1-$\sigma$ error bars. \new{The retention of a larger core and other choices in the Preferred settings produce more massive BHs, consistent with observations.}}
    \label{fig:fiducial_vs_default}
\end{figure}

To identify BH-HMXB candidates, we implement a new option in COMPAS that outputs the state of the binary if one component is a BH or NS and its companion has a Roche lobe filling factor $f\geq f_\mathrm{crit}$ \citep{HiraiMandel:2021:AccretionDiskXRays}. 
We further filter these systems to include only those binaries containing a BH and a main sequence donor with $M_\mathrm{d} \geq 10$~M$_\odot$.
The Roche lobe filling factor is defined as $f\equiv R_\mathrm{d}/R_L$, where $R_\mathrm{d}$ is the donor stellar radius and $R_L$ is the Roche lobe radius. For eccentric binaries, the Roche lobe radius is approximated by that of a circular orbit with the separation equal to the periastron distance. The exact value of $f_\mathrm{crit}$ depends on the stellar and orbital properties, but generally take values in a relatively narrow range ($0.8$--$0.9$; see Figures~\ref{fig:ryo-prescription} and \ref{fig:ryo-prescription-nospin}). We set the default value as $f_\mathrm{crit}=0.8$, but also quote BH-HMXB yields using $f_\mathrm{crit}=0.9$ to quantify the impact of this choice (see Section~\ref{sec:results}).

\new{
We assume that the spin-up of a BH progenitor in a BH-HMXB can only occur through tides during stable Case A MT from the primary, an argument put forth by, e.g., \citet{Valsecchi:2010,Qin:2019:HMXBspin}.
This argument relies on efficient angular momentum transport within stellar interiors, allowing the core to spin up while the outer layers are stripped \citep{Qin:2019:HMXBspin}.
There are further uncertainties due to our neglect of the influence of tides and wind drag on the evolution of close binaries, which could be important for systems like Cygnus X-1 \citep{Schroder:2021:RadiationDrivenWinds}.
The high BH spins in observed HMXBs have significant measurement uncertainties and systematic errors \citep[see][and references therein]{MillerMiller:2015,Reynolds:2020,Belczynski:2021:AllApples}.
Hence, calibrating our simulations using the requirement that we obtain high BH spins can provide only tentative constraints on the physics that we investigate. 
}

\new{
Most of the binaries that produce BH-HMXB candidates through stable Case A MT in our simulations experience a brief subsequent episode of stable MT while the primary is on the Hertzsprung gap (Case AB MT), before the primary collapses to a BH. In our Preferred simulation, there are $\approx35$ times as many BH-HMXBs forming through Case AB than pure Case A MT.
Since the primary loses most of its envelope during Case A MT, any subsequent Case B phase is short-lived and early on the Hertzsprung gap.
In these binaries, Case B MT occurs at roughly the separation where the Case A MT ends, so there is limited donor expansion, and the resulting mass and angular momentum loss during Case B MT is minor. 
Therefore, we treat any BH-HMXB candidate forming via stable Case A MT---with or without a subsequent Case B MT phase---as likely to contain a highly-spinning BH.
We also note that the distinction between Case AB and pure Case A systems is somewhat arbitrary, as it is strongly related to how Case A MT is treated in our simulations. 
}

Noting that our results are likely to be sensitive to the assumptions outlined above, we assess the implications of the observed masses, separations, spins, and rates of known BH-HMXBs for the astrophysics behind BH-HMXB formation.

We vary four key parameters, detailed below, and evolve $0.5$ million binaries per simulation. 

\begin{enumerate}
    \item \textbf{Metallicity at birth ($Z$):} The metallicity of the progenitor of Cygnus X-1 at formation is challenging to ascertain: \citet{Shimanskii:2012:CygnusX1} measure the donor to have surface abundances of $\sim 2~Z_\odot$, which may reflect its birth metallicity, but could instead be influenced by an earlier episode of accretion from the BH progenitor. Stars in the vicinity of Cygnus X-1 have abundances slightly lower than $Z_\odot$ \citep{Daflon:2001:OBstars}. We vary $Z$ between \new{$\frac{1}{10}~Z_\odot$} and $2~Z_\odot$, where $Z_\odot=0.0142$. 
    
    \item \textbf{MT efficiency ($\beta$):} The MT efficiency, $\beta$, defines what fraction of mass lost from a donor is accreted by its companion. By default in COMPAS, $\beta = \beta_\mathrm{th}$, a quantity determined by the ratio of the thermal timescales of the donor and accretor, with MT becoming fully conservative when donor thermal timescales are no more than 10 times shorter than accretor thermal timescales \citep{COMPAS:2021}. However, if the accreting star (in this case, the secondary, before the primary evolves into a BH) is spun up by MT, $\beta$ may be substantially reduced \citep{Packet:1981:SpinUp,Qin:2019:HMXBspin}. We model this possibility by setting $\beta$ to a constant value to evaluate the impact of MT efficiency on BH-HMXBs binaries. We vary $\beta$ between $\beta=0.5$ to $\beta=0.0$, the latter representing completely non-conservative MT. 
    
    \item \textbf{Specific angular momentum lost in non-conservative MT ($f_\gamma$):}     When MT is non-conservative, $f_\gamma$ determines how much angular momentum the non-accreted mass carries with it.  By default, COMPAS treats ejected mass as carrying away the specific angular momentum of the accretor, which corresponds to $f_\gamma=0$ (``isotropic re-emission'').  However, \citet{Macleod:2020:MassLoss} argue that the ejected angular momentum may be higher.  At the other extreme, if material lost from the binary escapes from the L2 Lagrange point, $f_\gamma=1$.  The parameter $f_\gamma$ linearly interpolates in the distance from the binary centre of mass between the location of the accretor and the L2 point \citep{Willcox:inprep:fGamma}.  We vary $f_\gamma$ between $0$ and $0.4$. 
    
    \item \textbf{Wolf-Rayet wind mass loss efficiency ($f_\mathrm{WR}$):} Following the removal of the primary's hydrogen envelope, its evolution will depend on the wind mass loss rate of stripped (Wolf-Rayet) stars \citep[e.g.,][]{Vink:2017:WRWinds, Sander:2020:WRWinds, Sen:2021:PopSynthXray}. We use a mass loss prescription from \citet{HamannKoesterke:1998:WRWinds} with a metallicity scaling from \citet{VinkdeKoter:2005:m} and an additional scaling parameter $f_\mathrm{WR}$ allowing the wind strength to be modified \citep{Barrett:2018:WRWinds}, yielding the mass loss rate $f_\mathrm{WR} \times 10^{-13} (L / L_\odot)^{1.5}(Z/Z_\odot)^{0.86}$~M$_\odot$~yr$^{-1}$. \citet{Neijssel:2021:WRmassloss} showed that $f_\mathrm{WR}$ must be $\lesssim 0.2$ when $Z=Z_\odot$ in order to explain the observed properties of Cygnus X-1.  However, the metallicity of Cygnus X-1 at formation is uncertain (see above).  We vary $f_\mathrm{WR}$ between the COMPAS \new{v02.31.03} default of $1$ and a minimum of $f_\mathrm{WR}=0.05$. 
\end{enumerate}

\section{3. Results: Comparison of simulated BH-HMXB populations to observations}
\label{sec:results}

\begin{figure*}
    \centering
    \includegraphics[width=0.9\textwidth]{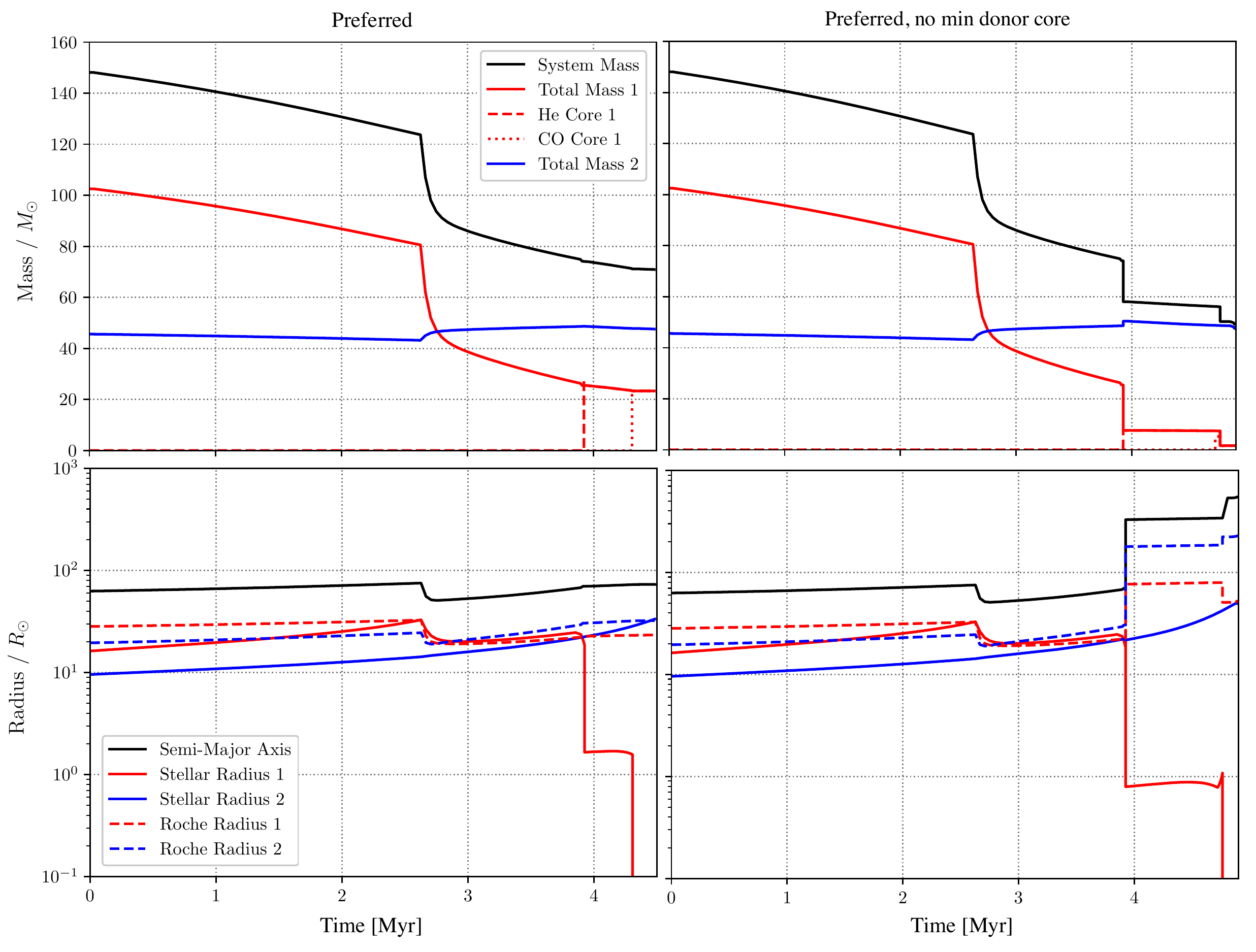}
    \caption{\new{Evolution of the binary masses (top row) and separations/radii (bottom row) for one example system, evolved using the Preferred model including our new prescription for core mass growth during the MS (left column) and the Preferred model without this change (right column). In the updated model, the minimum helium core mass $M_{c, \mathrm{min}}$ of the primary following Case A MT (see Section \ref{sec:method}) exceeds its total mass at TAMS, so the star becomes a stripped helium star. In comparison, the model without the update forms a much smaller helium core at TAMS following Case A MT, and the mass loss drives the binary to higher separations. In the model with the update, this system becomes a BH-HMXB, while in the no-minimum-core-mass model, the system becomes a wide NS-MS binary and merges while the secondary crosses the HG.}}
    \label{fig:detailed_evolution_comparison}
\end{figure*}

\begin{figure*}
    \centering
    \includegraphics[width=\textwidth]{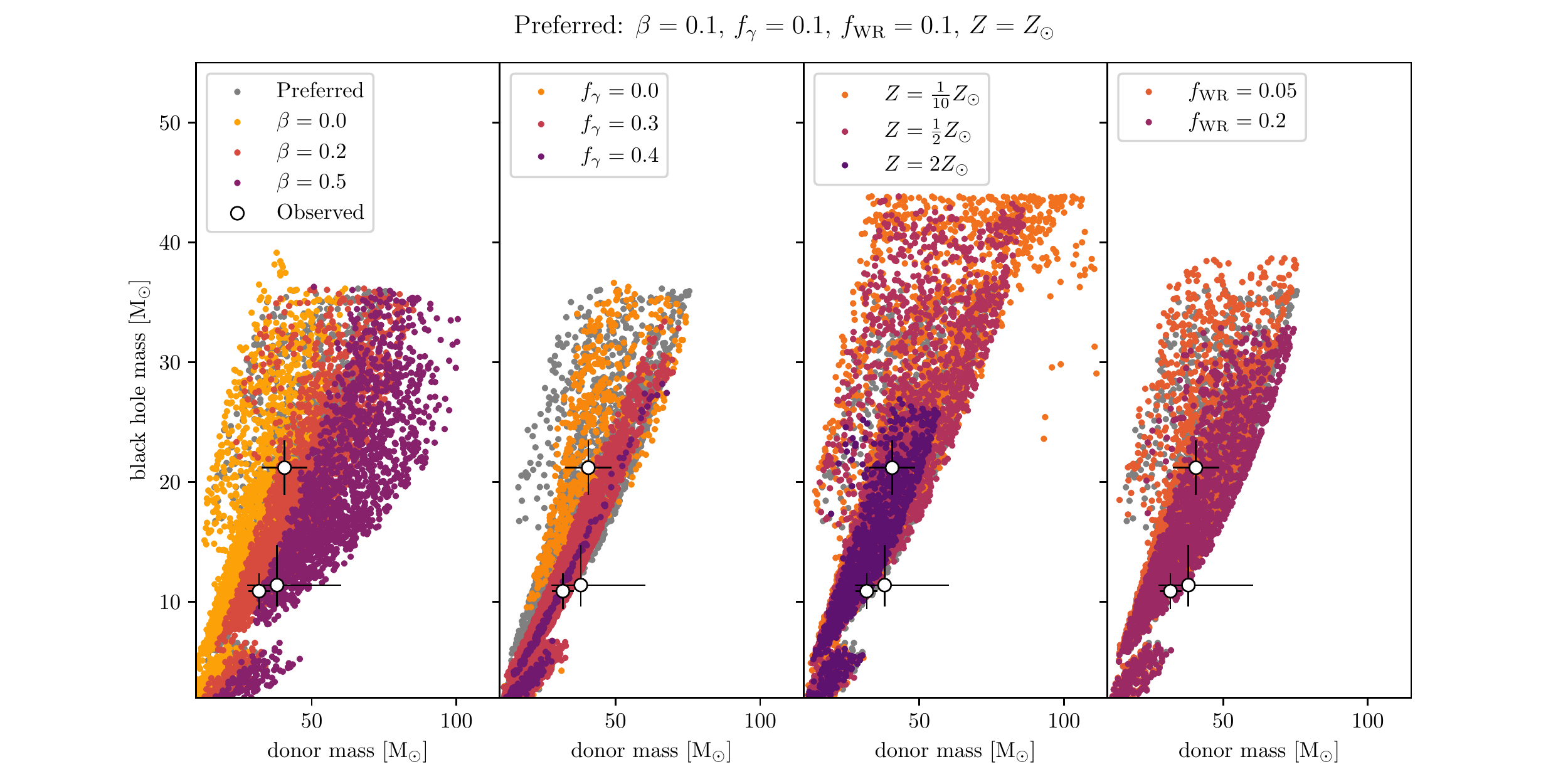}
    \caption{BH mass versus companion mass for BH-HMXB candidates evolving through stable Case A or Case AB MT from the primary. The simulations are split into four panels by the physics that is varied in each run (from left to right: $\beta$, $f_\gamma$, $f_\mathrm{WR}$, $Z$). 
    All unspecified settings match the Preferred settings.
    Each scatter point represents the BH-HMXB at its point of smallest semi-latus rectum. 
   Increasing $\beta$ tends to shift the binary to higher donor masses (and wider separations). Increasing $f_\gamma$ increases the angular momentum losses during non-conservative MT, which reduces the yield of BH-HMXBs. Increasing $f_\mathrm{WR}$ only reduces the maximum BH mass, while increasing $Z$ reduces both the maximum donor and BH masses.}
    \label{fig:mass_scatter_plot_all}
\end{figure*}

\subsection{\new{3.1. Effects of new treatment for core growth during Case A MT}}
\label{sec:core-growth}

\new{We assess the impact of the new Case A MT treatment in Figure \ref{fig:fiducial_vs_default}. The BH-HMXBs formed via the new Case A MT treatment (plotted in dark grey and dark pink) exhibit a wider range of BH masses than those without this change (lighter dots). With the Default settings (pink dots), the maximum BH mass with the new Case A MT treatment reaches $\sim12$~M$_\odot$, while the maximum under the traditional treatment is $\sim5$~M$_\odot$. Meanwhile, the maximum BH-HMXB donor mass shrinks with the new treatment. 
With our Preferred settings, donor masses are lighter than in the Default case because of the reduced MT efficiency (see Section \ref{sec:masses}).} 

\new{We compare the history of one typical example system, evolved with and without this updated prescription for core mass growth, in Figure \ref{fig:detailed_evolution_comparison}. The system starts as a $m_1\sim100$~M$_\odot$, $m_2\sim45$~M$_\odot$ binary with a separation of $\sim60$~R$_\odot$. In both models, Case A MT commences at $\sim2.5$~Myr. In the updated model, the minimum primary core mass $M_{c, \mathrm{min}}$ is at least as large as its total mass at TAMS, so the entire primary becomes a He star at TAMS.  In the other model, a rapid phase of Case B MT occurs instead; after the star is striped to a He star, it has a much lower mass. This phase of rapid mass loss in the no-minimum-donor-core model drives the binary apart, and the system ends up as a wide NS+MS binary, before eventually merging after the secondary evolves onto the Hertszprung gap (HG). In the updated model, the system becomes an observable BH-HMXB. This example highlights the two-fold impact of our new Case A MT treatment that is favourable for creating Cygnus X-1-like systems: increasing the final BH mass and reducing mass loss, leading to tighter orbits.} 

\new{An additional new treatment in this work is identifying observable BH-HMXBs as BH-MS pairs where the MS fills its Roche lobe by more than a critical fraction $f_\mathrm{crit}$, which is between $0.8$ and $0.9$. Applying this criterion does not strongly influence the distribution of system properties shown in Figure \ref{fig:fiducial_vs_default}. The strongest consequence of applying the Roche lobe filling factor condition is to change the duration of the observable BH-HMXB phase for a given binary, which directly impacts the number of BH-HMXBs we would expect to be visible in a Milky Way-like galaxy today. The yield of BH-HMXBs decreases as $f_\mathrm{crit}$ increases because some donors evolve onto the Hertszprung gap (HG) before expanding to fill more than $90\%$ of their Roche lobe. We discuss this further in Section \ref{sec:rates}.} 

\subsection{3.2. Choice of Preferred model}
\begin{figure*}
    \centering
    \includegraphics[width=\textwidth]{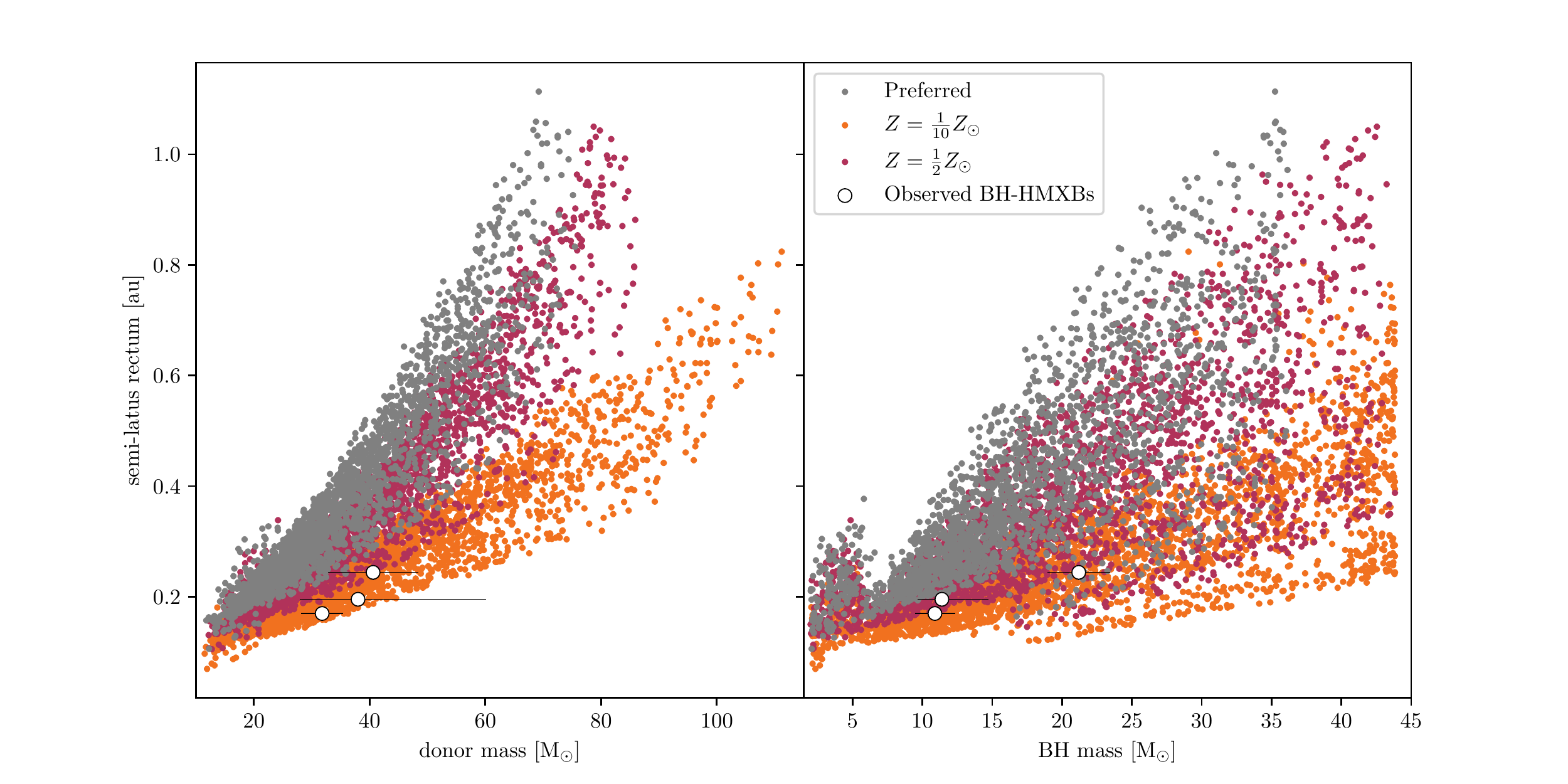}
    \caption{Smallest BH-HMXB semi-latus rectum ($\rho=a(1-e^2)$, where $a$ is the semi-major axis and $e$ is the eccentricity) as a function of donor mass (left-hand panel) and BH mass (right-hand panel) for \new{the three simulations that produce systems with the closest matches to the observed BH-HMXBs.}
    Observed BH-HMXBs are plotted in black, with horizontal lines denoting the uncertainty on their mass and vertical lines (so small as to be unresolvable) showing uncertainty on their separations. 
    Lower metallicities enable tighter and more massive BH-HMXBs with settings otherwise matching our Preferred run.  \new{Alternatively, the tight observed binaries could point to stellar radii being lower than in COMPAS models.}}
    \label{fig:semi-latus-selected}
\end{figure*}

The BH-HMXB populations produced with varying $Z$, $\beta$, $f_\gamma$ and $f_\mathrm{WR}$ are plotted in Figures \ref{fig:fiducial_vs_default} and \ref{fig:mass_scatter_plot_all}. 
Reducing $Z$ results in larger masses for both binary components and reduces their orbital separation in the HMXB phase (see Figure \ref{fig:semi-latus-selected}).
Increasing $\beta$ leads to more massive and more distant stellar companions for systems evolving via stable MT from the primary, since more mass is transferred to the secondary during MT episodes.
MT hardens the binary more efficiently as $f_\gamma$ increases, reducing the yield of HMXBs for larger values of $f_\gamma$ as a greater fraction of binaries merge during MT \citep{Willcox:inprep:fGamma}. The thin line of systems dominating the yield of BH-HMXBs when $f_\gamma=0.4$ are actually a different ZAMS population than those that become BH-HMXBs with lower $f_\gamma$, starting life as generally more widely-separated and equal-mass binaries.
When $f_\gamma\gtrsim0.5$, no BH-HMXBs are formed through stable Case A MT.  
Increasing $f_\mathrm{WR}$ reduces the number of BH-HMXB candidates evolving through Case A MT and decreases the maximum BH mass produced in our simulations.

Our Preferred model settings are $\beta=0.1$, $f_\gamma=0.1$, $f_\mathrm{WR}=0.1$, $Z=Z_\odot$. 
These parameters yield a BH-HMXB population similar in masses and separations to those observed, through a channel likely to give rise to highly-spinning BH primaries.  However, it should be noted that these are not the only parameters that produce binaries consistent with the observed population; for example, as shown in Figure \ref{fig:mass_scatter_plot_all}, several variations on these parameters can produce binaries through Case A MT with component masses observed in BH-HMXBs.

\new{While lower metallicities produce BH-HMXBs with smaller separations more consistent with observations, we use solar metallicity in our Preferred simulation because Cygnus X-1 is thought to have $Z \gtrsim Z_\odot$ \citep[see discussion in][]{Neijssel:2021:WRmassloss}. In all of our simulations, the distribution of binary age at the onset of the BH-HMXB phase is strongly peaked around $5$~Myr, and the typical duration of the phase is short, $\lesssim 0.2$~Myr. Therefore, any BH-HMXBs observable in the local Universe today are young systems whose metallicities are determined by those in local star-forming regions.  Finally, it is likely that separations during the BH-HMXB phase are overpredicted by COMPAS; see Section \ref{sec:separations} for further discussion of this.} 

Figure \ref{fig:evolution-example} shows a typical evolutionary pathway for a BH-HMXB forming via dynamically stable Case A MT \new{in our Preferred simulation}.
At each illustrated step in the evolution, we provide the fraction of binaries that reach that stage in our Preferred model.
In Figure \ref{fig:fiducial_vs_default}, we show how the BH-HMXB masses produced in our Preferred run compare to those produced when we use the COMPAS \new{v02.31.03} Default settings: $\beta = \beta_\mathrm{th}$, $f_\gamma = 0$, $f_\mathrm{WR} = 1.0$, $Z = Z_\odot$.  \new{Unlike the Default model, the Preferred model contains binaries with masses and separations similar to those observed.}

\begin{figure*}
    \centering
    \includegraphics[width=0.75\textwidth]{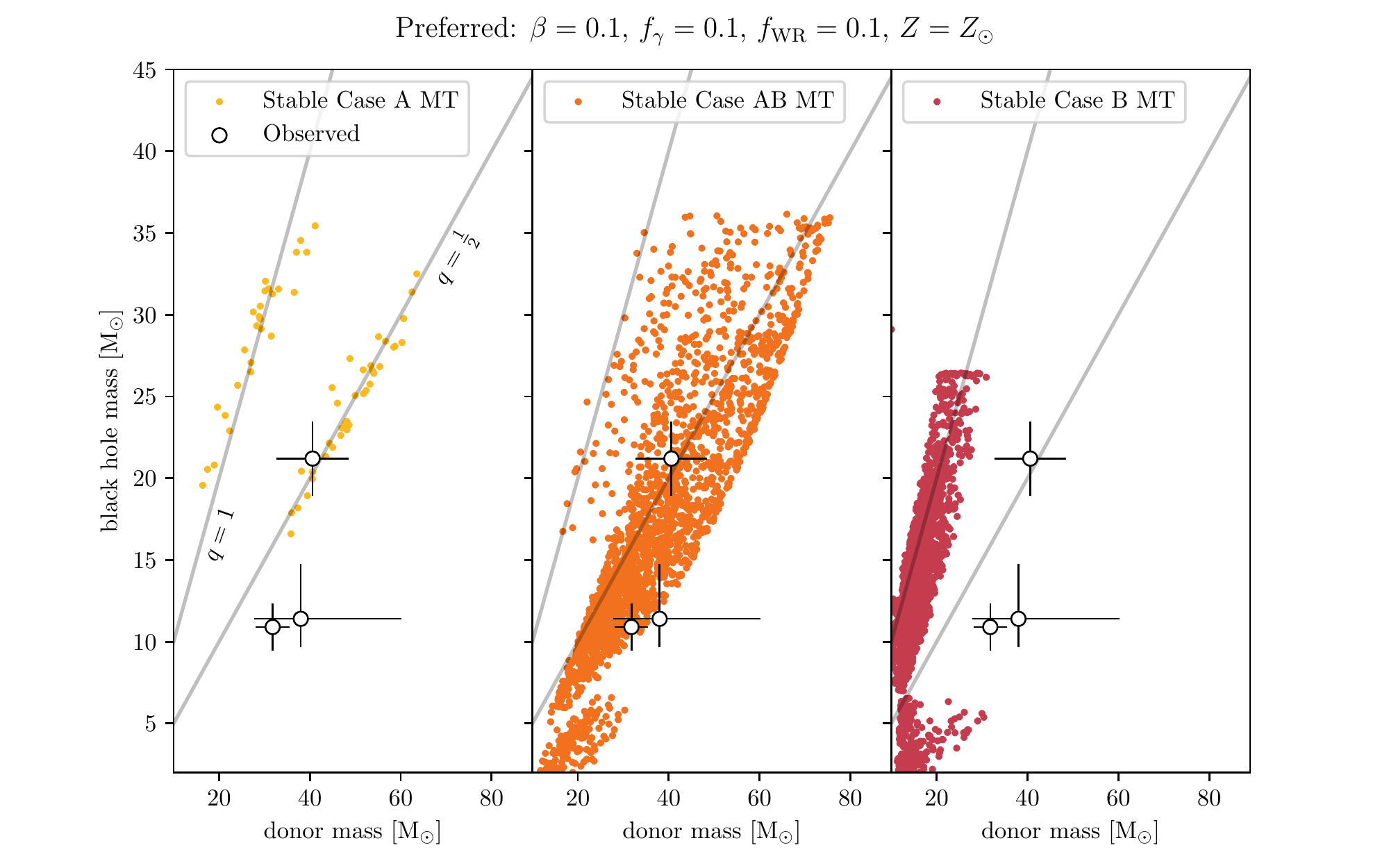}
    \caption{BH versus donor mass for BH-HMXB candidates produced through all evolutionary pathways in our Preferred simulation. We show BH-HMXB candidates that evolve through pure stable Case A MT (left-hand panel), stable Case A followed by stable Case B MT (middle panel), and pure stable Case B MT (right-hand panel). 
    Only the point with the smallest separation (measured at the semi-latus rectum) is plotted. The masses of observed BH-HMXBs are shown with white circles and black error bars.  BH to donor mass ratios $q$ of 1 and 0.5 are indicated with faint grey lines.}
    \label{fig:all_hmxbs_fiducial_model}
\end{figure*}

\subsection{3.3. Masses}
\label{sec:masses}

Cygnus X-1 contains an main sequence wind donor of $\sim40$~M$_\odot$ and a $\sim21$~M$_\odot$ BH accretor \citep{MillerJones:2021:CygX1}. To obtain similar masses, we find that it is crucial to retain a higher core mass at the end of Case A MT than is retained by the default treatment in COMPAS \new{v02.31.03}, as described in Section \ref{sec:method} \new{and illustrated in Figure \ref{fig:fiducial_vs_default}.}
While we form a wide range of BH masses through Case A MT in all simulations, the mass range of main sequence wind donors is sensitive to model assumptions. 
We do not directly restrict mass accretion efficiency for accretors with high rotational velocities, as in \citet{Qin:2019:HMXBspin}; instead, we study the effect of varying the fixed value of $\beta$ between different runs. 
We find that we produce companions in the correct mass range via stable Case A MT from the primary only when $\beta$ is low ($\beta\lesssim0.2$), representing largely non-conservative MT, consistent with expectations for rapidly-spinning accretors.

There is a distinct subpopulation of low-mass BHs visible in the mass distributions in all panels of Figure \ref{fig:mass_scatter_plot_all}.  This is a consequence of the stochastic remnant mass and kick prescription of \citet{MandelMuller:2020:kicks}, as low-mass BHs can be produced through either complete fallback with no kicks from progenitors with lower-mass carbon-oxygen cores or through partial fallback with natal kicks from progenitors with more massive cores. Equivalently, progenitors with the same carbon-oxygen core mass between 3 and 8 M$_\odot$ can give rise to more massive BHs through complete fallback or lower-mass BHs through partial fallback (or even yield neutron stars).

In our Preferred simulation, binaries that evolve through only Case A MT have high-mass primaries at birth ($M_{1, \mathrm{ZAMS}} \gtrsim 70$~M$_\odot$), initial mass ratios $q_\mathrm{ZAMS} = M_{1, \mathrm{ZAMS}} / M_{2, \mathrm{ZAMS}} \approx 2$, and small initial separations $a_\mathrm{ZAMS}\approx0.2$--$0.5$~au.
The primary loses $\approx15\%$ of its ZAMS mass in winds before commencing Case A MT.
This MT episode happens relatively late on the main sequence, and depletes the majority of the primary's envelope (and $\approx 60\%$ of its ZAMS mass), preventing further MT from the primary.
The systems that do undergo a subsequent early Case B MT phase have a wider range of primary masses at birth ($M_{1, \mathrm{ZAMS}} \gtrsim 15$~M$_\odot$), mass ratios $q_\mathrm{ZAMS} \approx 1.25$--$2$, and a slightly larger range of initial separations $a_\mathrm{ZAMS}\approx0.1$--$0.6$~au.

As shown in the leftmost panel of Figure \ref{fig:all_hmxbs_fiducial_model}, BH-HMXB candidates forming through only stable Case A MT, when plotted at their point of smallest semi-latus rectum, follow two distinct trends in mass ratio $q = M_\mathrm{BH} / M_\mathrm{d}$: $q\approx0.5$ and $q\approx1$.
In fact, the systems at $q\approx1$ are going through their second BH-HMXB phase.  
All BH-HMXBs that evolve via only stable Case A MT have their first BH-HMXB phase as the secondary is in the process of expanding to overflow its Roche lobe.
Subsequently, after the secondary loses mass through Roche lobe overflow, it may detach and engage in a second BH-HMXB phase with a donor that has lost $\sim 50\%$ of its mass, just before the donor evolves off the main sequence.
While these binaries going through a second BH-HMXB phase only appear once in Figure \ref{fig:all_hmxbs_fiducial_model}---on the $q=1$ branch, since this is when their point of smallest semi-latus rectum occurs---\new{these binaries previously experienced a BH-HMXB phase on the $q=0.5$ branch.}
Once $q\approx1$, the secondary becomes detached, as expected \citep{Podsiadlowski:2002:XRayBinaryEvolution, Podsiadlowski:2003:CygnusX1}.
However, very massive ($M_\mathrm{ZAMS} \gtrsim 50$~M$_\odot$) main sequence donors may detach from MT before this point: once these stars are stripped, the exposed core is effectively chemically homogeneous, and therefore smaller in size than we are accounting for here.
As a consequence, they detach from MT even though the binary has been driven closer together.
If this is the case, some of the more massive Case A and Case AB binaries close to $q=1$ would be pushed to slightly more extreme mass ratios at their point of smallest semi-latus rectum while in an observable BH-HMXB state.

\subsection{3.4. Separations}
\label{sec:separations}

The orbital period of Cygnus X-1 is $5.6$~days and its near-circular orbit (eccentricity $e \sim 0.02$) has a semi-major axis of $a \sim0.24$~au \citep{MillerJones:2021:CygX1}. 
Here, we take the semi-latus rectum $\rho\equiv a (1-e^2)$ as a proxy for binary separation because this value should be approximately conserved as the binary circularises through tides.

In Figure \ref{fig:semi-latus-selected}, we plot $\rho$ against $M_\mathrm{d}$ for binaries forming through stable Case A or AB MT in our Preferred simulation, as well as \new{simulations with lower metallicity}.  The observed BH-HMXB separations are on the lower edge of those predicted by our \new{Preferred model, and within the range predicted by the simulation with $Z=\frac{1}{10}Z_\odot$}.  The location of \new{the lower edge of the semi-latus rectum distribution} is determined largely by the radial evolution of main sequence donors, \new{which} must fit within their Roche lobes for a given binary separation. 
Since lower-metallicity stars have smaller radii (and also experience reduced wind mass loss, which widens binaries), reducing the metallicity allows for smaller binary separations at the same masses.

\new{However, some simplifying assumptions in the evolution of close binaries in COMPAS are likely to lead to predicted binary separations that are too wide.
Firstly, wind interactions are not modelled in COMPAS. The loss of orbital energy via wind drag on the companion may lead to smaller separations than are currently predicted \citep[see][]{BrookshawTavani:1993,Schroder:2021:RadiationDrivenWinds}.
Secondly, we do not account for tidal dissipation of orbital energy in COMPAS.
The increasing moment of inertia of an expanding, tidally-locked donor extracts angular momentum from the orbit, causing the orbital separation to shrink.
Neglecting this effect leads our current orbital separation predictions to be too large.
For $q = M_\mathrm{BH} / M_\mathrm{d} = 0.5$, the donor has a Roche lobe of $0.44~a$, where $a$ is the separation.  If the donor is $\sim90\%$ Roche lobe filling, $a = 2.5~R_\mathrm{d}$, where $R_\mathrm{d}$ is the radius of the donor. The angular momentum ratio is then $I_\mathrm{d}/I_\mathrm{binary} \sim (M_\mathrm{d} R_\mathrm{d}^2 / 10) / (M_\mathrm{d} a^2 / 3) \sim 0.05$.  A $10\%$ increase in the stellar radius through homologous expansion increases the star's moment of inertia by $\sim20\%$.  If the star's moment of inertia is indeed $\sim5\%$ of that of the binary, this would remove $\sim1\%$ of the binary's angular momentum, causing the binary to spiral in by $\sim2\%$. We therefore estimate that the orbital separation is overpredicted by a few percent due to neglecting tidal driving of the orbit.  Ultimately, however, reducing the separation while avoiding Roche lobe overflow requires decreasing the radius of the optical companion, e.g., through reduced metallicity or corrections to single-star evolutionary models.
}
 
\subsection{3.5. Rates}
\label{sec:rates}

Three confident BH-HMXBs are known observationally, with only one---Cygnus X-1---in the Milky Way.
In order to convert yields from our simulations to observable rates in the Milky Way today, we take into account the time spent by each candidate as an observable HMXB, $T_\mathrm{HMXB}$.
To convert our simulated yield to an observable number $N$ of BH-HMXBs in the Galaxy, we use the conversion function
\begin{equation}
    \label{eq:N}
    N = \sum^n_i{T_\mathrm{HMXB, i}} \times \frac{\dot{M}_\mathrm{new}}{M_\mathrm{rep}},
\end{equation}
where the assumed (constant) Milky Way star formation rate is $\dot{M}_\mathrm{new} = 2$~M$_\odot$ per year, $T_\mathrm{HMXB, i}$ is the observable lifetime of the $i$'th BH-HMXB produced in the simulation, and $M_\mathrm{rep}$  is the total star forming mass represented by a COMPAS simulation, in this case $M_\mathrm{rep} = 2.65 \times 10^8$~M$_\odot$. 
We use this conversion function to compute the expected number of observable Galactic BH-HMXBs from our Preferred simulation with $\beta=0.1$, $f_\gamma=0.1$, $f_\mathrm{WR}=0.1$, $Z=Z_\odot$, and a minimum core mass retained after Case A MT. 

\new{Since the precise value of $f_\mathrm{crit}$ is system-dependent, we quote expected yields of BH-HMXBs using both $f_\mathrm{crit}=0.8$ and $0.9$. 
Changing this value does not strongly influence the masses or separations of binaries.
A lower value of $f_\mathrm{crit}$ naturally leads to a longer observable BH-HMXB phase for the same binary, while a higher value of $f_\mathrm{crit}$ filters out some systems with MS stars that never fill their Roche lobes to this higher fraction.
In many cases, donors evolve to become HG stars before they fill more than $90\%$ of their Roche lobe, so no longer meet our stringent BH-HMXB identification criteria (which require that the donor is a MS star).}
The observable numbers of binaries \new{per Milky Way-like galaxy today} identified as BH-HMXBs with Roche lobe filling factors $\geq 0.8$ ($\geq 0.9$) evolving through stable MT are as follows (for Stable Case A MT only, all donors are $\geq 0.9$ Roche lobe filling):
\begin{itemize}
    \item \textit{Stable Case A MT only}---Median average $T_\mathrm{HMXB} = 0.17$~Myr, $N = 0.07$ BH-HMXBs observable in the Milky Way today.
    \item \textit{Stable Case AB MT}---Median average $T_\mathrm{HMXB} = 0.13$ ($0.17$)~Myr, $N = 2.46$ ($2.01$) BH-HMXBs observable today.
\end{itemize}
Despite the distinction above, we treat BH-HMXBs evolving via Case A only or Case AB MT as effectively coming from the same channel, since the distinction between them is sensitive to slight changes to our prescription for the core vs.~envelope mass following Case A MT.

\new{To assess the influence of employing the \citet{HiraiMandel:2021:AccretionDiskXRays} identification criterion for BH-HMXBs ($f_\mathrm{crit} \approx 0.8$), we compare these observable yields to those we would expect for $f_\mathrm{crit}=0$.
In this case, we effectively designate as BH-HMXBs all BH-MS binaries with $m_2 \geq 10$~M$_\odot$ which evolved through Case A MT.
When we reduce $f_\mathrm{crit}$ to $0$, BH-HMXBs have a median average observable lifetime of $\approx0.9$~Myr.
Folding this into Equation \ref{eq:N}, we would expect to see $\approx137$ observable BH-HMXBs per Milky Way-like galaxy today.
This yield is obviously far too high given the low observed rate of BH-HMXBs in the Galaxy, even when accounting for the selection effects that we consider below. 
Clearly, the higher $f_\mathrm{crit}$ calculated by \citet{HiraiMandel:2021:AccretionDiskXRays} -- based on the requirement of having sufficient angular momentum in the accreted material to form a disk --  is more in line with observations.}

\begin{figure*}
    \centering
    \includegraphics[width=0.8\textwidth]{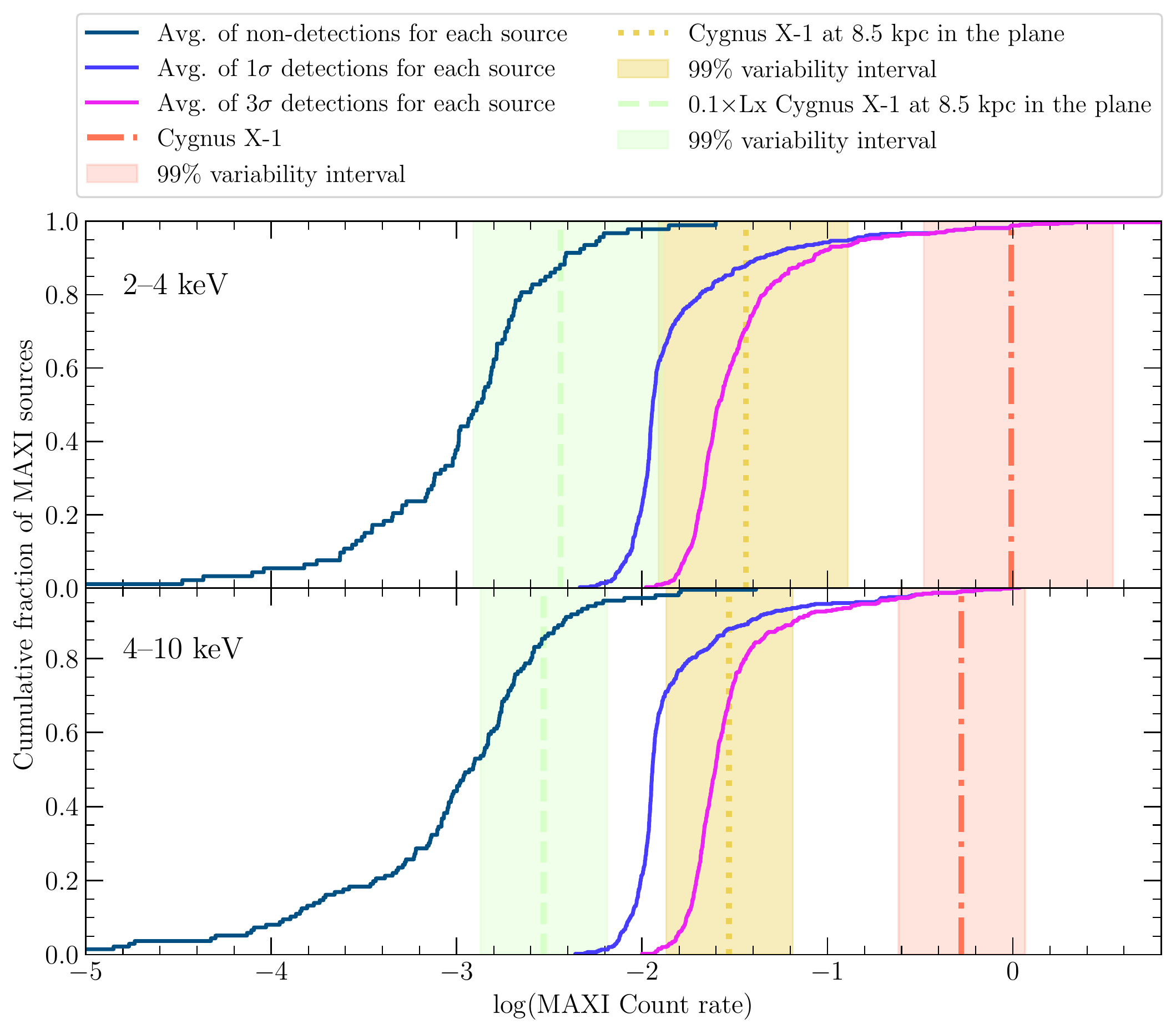}
    \caption{Observability of an XRB as bright as Cygnus X-1 in the X-rays detected by \textit{MAXI}/GSC. Dark blue, light blue and pink solid lines represent the cumulative distribution of weighted averages of \textit{MAXI} daily count rates (averaged per source) for non-detections, 1-$\sigma$, and 3-$\sigma$ detections respectively for all \new{X-ray sources recorded by} \textit{MAXI} \new{since its science phase began in August 2009}. These distributions map approximate \textit{MAXI} sensitivity to its observed count rates: for example, a source with a count rate of 0.01 (in either 2--4 or 4--10 keV bands) has never been detected at 3-$\sigma$. \new{The vertical lines represent Cygnus X-1's observed absorbed count rate under different assumptions, while the shaded intervals represent the range of variability that Cygnus X-1 shows in brightness (day-to-day/week-to-week).} The vertical orange dash-dotted line and the faded orange band represent the \new{true} average count rate of Cygnus X-1 and the extent of the source variability, respectively, as observed by \textit{MAXI} \new{over the last 14 years}. Vertical yellow dotted line and the faded yellow band represent the average and the variable count rate of Cygnus X-1 placed $8.5$~kpc from Earth and in the Galactic plane (assuming a hydrogen column density of $5\times10^{22}$~cm$^{-2}$). The vertical light green dashed line and the faded green band represent the average and the variable count rate of a source $10\%$ as luminous as Cygnus X-1, at $8.5$~kpc away in the Galactic plane. \new{This last example is included because many BH-HMXBs are likely to be at lower luminosities than Cygnus X-1, so can be easily missed by current all-sky monitors.}
    }
    \label{fig:maxi_counts}
\end{figure*}

\new{
We estimate the selection effects for BH-HMXBs in the Galaxy as follows. We consider an X-ray source with the same X-ray spectrum and luminosity as Cygnus X-1, as observed by the Monitor of All-sky X-ray Image \citep[\textit{MAXI};][]{MAXI2009}. For this purpose, we first converted \textit{MAXI}'s recorded daily count rates from August 2009 to February 2023 for Cygnus X-1 (to account for variability of Cygnus X-1) to X-ray luminosity using an absorbed power-law spectral model with a photon index of 2.2 and a hydrogen column density of $6\times10^{21}$~cm$^{-2}$ (based on the modelling of the X-ray spectrum of Cygnus X-1 by, e.g., \citealt{Schulz2002}), then transformed the X-ray luminosities back to \textit{MAXI} count rates, using the \textsc{pimms} \citep{Mukai1993} software (which considers detector sensitivities), assuming a distance of 8.5~kpc (the distance to the Galactic bulge/centre, where an overabundance of X-ray binaries is expected), and a conservative hydrogen column density corresponding to that distance at Galactic mid-plane (based on estimates of extinction and hydrogen column density by, e.g., \citealt {HI4PI2016, Green2019}).
}
The effect of this on the X-ray observability in the $2$-$4$~keV and $4$-$10$~keV bands is shown in Figure \ref{fig:maxi_counts}.

While Figure \ref{fig:maxi_counts} indicates that an unfortunately-placed Cygnus X-1 would still be detected by \textit{MAXI} most of the time, this may be over-optimistic.
The X-ray background is significantly elevated in the Galactic bulge, both due to extended X-ray emission and crowding, which particularly impacts all-sky monitors with large point-spread functions \citep[e.g., FWHM of $\sim1.5^\circ$ for \textit{MAXI}/GSC;][]{Hori2018}. An unfortunately-placed Cygnus X-1 will, at best, be a relatively faintly-detected \textit{MAXI} source, which may or may not be identified as an XRB immediately. 
Since detection of optical and UV emission would be hampered by the interstellar medium, mass and distance measurements would be difficult and have large uncertainties, making such a source hard to identify as a BH-HMXB.
If the source was a factor of $10$ less luminous than Cygnus X-1 (as is common in XRBs outbursting in the hard state), it would be completely undetectable by \textit{MAXI} at 8.5 kpc, although it could be picked up by hard-band monitors such as \textit{Swift}/BAT \citep{Barthelmy2005}, or pointed surveys with \textit{Swift}/XRT \citep{Burrows:2005:Swift}. Given these results, we may expect that $\mathcal{O}(50\%)$ of Cygnus X-1-like systems in the Galaxy would be detected and confidently identified as BH-HMXBs.  Therefore, the total of $N \sim 2$ predicted Galactic BH-HMXBs from the stable Case A only and Case AB channels is a good match to one observed Galactic BH-HMXB, Cygnus X-1.  

The observable duration ($\mathcal{O}(0.1)$~Myr) of BH-HMXBs formed through stable Case A and AB MT in our Preferred simulation agrees with observational constraints \citep[e.g.,][]{Mineo:2012:XRayBinaryDuration}.
We note that, since we do not account for tidal driving of orbital evolution, these timescales may be slightly overestimated: the increasing moment of inertia of the expanding, tidally locked donor will extract angular momentum from the orbit, causing the orbital separation to shrink and hastening Roche lobe overflow by $\mathcal{O}(10\%)$. \new{Moreover, rapid rotation induced by tides will impact both the evolutionary timescale and the radius of the optical companion, further complicating more precise estimates of the duration of the BH-HMXB phase.}

\section{4. Formation of merging compact-object binaries from BH-HMXBs}
\label{sec:mergers}

\begin{figure*}
    \centering
    \includegraphics[width=0.8\textwidth]{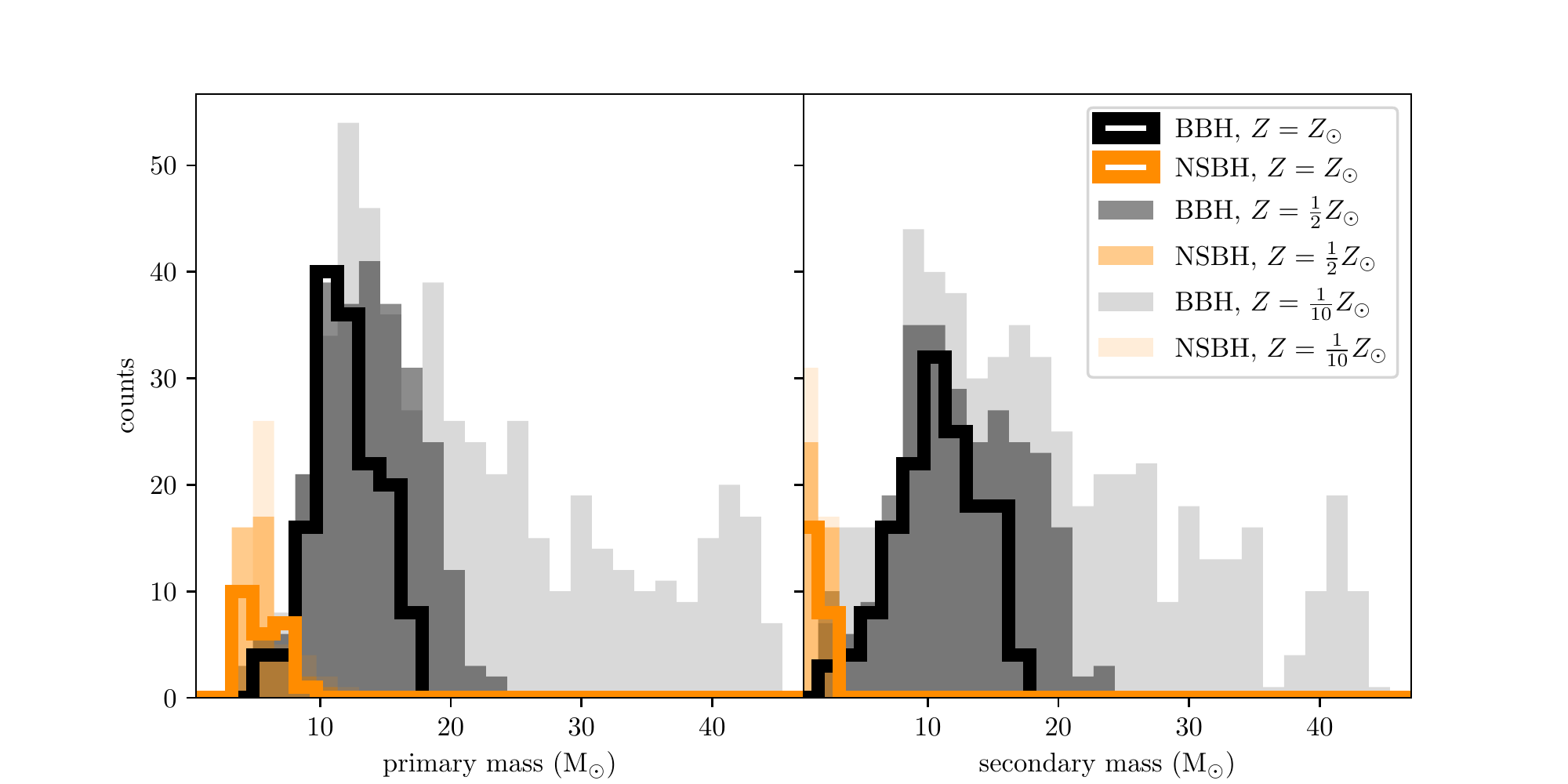}
    \caption{Histograms of primary (left) and secondary (right) masses of the merging BBH and NSBH that evolve from BH-HMXBs formed in the Preferred simulation and its \new{lower-metallicity variants}. All masses represented in the left-hand panel are BHs, while those plotted in orange in the right-hand panel are NSs.}
    \label{fig:merger_masses}
\end{figure*}

A non-negligible \new{fraction} of BH-HMXBs forming through stable Case AB MT proceed to merge as BBHs or neutron star-BHs (NSBHs) within a Hubble time in our simulations.
In our Preferred simulation, the \new{fraction} of BH-HMXBs evolving through Case AB MT that go on to merge as either BBHs or NSBHs is $8\%$.
\new{When the metallicity is reduced to $\frac{1}{2}~Z_\odot$, the fraction rises to $12\%$.}
\new{At the lowest metallicity we consider, $Z=\frac{1}{10}~Z_\odot$, $21\%$ of BH-HMXBs formed through Case AB MT yield BBH or NSBH mergers within a Hubble time.}
In the Preferred simulation, $1.1\%$ of BH-HMXBs become NSBH mergers and $7\%$ become BBH mergers\new{; in the $Z=\frac{1}{2}~Z_\odot$ simulation the respective fractions are $1.5\%$ and $10\%$, and when $Z=\frac{1}{10}~Z_\odot$ the respective fractions are $1.7\%$ and $19\%$}. 
In Figure \ref{fig:merger_masses}, we plot the distribution of primary and secondary masses for BBH and NSBH mergers \new{subsequent to a BH-HMXB phase} in the Preferred simulation \new{and simulations with lower metallicity.}
\new{When $Z =Z_\odot$}, the merger masses are lower than the (selection-biased) peak of gravitational-wave observations, with both $M_1$ and $M_2\lesssim 18$~M$_\odot$ in our Preferred simulation; the upper limit of the mass range increases to $25$~M$_\odot$ for $Z=\frac{1}{2}~Z_\odot$.  
\new{When $Z=\frac{1}{10}Z_\odot$, the range of masses is more consistent with GW observations of merging BBHs and NSBHs.}

\new{Binary compact mergers forming in isolation are expected to have close-to-aligned spins that are nearly aligned with the orbital angular momentum, and negligible orbital eccentricity at detection \citep[e.g.,][]{Mapelli:2020:Review}. 
However, a} natal kick associated with core collapse supernova of the secondary can tilt the binary's orbit, misaligning it with the spin axis of the first-born BH.  This will cause the BH spin and orbital plane to precess around the total angular momentum vector, leaving a characteristic imprint on the gravitational-wave signal of the merging binary \citep{Apostolatos:1994:Precession}.  Our simulations include a few NSBH systems with spin-orbit misalignments up to $\sim 100$ degrees, typically when the secondary produces an NS which receives a significant natal kick. Supernova natal kicks can also impart a significant eccentricity, $e \gtrsim 0.8$ in some simulated binaries.  However, the eccentricity imparted by the second supernova will be damped to undetectable levels by the time the binary's gravitational-wave emission is observed by ground-based instruments \citep{Lower:2018:Eccentricity, Romero-Shaw:2019:SearchingForEccentricity}.



The Preferred model yields Galactic BBH (NSBH) merger rates of previously observable BH-HMXBs of $1.8$ ($0.2$) per Myr, assuming a Milky Way star formation rate of $2
$~M$_\odot$ yr$^{-1}$.  Without accounting for metallicity variations, and assuming a Milky Way equivalent galaxy density of $\sim 10^7$ per Gpc$^3$ \citep{Abadie:2010:Rates}, this corresponds to local merger rates of $\sim 18$ BBH~Gpc$^{-3}$~yr$^{-1}$ and $\sim 2$ NSBH~Gpc$^{-3}$~yr$^{-1}$. 

\new{In fact, the stellar progenitors of the mergers we observe today would have been born with a range of metallicities, mostly lower than in our Preferred simulation.
We perform another simulation of $0.5$ million binaries with $\beta=0.1$, $f_\gamma=0.1$, $f_\mathrm{WR}=0.1$, and a log-uniform distribution of metallicities, ranging between the COMPAS default minimum and maximum of $0.007 Z_\odot$ and $2.11 Z_\odot$.
We then compute the observable merger rate of BBH and NSBH by convolving the COMPAS models with a metallicity-specific star formation rate (MSSFR), which we choose to be the preferred model described by \citet{Neijssel:2019:MSSFR}.
We use detector sensitivities consistent with the third observing run (O3), set a single-detector signal-to-noise ratio threshold to $8$ as a proxy for detectability by a network and set a hard detection limit at a horizon redshift of $z=2$.  Without filtering for binaries that experienced a BH-HMXB phase, we estimate that we would expect to observe $330$ BBHs per year or $6.3$ per week of continuous observing at O3 sensitivity.  If filtering for binaries that experienced a BH-HMXB phase we would expect to detect $26$ BBH mergers per year ($0.5$ per week), assuming continuous observing with a 100\% duty cycle.}

\new{These detection rates are clearly over-predictons: the actual rate of BBH observations per observing week was $\sim1.82$ in the first half of O3, which ran for six months with two or more detectors observing $81.9\%$ of the time \citep{GWTC-2}, and $\sim1.95$ detections per week in the second half of O3, which ran for a little under five months with two or more detectors observing $85.3\%$ of the time.  These over-estimates could be due to a broad range of uncertainties in the treatment of binary evolution, including chemically homogeneous evolution, common-envelope modelling and natal kick prescriptions, as well as in the low-metallicity end of the metallicity-specific star formation rate \citep[e.g.,][]{Neijssel:2019:MSSFR, Neijssel:2021:WRmassloss,StevensonClarke:2022:COMPASConstraints,vanSon:2022}.}

\new{Previous BH-HMXBs form $7.9\%$ of the expected COMPAS BBH yield.  If we assume that causes of overprediction influence the population uniformly, and that all BBH observed by the LVK form from field binaries rather than, e.g., through dynamical interactions in densely-populated environments or field triples, then our estimate means that $7.9\%$ of the BBH mergers observed by the LVK should come from binaries that previously underwent a BH-HMXB phase. 
Since there are $69$ BBH mergers in GWTC-3 with a false alarm rate lower than $1$~yr$^{-1}$~\citep{LVK:2022:GWTC-3}, we estimate that $\approx5$ of these may have come from previous BH-HMXBs, if we take $f_\mathrm{crit}=0.8$.  Using the higher $f_\mathrm{crit}$ of $0.9$, the fraction of BBHs from previously observable BH-HMXBs decreases to $3.4\%$, contributing $\approx2$ LVK detections.}


\new{In order to predict the final fate of Cygnus X-1, we consider only the BH-HMXBs with masses inside the uncertainty range of Cygnus X-1 in our Preferred simulation.
None of the $210$ sampled binaries that fall into this range become compact object mergers, consistent with previous works.
Instead of merging as double compact objects, these binaries either merge following the BH-HMXB phase before the companion collapses into a black hole, or end their lives as as wide BH-BH binaries.
We note that our estimates are sensitive to models of the evolution of these binaries after the BH-HMXB phase, including assumptions about the natal kicks received by the secondary.  
For example, \citet{Neijssel:2021:WRmassloss} found that Cygnus X-1 only has a $0.1\%$ probability of merging as an NSBH unless higher core mass at the end of Case A MT is taken into account, in which case it forms a BBH that is too wide to merge without natal kicks, but will merge $\sim 4\%$ of the time if the \citet{FryerKicks} BH kick prescription is used. 
We discuss further differences between our predictions and those of other works in Section \ref{sec:discussion}.}

\section{5. Alternative evolutionary recipes for BH-HMXB candidates}
\label{sec:pathways}

Our Preferred model simulation produces many BH-HMXB candidates via stable Case B MT from the primary to the secondary in addition to the systems that evolve via stable Case A or AB MT.
BH-HMXB candidates evolving through Case B MT have primaries that only expand sufficiently to instigate Roche lobe overflow after core hydrogen depletion.  These binaries start at wider separations than those that go through Case A MT, with $a_\mathrm{ZAMS} \approx 0.2$--$4$~au, and typical initial masses  $M_{1, \mathrm{ZAMS}}\lesssim 110$~M$_\odot$ and $M_{2, \mathrm{ZAMS}}\lesssim 30$~M$_\odot$.  These binaries require significant hardening from MT in order to become BH-HMXBs, so $q_\mathrm{ZAMS}$ is typically greater than 2.

In our Preferred simulation, the predicted rate of BH-HMXBs identified with the Roche lobe filling factor threshold $f_\mathrm{crit} \geq 0.8$ ($\geq 0.9$)  is:

\begin{itemize}
    \item \textit{Stable Case B MT only}---Median average $T_\mathrm{HMXB} = 0.91$ ($1.09$)~Myr, $N = 20.77$ ($20.18$) BH-HMXBs observable today.
\end{itemize}

When a BH-HMXB evolves through stable Case B MT only, we expect the BH to be non-spinning: since the compact core is likely decoupled from the extended envelope at the onset of MT, there is no spin-up of the BH progenitor through tidal locking. The HMXB observability criterion of \citet{HiraiMandel:2021:AccretionDiskXRays} slightly increases the value of $f_\mathrm{crit}$ when the BH is not spinning (see Figure \ref{fig:ryo-prescription-nospin} in the Appendix).  Furthermore, while an accretion disk can be formed around a non-spinning BH, the X-ray luminosity can be up to a factor of 6 lower than for prograde accretion onto a maximally spinning BH due to the larger radius of the innermost stable circular orbit in the non-spinning case.   Moreover, BH-HMXB candidates forming through purely stable Case B MT typically have lower donor masses than those formed through Case A MT.  These will have lower wind mass loss rates, reducing the observability of BH-HMXBs evolving through Case B MT.

Nevertheless, the predicted yield from the Case B channel is far higher than expected given the single observed BH-HMXB in the Galaxy and the overall lack of observations of HMXBs with non-spinning BHs.  Additionally, as shown in Figure \ref{fig:all_hmxbs_fiducial_model}, the donor masses predicted from this channel are at odds with the observed BH-HMXBs.  This suggests that Case B MT rarely produces BH-HMXB systems after all.  

One possible explanation for the overproduction of Case B BH-HMXBs in our simulations is the treatment of MT efficiency.  While we assumed a fixed MT efficiency of $\beta=0.1$ for all stable MT in the Preferred model, MT efficiency is very likely different between Case A and Case B MT. 
During Case B MT, the accretor fills a smaller fraction of its Roche lobe than it does during Case A MT, and hence the angular momentum of the accretion stream can lead to the formation of an accretion disk instead of directly impacting the surface of the accretor.  The disk can cool the accreting matter through viscous dissipation, possibly reducing the entropy of accreted material compared to direct accretion which can only cool through shocks.  

There is also more space around the accretor to expand into for Case B MT. When the size of the Roche lobe around the accretor is relatively small, as in typical Case A MT, the accretor can quickly overflow its own Roche lobe, spilling the remaining transferred matter out of the system. With more space, the star can delay overfilling its Roche lobe or avoid it altogether.  We speculate, therefore, that stable Case B MT could have a higher $\beta$ value than in stable Case A MT.

The requirement for a higher mass retention rate for Case B MT is also in line with findings based on the mass distribution of Be X-ray binary donors \citep{Vinciguerra:2020:MTEfficiency}.
A COMPAS simulation with $\beta=1.0$ ($0.5$) produces $\approx0.6$ ($2$) observable BH-HMXB candidates in the Milky Way today that evolved through stable Case B MT, using an $f_\mathrm{crit}$ of $0.8$. 
This is a significant reduction compared to the $\approx20$ systems when $\beta=0.1$, and more consistent with the lack of observed HMXBs with non-spinning BH primaries. 

We note that highly efficient accretion during stable Case B MT is unlikely to change our predicted yields of BH-HMXB candidates that evolve via Case AB MT: since our argument depends on the Roche lobe filling fraction of the accretor, early Case B MT likely proceeds similarly to Case A MT in this scenario.

We also produce BH-HMXB candidates through common envelope (CE) evolution in our Preferred model. However, since CE prescriptions in population synthesis are notoriously uncertain, we treat the predictions of this channel with extreme scepticism.  The two-stage CE model of \citet{HiraiMandel:2022:CommonEnvelope} suggests that the CE channel would lead to wider separations than observed BH-HMXBs; it is also likely to lead to non-spinning BH primaries \citep[see][and references therein]{Bavera:2020:BHSpinsCE}.  We check our intuition by changing the value of $\alpha_\mathrm{CE}$ in COMPAS from the default value ($1$ \new{in COMPAS v02.31.03}) to a value ($50$) more in line with the prescription of \citet{HiraiMandel:2022:CommonEnvelope} for high-mass binaries with comparable mass ratios.
This drastically reduces the expected observable yield of BH-HMXBs from CE evolution to $< 0.6$ per Milky Way-like galaxy.

\section{6. Discussion and Conclusion}
\label{sec:discussion}
We carry out a binary population synthesis study of the formation of wind-fed BH-HMXBs similar to Cygnus X-1.
We explore the impact of the MT efficiency $\beta$, the Wolf-Rayet wind mass loss rate \new{parametrized with} $f_\mathrm{WR}$, the specific angular momentum loss \new{parametrized with} $f_\gamma$, and the metallicity $Z$ on the yield of BH-HMXBs. 
\new{We find that accounting for the build-up of helium during the main sequence prior to mass transfer, thereby increasing the core mass retained by donors following Case A MT, is essential for producing massive BHs in BH-HMXBs.
With this modification, we find that we are able to produce binaries with masses, spins, and separations like those of Cygnus X-1 when our simulation involves significantly non-conservative MT, low specific angular momentum loss, and low Wolf-Rayet wind mass loss}.
\new{Employing the \citet{HiraiMandel:2021:AccretionDiskXRays} criterion for identifying observable BH-HMXBs as BH-MS pairs with Roche lobe filling factors $f_\mathrm{crit} \gtrsim 0.8$ leads us to expect $\sim1-2$ observable BH-HMXBs per Milky Way-like galaxy for our Preferred model, consistent with observed rates.}
Our Preferred model for producing BH-HMXBs via dynamically stable Case A MT at a rate that is consistent with observations \new{has} MT efficiency $\beta=0.1$, specific angular momentum loss $f_\gamma=0.1$, Wolf-Rayet wind mass loss rate $f_\mathrm{WR}=0.1$, and metallicity $Z=Z_\odot$.
We find that most BH-HMXBs undergo both Case A and early Case B MT from the BH progenitor, although the fraction of BH-HMXBs that evolved through Case AB MT is sensitive to our treatment of Case A MT.

\new{The expected yield of high-mass BH-MS pairs per Milky Way-like galaxy that have rapidly-spinning BH primaries but are \textit{not} observable as BH-HMXBs is about $\approx135$ in our Preferred simulation (see Section \ref{sec:rates}). There are plenty of observations of binaries that may fit into this category. Binaries consisting of likely BHs with probable MS companions have been identified via both spectral studies \citep[e.g.,][]{Thompson:2019:BH-MS, Thompson:2020:BH-MS, Price-Whelan:2020:BH-MS, Tomer:2022:BH-MS, Shenar:2022:BH-MS, Mahy:2022:HD130298, Saracino:2022:BH-MS, El-BadryBurge:2022:BH-MS, Saracino:2023:BH-MS} and astrometrically \citep[e.g.,][]{Shion:2022:BH-MS, Andrews:2022:BH-MS, El-Badry:2022:BH-MS, Tanikawa:2023:BH-MS, El-Badry:2023A:BH-MS}. 
The companion star has not been confidently identified as MS in all of these pairs, and any binaries within globular clusters \citep[e.g.,][]{Giesers:2018:BH-MS, Giesers:2019:BH-MS} may have distinctly different evolutionary histories than the isolated field evolution simulated with COMPAS. Nonetheless, some of these systems could be dormant BH-MS pairs that evolved through Case A MT, which would be visible as BH-HMXBs were it not for the Roche lobe filling criterion identified by \citet{HiraiMandel:2021:AccretionDiskXRays}.}

\new{In our Preferred simulation, $8\%$ of BH-HMXBs end their lives as BBH or NSBH mergers. Using a log-uniform distribution of metallicities and employing the \texttt{CosmicIntegrator} class in COMPAS, we predict that $\approx2$--$5$ of the BBH detections made by the LVK in O3 should come from binaries that were previously BH-HMXBs.
Comparing these results to other previous work is not simple, since there are many differences between our simulations and these studies.
Qualitatively, our results are similar to those of \citet{Liotine:2022:SelectionEffects}, \citet{GallegosGarcia:2022:HMXBs} and \citet{Neijssel:2021:WRmassloss}: all find that BH-HMXBs rarely lead to observable merging BBHs.}  

There are discrepancies between the predicted number of likely non-spinning BH-HMXBs evolving through purely stable Case B MT or CE in our model and the observations.
There are several possible reasons for the overproduction of BH-HMXBs through these channels.  We \new{suggest in Section \ref{sec:pathways}} that mass transfer efficiency may be higher in Case B MT than in Case A MT for massive stars\new{, while CEs will leave behind very wide binaries over the relevant parameter space \citep{HiraiMandel:2022:CommonEnvelope}.  If these two proposals hold, then rates of BH-HMXBs produced via these channels are suppressed close to zero.}

Low-luminosity BH-HMXB candidates have been discovered in radial velocity surveys of bright stars; HD96670 is one such example of a Galactic BH-HMXB candidate $\sim10^6$ times fainter than Cygnus X-1, with a similar orbital period \citep{GomezGrindlay:2021:HD96670} and masses ($M_d \sim 23$~M$_\odot$, $M_\mathrm{BH} \sim 6$~M$_\odot$) consistent with the BH-HMXBs evolving via Case AB MT in our simulations.
There are three additional unconfirmed candidates for BH-HMXBs in the Galaxy: SS 433 \citep{Seifina:2010:SS433}, Cygnus X-3 \citep{Zdziarski:2018:CygnusX3} and MWC 656 \citep{Grudzinska:2015:MWC656}.  HD 130298 is  a non-mass transferring Galactic binary whose components are strongly suggested to be a main sequence star and a BH, and is therefore a possible BH-HMXB progenitor \citep{Mahy:2022:HD130298}.  Our estimates for the number of in-principle observable BH-HMXBs in the Milky Way, with only a fraction of these confidently identifiable depending on their location, are consistent with these observations.

\new{\citet{Lehmer:2021:HMXB-L-Z-relation} find that the number of high-luminosity HMXBs increases with decreasing metallicity, while the number of lower-luminosity ($L<10^{38}$~erg~s$^{-1}$) sources is insensitive to metallicity.
While we explore only BH-HMXBs, and do not model luminosities, our results are broadly consistent with a trend of weakly increasing BH-HMXB abundance with decreasing metallicity.
Without accounting for selection effects, we predict $\approx2$ BH-HMXBs per Milky Way-like galaxy at $Z=Z_\odot$, $\approx3$ at $Z=\frac{1}{2}Z_\odot$, and $\approx4$ at $Z=\frac{1}{10}Z_\odot$.}

The collapse of rapidly-spinning stellar cores to BH has been proposed as a progenitor for long gamma-ray bursts \citep[LGRBs;][]{Woosley:1993:GRBs, Paczynski:1998:GRBs, vandenHeuvel:2007:LGRBs, Detmers:2008:GRBs, Chrimes:2020:LGRBs, Bavera:2022:LGRB}.
If this is the case, then the observed BH-HMXBs may have powered LGRBs when the BH formed.
Our Preferred model produces $\sim 2$ BH-HMXBs with a rapidly-spinning BH in the Milky Way, observable for a median lifetime of $\sim 0.1$~Myr.  Assuming a beaming fraction of $f_\mathrm{B}=0.05$, our model yields a LGRB rate from BH-HMXB formation of $\mathcal{O}(10)$~Gpc$^{-3}$~yr$^{-1}$ at $z=0$, far above the observed rate of less than $0.6$~Gpc$^{-3}$~yr$^{-1}$.  
This implies that the collapse of rapidly-rotating primaries into the spinning BH seen in BH-HMXBs is not responsible for the observed set of LGRBs, unless the beaming fraction is an order of magnitude lower than assumed or unless only a fraction of rapidly spinning BHs power a LGRB. This is consistent with the fact that Cygnus X-1 has a low eccentricity ($e\sim0.02$) and a low velocity relative to its host association ($\lesssim 10$ km s$^{-1}$), implying that little to no mass loss occurred upon the collapse to a BH.  On the other hand, estimates of beaming angles range broadly across GRBs, with some beaming fractions possibly being as low as $f_\mathrm{B} \approx 0.001$  \citep[e.g.,][]{Goldstein:2016}.\\

\section{7. Acknowledgements}
We thank \new{Monica Gallegos-Garcia,} Camille Liotine, Pablo Marchant, and Jeff Riley for useful discussions.
\new{We also thank our anonymous referee for their comments, which improved the manuscript.}
IMR-S acknowledges support received from the Herchel Smith Postdoctoral Fellowship Fund.
\new{RH, RW and IM} acknowledge support from the Australian Research Council Centre of Excellence for Gravitational  Wave  Discovery  (OzGrav), through project number CE17010004.  IM is a recipient of the Australian Research Council Future Fellowship FT190100574. 

\section{8. Data Availability}
Any data used in this article is available upon reasonable request submitted to the corresponding author. 

\appendix

\section{Non-spinning black hole observability criteria}

\begin{figure}
    \centering
    \includegraphics[width=0.48\textwidth]{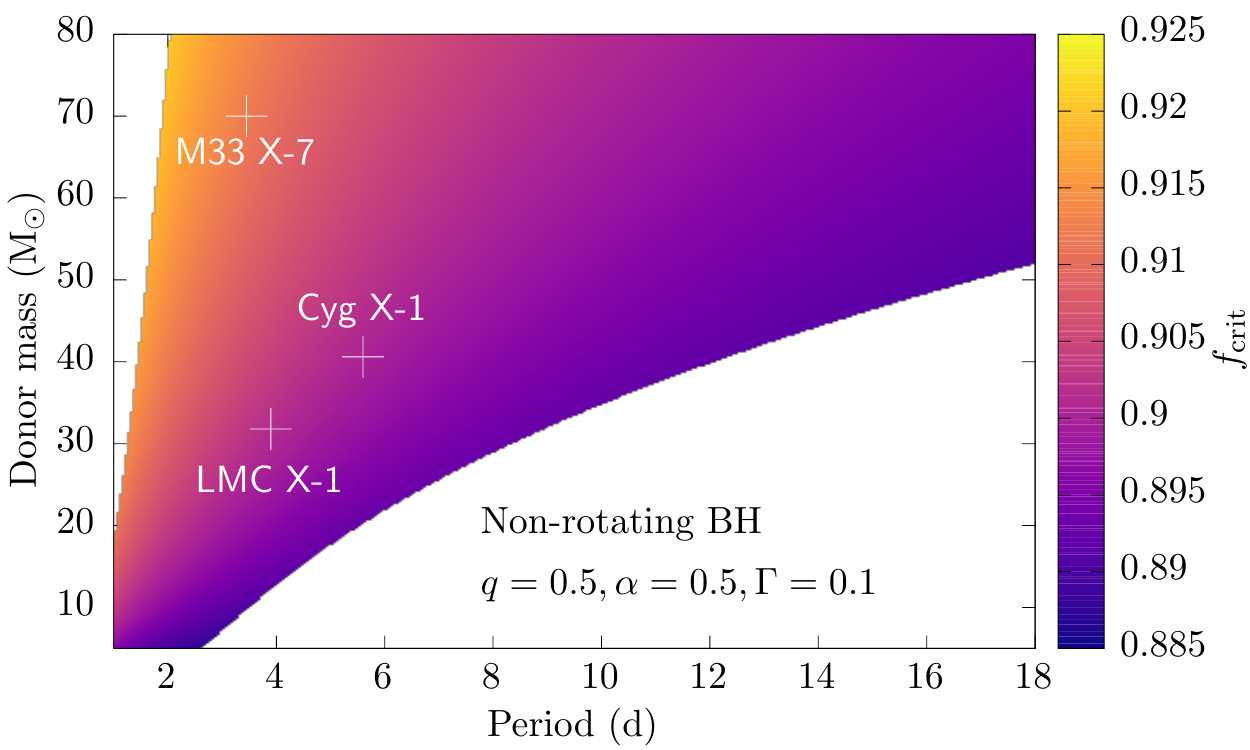}
    \caption{The critical Roche lobe filling factor to create an accretion disk as a function of orbital period and donor mass as predicted by the model of \citet{HiraiMandel:2021:AccretionDiskXRays} for a non-spinning BH; otherwise, the notation is the same as in Fig. \ref{fig:ryo-prescription}.}
    \label{fig:ryo-prescription-nospin}
\end{figure}

In Figure \ref{fig:ryo-prescription-nospin}, we show the equivalent of Figure \ref{fig:ryo-prescription} for identical binary parameters but with a non-spinning BH accretor. Although the radius and specific angular momentum of the innermost stable circular orbit are larger in this case, requiring a larger $f_\mathrm{crit}$, it is possible to produce accretion disks around even non-rotating BH using the prescription of \citet{HiraiMandel:2021:AccretionDiskXRays}.

\bibliography{bib}

\begin{thebibliography}{}
\expandafter\ifx\csname natexlab\endcsname\relax\def\natexlab#1{#1}\fi
\providecommand{\url}[1]{\href{#1}{#1}}
\providecommand{\dodoi}[1]{doi:~\href{http://doi.org/#1}{\nolinkurl{#1}}}
\providecommand{\doeprint}[1]{\href{http://ascl.net/#1}{\nolinkurl{http://ascl.net/#1}}}
\providecommand{\doarXiv}[1]{\href{https://arxiv.org/abs/#1}{\nolinkurl{https://arxiv.org/abs/#1}}}

\bibitem[{{Abadie} {et~al.}(2010){Abadie}, {Abbott}, {Abbott}, {Abernathy},
  {Accadia}, {Acernese}, {Adams}, {Adhikari}, {Ajith}, {Allen}, {Allen},
  {Amador Ceron}, {Amin}, {Anderson}, {Anderson}, {Antonucci}, {Aoudia},
  {Arain}, {Araya}, {Aronsson}, {Arun}, {Aso}, {Aston}, {Astone}, {Atkinson},
  {Aufmuth}, {Aulbert}, {Babak}, {Baker}, {Ballardin}, {Ballmer}, {Barker},
  {Barnum}, {Barone}, {Barr}, {Barriga}, {Barsotti}, {Barsuglia}, {Barton},
  {Bartos}, {Bassiri}, {Bastarrika}, {Bauchrowitz}, {Bauer}, {Behnke}, {Beker},
  {Belczynski}, {Benacquista}, {Bertolini}, {Betzwieser}, {Beveridge},
  {Beyersdorf}, {Bigotta}, {Bilenko}, {Billingsley}, {Birch}, {Birindelli},
  {Biswas}, {Bitossi}, {Bizouard}, {Black}, {Blackburn}, {Blackburn}, {Blair},
  {Bland}, {Blom}, {Blomberg}, {Boccara}, {Bock}, {Bodiya}, {Bondarescu},
  {Bondu}, {Bonelli}, {Bork}, {Born}, {Bose}, {Bosi}, {Boyle}, {Braccini},
  {Bradaschia}, {Brady}, {Braginsky}, {Brau}, {Breyer}, {Bridges}, {Brillet},
  {Brinkmann}, {Brisson}, {Britzger}, {Brooks}, {Brown}, {Budzy{\'n}ski},
  {Bulik}, {Bulten}, {Buonanno}, {Burguet-Castell}, {Burmeister}, {Buskulic},
  {Byer}, {Cadonati}, {Cagnoli}, {Calloni}, {Camp}, {Campagna}, {Campsie},
  {Cannizzo}, {Cannon}, {Canuel}, {Cao}, {Capano}, {Carbognani}, {Caride},
  {Caudill}, {Cavagli{\`a}}, {Cavalier}, {Cavalieri}, {Cella}, {Cepeda},
  {Cesarini}, {Chalermsongsak}, {Chalkley}, {Charlton}, {Chassande Mottin},
  {Chelkowski}, {Chen}, {Chincarini}, {Christensen}, {Chua}, {Chung}, {Clark},
  {Clark}, {Clayton}, {Cleva}, {Coccia}, {Colacino}, {Colas}, {Colla},
  {Colombini}, {Conte}, {Cook}, {Corbitt}, {Corda}, {Cornish}, {Corsi},
  {Costa}, {Coulon}, {Coward}, {Coyne}, {Creighton}, {Creighton}, {Cruise},
  {Culter}, {Cumming}, {Cunningham}, {Cuoco}, {Dahl}, {Danilishin},
  {Dannenberg}, {D'Antonio}, {Danzmann}, {Dari}, {Das}, {Dattilo}, {Daudert},
  {Davier}, {Davies}, {Davis}, {Daw}, {Day}, {Dayanga}, {De Rosa}, {DeBra},
  {Degallaix}, {del Prete}, {Dergachev}, {DeRosa}, {DeSalvo}, {Devanka},
  {Dhurandhar}, {Di Fiore}, {Di Lieto}, {Di Palma}, {Emilio}, {Di Virgilio},
  {D{\'\i}az}, {Dietz}, {Donovan}, {Dooley}, {Doomes}, {Dorsher}, {Douglas},
  {Drago}, {Drever}, {Driggers}, {Dueck}, {Dumas}, {Eberle}, {Edgar},
  {Edwards}, {Effler}, {Ehrens}, {Engel}, {Etzel}, {Evans}, {Evans}, {Fafone},
  {Fairhurst}, {Fan}, {Farr}, {Fazi}, {Fehrmann}, {Feldbaum}, {Ferrante},
  {Fidecaro}, {Finn}, {Fiori}, {Flaminio}, {Flanigan}, {Flasch}, {Foley},
  {Forrest}, {Forsi}, {Fotopoulos}, {Fournier}, {Franc}, {Frasca}, {Frasconi},
  {Frede}, {Frei}, {Frei}, {Freise}, {Frey}, {Fricke}, {Friedrich},
  {Fritschel}, {Frolov}, {Fulda}, {Fyffe}, {Gammaitoni}, {Garofoli}, {Garufi},
  {Gemme}, {Genin}, {Gennai}, {Gholami}, {Ghosh}, {Giaime}, {Giampanis},
  {Giardina}, {Giazotto}, {Gill}, {Goetz}, {Goggin}, {Gonz{\'a}lez},
  {Gorodetsky}, {Go{\ss}ler}, {Gouaty}, {Graef}, {Granata}, {Grant}, {Gras},
  {Gray}, {Greenhalgh}, {Gretarsson}, {Greverie}, {Grosso}, {Grote},
  {Grunewald}, {Guidi}, {Gustafson}, {Gustafson}, {Hage}, {Hall}, {Hallam},
  {Hammer}, {Hammond}, {Hanks}, {Hanna}, {Hanson}, {Harms}, {Harry}, {Harry},
  {Harstad}, {Haughian}, {Hayama}, {Heefner}, {Heitmann}, {Hello}, {Heng},
  {Heptonstall}, {Hewitson}, {Hild}, {Hirose}, {Hoak}, {Hodge}, {Holt},
  {Hosken}, {Hough}, {Howell}, {Hoyland}, {Huet}, {Hughey}, {Husa}, {Huttner},
  {Huynh-Dinh}, {Ingram}, {Inta}, {Isogai}, {Ivanov}, {Jaranowski}, {Johnson},
  {Jones}, {Jones}, {Jones}, {Ju}, {Kalmus}, {Kalogera}, {Kandhasamy},
  {Kanner}, {Katsavounidis}, {Kawabe}, {Kawamura}, {Kawazoe}, {Kells},
  {Keppel}, {Khalaidovski}, {Khalili}, {Khazanov}, {Kim}, {Kim}, {King},
  {Kinzel}, {Kissel}, {Klimenko}, {Kondrashov}, {Kopparapu}, {Koranda},
  {Kowalska}, {Kozak}, {Krause}, {Kringel}, {Krishnamurthy}, {Krishnan},
  {Kr{\'o}lak}, {Kuehn}, {Kullman}, {Kumar}, {Kwee}, {Landry}, {Lang}, {Lantz},
  {Lastzka}, {Lazzarini}, {Leaci}, {Leong}, {Leonor}, {Leroy}, {Letendre},
  {Li}, {Li}, {Lin}, {Lindquist}, {Lockerbie}, {Lodhia}, {Lorenzini},
  {Loriette}, {Lormand}, {Losurdo}, {Lu}, {Luan}, {Lubi{\'n}ski}, {Lucianetti},
  {L{\"u}ck}, {Lundgren}, {Machenschalk}, {MacInnis}, {Mackowski},
  {Mageswaran}, {Mailand}, {Majorana}, {Mak}, {Man}, {Mandel}, {Mandic},
  {Mantovani}, {Marchesoni}, {Marion}, {M{\'a}rka}, {M{\'a}rka}, {Maros},
  {Marque}, {Martelli}, {Martin}, {Martin}, {Marx}, {Mason}, {Masserot},
  {Matichard}, {Matone}, {Matzner}, {Mavalvala}, {McCarthy}, {McClelland},
  {McGuire}, {McIntyre}, {McIvor}, {McKechan}, {Meadors}, {Mehmet}, {Meier},
  {Melatos}, {Melissinos}, {Mendell}, {Men{\'e}ndez}, {Mercer}, {Merill},
  {Meshkov}, {Messenger}, {Meyer}, {Miao}, {Michel}, {Milano}, {Miller},
  {Minenkov}, {Mino}, {Mitra}, {Mitrofanov}, {Mitselmakher}, {Mittleman},
  {Moe}, {Mohan}, {Mohanty}, {Mohapatra}, {Moraru}, {Moreau}, {Moreno},
  {Morgado}, {Morgia}, {Morioka}, {Mors}, {Mosca}, {Moscatelli}, {Mossavi},
  {Mours}, {MowLowry}, {Mueller}, {Mukherjee}, {Mullavey},
  {M{\"u}ller-Ebhardt}, {Munch}, {Murray}, {Nash}, {Nawrodt}, {Nelson}, {Neri},
  {Newton}, {Nishizawa}, {Nocera}, {Nolting}, {Ochsner}, {O'Dell}, {Ogin},
  {Oldenburg}, {O'Reilly}, {O'Shaughnessy}, {Osthelder}, {Ottaway}, {Ottens},
  {Overmier}, {Owen}, {Page}, {Pagliaroli}, {Palladino}, {Palomba}, {Pan},
  {Pankow}, {Paoletti}, {Papa}, {Pardi}, {Pareja}, {Parisi}, {Pasqualetti},
  {Passaquieti}, {Passuello}, {Patel}, {Pedraza}, {Pekowsky}, {Penn},
  {Peralta}, {Perreca}, {Persichetti}, {Pichot}, {Pickenpack}, {Piergiovanni},
  {Pietka}, {Pinard}, {Pinto}, {Pitkin}, {Pletsch}, {Plissi}, {Poggiani},
  {Postiglione}, {Prato}, {Predoi}, {Price}, {Prijatelj}, {Principe},
  {Privitera}, {Prix}, {Prodi}, {Prokhorov}, {Puncken}, {Punturo}, {Puppo},
  {Quetschke}, {Raab}, {Rabaste}, {Rabeling}, {Radke}, {Radkins}, {Raffai},
  {Rakhmanov}, {Rankins}, {Rapagnani}, {Raymond}, {Re}, {Reed}, {Reed},
  {Regimbau}, {Reid}, {Reitze}, {Ricci}, {Riesen}, {Riles}, {Roberts},
  {Robertson}, {Robinet}, {Robinson}, {Robinson}, {Rocchi}, {Roddy},
  {R{\"o}ver}, {Rogstad}, {Rolland}, {Rollins}, {Romano}, {Romano}, {Romie},
  {Rosi{\'n}ska}, {Rowan}, {R{\"u}diger}, {Ruggi}, {Ryan}, {Sakata}, {Sakosky},
  {Salemi}, {Sammut}, {Sancho de la Jordana}, {Sandberg}, {Sannibale},
  {Santamar{\'\i}a}, {Santostasi}, {Saraf}, {Sassolas}, {Sathyaprakash},
  {Sato}, {Satterthwaite}, {Saulson}, {Savage}, {Schilling}, {Schnabel},
  {Schofield}, {Schulz}, {Schutz}, {Schwinberg}, {Scott}, {Scott}, {Searle},
  {Seifert}, {Sellers}, {Sengupta}, {Sentenac}, {Sergeev}, {Shaddock},
  {Shapiro}, {Shawhan}, {Shoemaker}, {Sibley}, {Siemens}, {Sigg}, {Singer},
  {Sintes}, {Skelton}, {Slagmolen}, {Slutsky}, {Smith}, {Smith}, {Smith},
  {Somiya}, {Sorazu}, {Speirits}, {Stein}, {Stein}, {Steinlechner},
  {Steplewski}, {Stochino}, {Stone}, {Strain}, {Strigin}, {Stroeer}, {Sturani},
  {Stuver}, {Summerscales}, {Sung}, {Susmithan}, {Sutton}, {Swinkels},
  {Talukder}, {Tanner}, {Tarabrin}, {Taylor}, {Taylor}, {Thomas}, {Thorne},
  {Thorne}, {Thrane}, {Th{\"u}ring}, {Titsler}, {Tokmakov}, {Toncelli},
  {Tonelli}, {Torres}, {Torrie}, {Tournefier}, {Travasso}, {Traylor}, {Trias},
  {Trummer}, {Tseng}, {Ugolini}, {Urbanek}, {Vahlbruch}, {Vaishnav}, {Vajente},
  {Vallisneri}, {van den Brand}, {Van Den Broeck}, {van der Putten}, {van der
  Sluys}, {van Veggel}, {Vass}, {Vaulin}, {Vavoulidis}, {Vecchio}, {Vedovato},
  {Veitch}, {Veitch}, {Veltkamp}, {Verkindt}, {Vetrano}, {Vicer{\'e}},
  {Villar}, {Vinet}, {Vocca}, {Vorvick}, {Vyachanin}, {Waldman}, {Wallace},
  {Wanner}, {Ward}, {Was}, {Wei}, {Weinert}, {Weinstein}, {Weiss}, {Wen},
  {Wen}, {Wessels}, {West}, {Westphal}, {Wette}, {Whelan}, {Whitcomb}, {White},
  {Whiting}, {Wilkinson}, {Willems}, {Williams}, {Willke}, {Winkelmann},
  {Winkler}, {Wipf}, {Wiseman}, {Woan}, {Wooley}, {Worden}, {Yakushin},
  {Yamamoto}, {Yamamoto}, {Yeaton-Massey}, {Yoshida}, {Yu}, {Yvert}, {Zanolin},
  {Zhang}, {Zhang}, {Zhao}, {Zotov}, {Zucker}, {Zweizig}, {LIGO Scientific
  Collaboration}, \& {Virgo Collaboration}}]{Abadie:2010:Rates}
{Abadie}, J., {Abbott}, B.~P., {Abbott}, R., {et~al.} 2010, Classical and
  Quantum Gravity, 27, 173001, \dodoi{10.1088/0264-9381/27/17/173001}

\bibitem[{Abbott {et~al.}(2018)}]{Aasi13}
Abbott, B.~P., {et~al.} 2018, Living Rev. Rel., 21, 3,
  \dodoi{10.1007/s41114-018-0012-9, 10.1007/lrr-2016-1}

\bibitem[{{Abbott} {et~al.}(2021{\natexlab{a}}){Abbott}, {Abbott}, {Acernese},
  {Ackley}, {Adams}, {Adhikari}, {Adhikari}, {Adya}, {Affeldt}, {Agarwal},
  {Agathos}, {Agatsuma}, {Aggarwal}, {Aguiar}, {Aiello}, {Ain}, {Ajith},
  {Akcay}, {Akutsu}, {Albanesi}, {Allocca}, {Altin}, {Amato}, {Anand}, {Anand},
  {Ananyeva}, {Anderson}, {Anderson}, {Ando}, {Andrade}, {Andres},
  {Andri{\'c}}, {Angelova}, {Ansoldi}, {Antelis}, {Antier}, {Appert}, {Arai},
  {Arai}, {Arai}, {Araki}, {Araya}, {Araya}, {Areeda}, {Ar{\`e}ne}, {Aritomi},
  {Arnaud}, {Arogeti}, {Aronson}, {Arun}, {Asada}, {Asali}, {Ashton}, {Aso},
  {Assiduo}, {Aston}, {Astone}, {Aubin}, {Austin}, {Babak}, {Badaracco},
  {Bader}, {Badger}, {Bae}, {Bae}, {Baer}, {Bagnasco}, {Bai}, {Baiotti},
  {Baird}, {Bajpai}, {Ball}, {Ballardin}, {Ballmer}, {Balsamo}, {Baltus},
  {Banagiri}, {Bankar}, {Barayoga}, {Barbieri}, {Barish}, {Barker}, {Barneo},
  {Barone}, {Barr}, {Barsotti}, {Barsuglia}, {Barta}, {Bartlett}, {Barton},
  {Bartos}, {Bassiri}, {Basti}, {Bawaj}, {Bayley}, {Baylor}, {Bazzan},
  {B{\'e}csy}, {Bedakihale}, {Bejger}, {Belahcene}, {Benedetto}, {Beniwal},
  {Bennett}, {Bentley}, {BenYaala}, {Bergamin}, {Berger}, {Bernuzzi}, {Berry},
  {Bersanetti}, {Bertolini}, {Betzwieser}, {Beveridge}, {Bhandare}, {Bhardwaj},
  {Bhattacharjee}, {Bhaumik}, {Bilenko}, {Billingsley}, {Bini}, {Birney},
  {Birnholtz}, {Biscans}, {Bischi}, {Biscoveanu}, {Bisht}, {Biswas}, {Bitossi},
  {Bizouard}, {Blackburn}, {Blair}, {Blair}, {Blair}, {Bobba}, {Bode}, {Boer},
  {Bogaert}, {Boldrini}, {Bonavena}, {Bondu}, {Bonilla}, {Bonnand}, {Booker},
  {Boom}, {Bork}, {Boschi}, {Bose}, {Bose}, {Bossilkov}, {Boudart},
  {Bouffanais}, {Bozzi}, {Bradaschia}, {Brady}, {Bramley}, {Branch},
  {Branchesi}, {Brandt}, {Brau}, {Breschi}, {Briant}, {Briggs}, {Brillet},
  {Brinkmann}, {Brockill}, {Brooks}, {Brooks}, {Brown}, {Brunett}, {Bruno},
  {Bruntz}, {Bryant}, {Bulik}, {Bulten}, {Buonanno}, {Buscicchio}, {Buskulic},
  {Buy}, {Byer}, {Cabourn Davies}, {Cadonati}, {Cagnoli}, {Cahillane},
  {Calder{\'o}n Bustillo}, {Callaghan}, {Callister}, {Calloni}, {Cameron},
  {Camp}, {Canepa}, {Canevarolo}, {Cannavacciuolo}, {Cannon}, {Cao}, {Cao},
  {Capocasa}, {Capote}, {Carapella}, {Carbognani}, {Carlin}, {Carney},
  {Carpinelli}, {Carrillo}, {Carullo}, {Carver}, {Casanueva Diaz}, {Casentini},
  {Castaldi}, {Caudill}, {Cavagli{\`a}}, {Cavalier}, {Cavalieri}, {Ceasar},
  {Cella}, {Cerd{\'a}-Dur{\'a}n}, {Cesarini}, {Chaibi}, {Chakravarti},
  {Chalathadka Subrahmanya}, {Champion}, {Chan}, {Chan}, {Chan}, {Chan},
  {Chan}, {Chandra}, {Chanial}, {Chao}, {Chapman-Bird}, {Charlton}, {Chase},
  {Chassande-Mottin}, {Chatterjee}, {Chatterjee}, {Chatterjee}, {Chaturvedi},
  {Chaty}, {Chatziioannou}, {Chen}, {Chen}, {Chen}, {Chen}, {Chen}, {Chen},
  {Chen}, {Chen}, {Cheng}, {Cheong}, {Cheung}, {Chia}, {Chiadini}, {Chiang},
  {Chiarini}, {Chierici}, {Chincarini}, {Chiofalo}, {Chiummo}, {Cho}, {Cho},
  {Choudhary}, {Choudhary}, {Christensen}, {Chu}, {Chu}, {Chu}, {Chua},
  {Chung}, {Ciani}, {Ciecielag}, {Cie{\'s}lar}, {Cifaldi}, {Ciobanu}, {Ciolfi},
  {Cipriano}, {Cirone}, {Clara}, {Clark}, {Clark}, {Clarke}, {Clearwater},
  {Clesse}, {Cleva}, {Coccia}, {Codazzo}, {Cohadon}, {Cohen}, {Cohen},
  {Colleoni}, {Collette}, {Colombo}, {Colpi}, {Compton}, {Constancio}, {Conti},
  {Cooper}, {Corban}, {Corbitt}, {Cordero-Carri{\'o}n}, {Corezzi}, {Corley},
  {Cornish}, {Corre}, {Corsi}, {Cortese}, {Costa}, {Cotesta}, {Coughlin},
  {Coulon}, {Countryman}, {Cousins}, {Couvares}, {Coward}, {Cowart}, {Coyne},
  {Coyne}, {Creighton}, {Creighton}, {Criswell}, {Croquette}, {Crowder},
  {Cudell}, {Cullen}, {Cumming}, {Cummings}, {Cunningham}, {Cuoco},
  {Cury{\l}o}, {Dabadie}, {Dal Canton}, {Dall'Osso}, {D{\'a}lya}, {Dana},
  {DaneshgaranBajastani}, {D'Angelo}, {Danila}, {Danilishin}, {D'Antonio},
  {Danzmann}, {Darsow-Fromm}, {Dasgupta}, {Datrier}, {Datta}, {Dattilo},
  {Dave}, {Davier}, {Davis}, {Davis}, {Daw}, {de Alarc{\'o}n}, {Dean}, {DeBra},
  {Deenadayalan}, {Degallaix}, {De Laurentis}, {Del{\'e}glise}, {Del Favero},
  {De Lillo}, {De Lillo}, {Del Pozzo}, {DeMarchi}, {De Matteis}, {D'Emilio},
  {Demos}, {Dent}, {Depasse}, {De Pietri}, {De Rosa}, {De Rossi}, {DeSalvo},
  {De Simone}, {Dhurandhar}, {D{\'\i}az}, {Diaz-Ortiz}, {Didio}, {Dietrich},
  {Di Fiore}, {Di Fronzo}, {Di Giorgio}, {Di Giovanni}, {Di Giovanni}, {Di
  Girolamo}, {Di Lieto}, {Ding}, {Di Pace}, {Di Palma}, {Di Renzo},
  {Divakarla}, {Dmitriev}, {Doctor}, {D'Onofrio}, {Donovan}, {Dooley},
  {Doravari}, {Dorrington}, {Drago}, {Driggers}, {Drori}, {Ducoin}, {Dupej},
  {Durante}, {D'Urso}, {Duverne}, {Dwyer}, {Eassa}, {Easter}, {Ebersold},
  {Eckhardt}, {Eddolls}, {Edelman}, {Edo}, {Edy}, {Effler}, {Eguchi},
  {Eichholz}, {Eikenberry}, {Eisenmann}, {Eisenstein}, {Ejlli}, {Engelby},
  {Enomoto}, {Errico}, {Essick}, {Estell{\'e}s}, {Estevez}, {Etienne}, {Etzel},
  {Evans}, {Evans}, {Ewing}, {Fafone}, {Fair}, {Fairhurst}, {Farah}, {Farinon},
  {Farr}, {Farr}, {Farrow}, {Fauchon-Jones}, {Favaro}, {Favata}, {Fays},
  {Fazio}, {Feicht}, {Fejer}, {Fenyvesi}, {Ferguson}, {Fernandez-Galiana},
  {Ferrante}, {Ferreira}, {Fidecaro}, {Figura}, {Fiori}, {Fishbach}, {Fisher},
  {Fittipaldi}, {Fiumara}, {Flaminio}, {Floden}, {Fong}, {Font}, {Fornal},
  {Forsyth}, {Franke}, {Frasca}, {Frasconi}, {Frederick}, {Freed}, {Frei},
  {Freise}, {Frey}, {Fritschel}, {Frolov}, {Fronz{\'e}}, {Fujii}, {Fujikawa},
  {Fukunaga}, {Fukushima}, {Fulda}, {Fyffe}, {Gabbard}, {Gabella}, {Gadre},
  {Gair}, {Gais}, {Galaudage}, {Gamba}, {Ganapathy}, {Ganguly}, {Gao},
  {Gaonkar}, {Garaventa}, {Garc{\'\i}a}, {Garc{\'\i}a-N{\'u}{\~n}ez},
  {Garc{\'\i}a-Quir{\'o}s}, {Garufi}, {Gateley}, {Gaudio}, {Gayathri}, {Ge},
  {Gemme}, {Gennai}, {George}, {George}, {Gerberding}, {Gergely}, {Gewecke},
  {Ghonge}, {Ghosh}, {Ghosh}, {Ghosh}, {Ghosh}, {Giacomazzo}, {Giacoppo},
  {Giaime}, {Giardina}, {Gibson}, {Gier}, {Giesler}, {Giri}, {Gissi},
  {Glanzer}, {Gleckl}, {Godwin}, {Goetz}, {Goetz}, {Gohlke}, {Golomb},
  {Goncharov}, {Gonz{\'a}lez}, {Gopakumar}, {Gosselin}, {Gouaty}, {Gould},
  {Grace}, {Grado}, {Granata}, {Granata}, {Grant}, {Gras}, {Grassia}, {Gray},
  {Gray}, {Greco}, {Green}, {Green}, {Gretarsson}, {Gretarsson}, {Griffith},
  {Griffiths}, {Griggs}, {Grignani}, {Grimaldi}, {Grimm}, {Grote}, {Grunewald},
  {Gruning}, {Guerra}, {Guidi}, {Guimaraes}, {Guix{\'e}}, {Gulati}, {Guo},
  {Guo}, {Gupta}, {Gupta}, {Gupta}, {Gustafson}, {Gustafson}, {Guzman}, {Ha},
  {Haegel}, {Hagiwara}, {Haino}, {Halim}, {Hall}, {Hamilton}, {Hammond}, {Han},
  {Haney}, {Hanks}, {Hanna}, {Hannam}, {Hannuksela}, {Hansen}, {Hansen},
  {Hanson}, {Harder}, {Hardwick}, {Haris}, {Harms}, {Harry}, {Harry},
  {Hartwig}, {Hasegawa}, {Haskell}, {Hasskew}, {Haster}, {Hattori}, {Haughian},
  {Hayakawa}, {Hayama}, {Hayes}, {Healy}, {Heidmann}, {Heidt}, {Heintze},
  {Heinze}, {Heinzel}, {Heitmann}, {Hellman}, {Hello}, {Helmling-Cornell},
  {Hemming}, {Hendry}, {Heng}, {Hennes}, {Hennig}, {Hennig}, {Hernandez},
  {Hernandez Vivanco}, {Heurs}, {Hild}, {Hill}, {Himemoto}, {Hines},
  {Hiranuma}, {Hirata}, {Hirose}, {Hochheim}, {Hofman}, {Hohmann}, {Holcomb},
  {Holland}, {Holley-Bockelmann}, {Hollows}, {Holmes}, {Holt}, {Holz}, {Hong},
  {Hopkins}, {Hough}, {Hourihane}, {Howell}, {Hoy}, {Hoyland}, {Hreibi},
  {Hsieh}, {Hsu}, {Huang}, {Huang}, {Huang}, {Huang}, {Huang}, {Huang},
  {H{\"u}bner}, {Huddart}, {Hughey}, {Hui}, {Hui}, {Husa}, {Huttner},
  {Huxford}, {Huynh-Dinh}, {Ide}, {Idzkowski}, {Iess}, {Ikenoue}, {Imam},
  {Inayoshi}, {Ingram}, {Inoue}, {Ioka}, {Isi}, {Isleif}, {Ito}, {Itoh},
  {Iyer}, {Izumi}, {JaberianHamedan}, {Jacqmin}, {Jadhav}, {Jadhav}, {James},
  {Jan}, {Jani}, {Janquart}, {Janssens}, {Janthalur}, {Jaranowski}, {Jariwala},
  {Jaume}, {Jenkins}, {Jenner}, {Jeon}, {Jeunon}, {Jia}, {Jin}, {Johns},
  {Johnson-McDaniel}, {Jones}, {Jones}, {Jones}, {Jones}, {Jones}, {Jonker},
  {Ju}, {Jung}, {Jung}, {Junker}, {Juste}, {Kaihotsu}, {Kajita}, {Kakizaki},
  {Kalaghatgi}, {Kalogera}, {Kamai}, {Kamiizumi}, {Kanda}, {Kandhasamy},
  {Kang}, {Kanner}, {Kao}, {Kapadia}, {Kapasi}, {Karat}, {Karathanasis},
  {Karki}, {Kashyap}, {Kasprzack}, {Kastaun}, {Katsanevas}, {Katsavounidis},
  {Katzman}, {Kaur}, {Kawabe}, {Kawaguchi}, {Kawai}, {Kawasaki},
  {K{\'e}f{\'e}lian}, {Keitel}, {Key}, {Khadka}, {Khalili}, {Khan}, {Khazanov},
  {Khetan}, {Khursheed}, {Kijbunchoo}, {Kim}, {Kim}, {Kim}, {Kim}, {Kim},
  {Kim}, {Kimball}, {Kimura}, {Kinley-Hanlon}, {Kirchhoff}, {Kissel}, {Kita},
  {Kitazawa}, {Kleybolte}, {Klimenko}, {Knee}, {Knowles}, {Knyazev}, {Koch},
  {Koekoek}, {Kojima}, {Kokeyama}, {Koley}, {Kolitsidou}, {Kolstein}, {Komori},
  {Kondrashov}, {Kong}, {Kontos}, {Koper}, {Korobko}, {Kotake}, {Kovalam},
  {Kozak}, {Kozakai}, {Kozu}, {Kringel}, {Krishnendu}, {Kr{\'o}lak}, {Kuehn},
  {Kuei}, {Kuijer}, {Kulkarni}, {Kumar}, {Kumar}, {Kumar}, {Kumar}, {Kume},
  {Kuns}, {Kuo}, {Kuo}, {Kuromiya}, {Kuroyanagi}, {Kusayanagi}, {Kuwahara},
  {Kwak}, {Lagabbe}, {Laghi}, {Lalande}, {Lam}, {Lamberts}, {Landry}, {Lane},
  {Lang}, {Lange}, {Lantz}, {La Rosa}, {Lartaux-Vollard}, {Lasky}, {Laxen},
  {Lazzarini}, {Lazzaro}, {Leaci}, {Leavey}, {Lecoeuche}, {Lee}, {Lee}, {Lee},
  {Lee}, {Lee}, {Lee}, {Lehmann}, {Lema{\^\i}tre}, {Leonardi}, {Leroy},
  {Letendre}, {Levesque}, {Levin}, {Leviton}, {Leyde}, {Li}, {Li}, {Li}, {Li},
  {Li}, {Li}, {Lin}, {Lin}, {Lin}, {Lin}, {Lin}, {Linde}, {Linker}, {Linley},
  {Littenberg}, {Liu}, {Liu}, {Liu}, {Liu}, {Llamas}, {Llorens-Monteagudo},
  {Lo}, {Lockwood}, {Loh}, {London}, {Longo}, {Lopez}, {Lopez Portilla},
  {Lorenzini}, {Loriette}, {Lormand}, {Losurdo}, {Lott}, {Lough}, {Lousto},
  {Lovelace}, {Lucaccioni}, {L{\"u}ck}, {Lumaca}, {Lundgren}, {Luo}, {Lynam},
  {Macas}, {MacInnis}, {Macleod}, {MacMillan}, {Macquet}, {Maga{\~n}a
  Hernandez}, {Magazz{\`u}}, {Magee}, {Maggiore}, {Magnozzi}, {Mahesh},
  {Majorana}, {Makarem}, {Maksimovic}, {Maliakal}, {Malik}, {Man}, {Mandic},
  {Mangano}, {Mango}, {Mansell}, {Manske}, {Mantovani}, {Mapelli},
  {Marchesoni}, {Marchio}, {Marion}, {Mark}, {M{\'a}rka}, {M{\'a}rka},
  {Markakis}, {Markosyan}, {Markowitz}, {Maros}, {Marquina}, {Marsat},
  {Martelli}, {Martin}, {Martin}, {Martinez}, {Martinez}, {Martinez},
  {Martinovic}, {Martynov}, {Marx}, {Masalehdan}, {Mason}, {Massera},
  {Masserot}, {Massinger}, {Masso-Reid}, {Mastrogiovanni}, {Matas},
  {Mateu-Lucena}, {Matichard}, {Matiushechkina}, {Mavalvala}, {McCann},
  {McCarthy}, {McClelland}, {McClincy}, {McCormick}, {McCuller}, {McGhee},
  {McGuire}, {McIsaac}, {McIver}, {McRae}, {McWilliams}, {Meacher}, {Mehmet},
  {Mehta}, {Meijer}, {Melatos}, {Melchor}, {Mendell}, {Menendez-Vazquez},
  {Menoni}, {Mercer}, {Mereni}, {Merfeld}, {Merilh}, {Merritt}, {Merzougui},
  {Meshkov}, {Messenger}, {Messick}, {Meyers}, {Meylahn}, {Mhaske}, {Miani},
  {Miao}, {Michaloliakos}, {Michel}, {Michimura}, {Middleton}, {Milano},
  {Miller}, {Miller}, {Miller}, {Millhouse}, {Mills}, {Milotti}, {Minazzoli},
  {Minenkov}, {Mio}, {Mir}, {Miravet-Ten{\'e}s}, {Mishra}, {Mishra}, {Mistry},
  {Mitra}, {Mitrofanov}, {Mitselmakher}, {Mittleman}, {Miyakawa}, {Miyamoto},
  {Miyazaki}, {Miyo}, {Miyoki}, {Mo}, {Modafferi}, {Moguel}, {Mogushi},
  {Mohapatra}, {Mohite}, {Molina}, {Molina-Ruiz}, {Mondin}, {Montani}, {Moore},
  {Moraru}, {Morawski}, {More}, {Moreno}, {Moreno}, {Mori}, {Morisaki},
  {Moriwaki}, {Morr{\'a}s}, {Mours}, {Mow-Lowry}, {Mozzon}, {Muciaccia},
  {Mukherjee}, {Mukherjee}, {Mukherjee}, {Mukherjee}, {Mukherjee}, {Mukund},
  {Mullavey}, {Munch}, {Mu{\~n}iz}, {Murray}, {Musenich}, {Muusse}, {Nadji},
  {Nagano}, {Nagano}, {Nagar}, {Nakamura}, {Nakano}, {Nakano}, {Nakashima},
  {Nakayama}, {Napolano}, {Nardecchia}, {Narikawa}, {Naticchioni}, {Nayak},
  {Nayak}, {Negishi}, {Neil}, {Neilson}, {Nelemans}, {Nelson}, {Nery},
  {Neubauer}, {Neunzert}, {Ng}, {Ng}, {Nguyen}, {Nguyen}, {Nguyen}, {Nguyen
  Quynh}, {Ni}, {Nichols}, {Nishizawa}, {Nissanke}, {Nitoglia}, {Nocera},
  {Norman}, {North}, {Nozaki}, {Nu{\~n}o Siles}, {Nuttall}, {Oberling},
  {O'Brien}, {Obuchi}, {O'Dell}, {Oelker}, {Ogaki}, {Oganesyan}, {Oh}, {Oh},
  {Oh}, {Ohashi}, {Ohishi}, {Ohkawa}, {Ohme}, {Ohta}, {Okada}, {Okutani},
  {Okutomi}, {Olivetto}, {Oohara}, {Ooi}, {Oram}, {O'Reilly}, {Ormiston},
  {Ormsby}, {Ortega}, {O'Shaughnessy}, {O'Shea}, {Oshino}, {Ossokine},
  {Osthelder}, {Otabe}, {Ottaway}, {Overmier}, {Pace}, {Pagano}, {Page},
  {Pagliaroli}, {Pai}, {Pai}, {Palamos}, {Palashov}, {Palomba}, {Pan}, {Pan},
  {Panda}, {Pang}, {Pang}, {Pankow}, {Pannarale}, {Pant}, {Panther},
  {Paoletti}, {Paoli}, {Paolone}, {Parisi}, {Park}, {Park}, {Parker},
  {Pascucci}, {Pasqualetti}, {Passaquieti}, {Passuello}, {Patel}, {Pathak},
  {Patricelli}, {Patron}, {Paul}, {Payne}, {Pedraza}, {Pegoraro}, {Pele},
  {Pe{\~n}a Arellano}, {Penn}, {Perego}, {Pereira}, {Pereira}, {Perez},
  {P{\'e}rigois}, {Perkins}, {Perreca}, {Perri{\`e}s}, {Petermann},
  {Petterson}, {Pfeiffer}, {Pham}, {Phukon}, {Piccinni}, {Pichot},
  {Piendibene}, {Piergiovanni}, {Pierini}, {Pierro}, {Pillant}, {Pillas},
  {Pilo}, {Pinard}, {Pinto}, {Pinto}, {Piotrzkowski}, {Piotrzkowski},
  {Pirello}, {Pitkin}, {Placidi}, {Planas}, {Plastino}, {Pluchar}, {Poggiani},
  {Polini}, {Pong}, {Ponrathnam}, {Popolizio}, {Porter}, {Poulton}, {Powell},
  {Pracchia}, {Pradier}, {Prajapati}, {Prasai}, {Prasanna}, {Pratten},
  {Principe}, {Prodi}, {Prokhorov}, {Prosposito}, {Prudenzi}, {Puecher},
  {Punturo}, {Puosi}, {Puppo}, {P{\"u}rrer}, {Qi}, {Quetschke},
  {Quitzow-James}, {Qutob}, {Raab}, {Raaijmakers}, {Radkins}, {Radulesco},
  {Raffai}, {Rail}, {Raja}, {Rajan}, {Ramirez}, {Ramirez}, {Ramos-Buades},
  {Rana}, {Rapagnani}, {Rapol}, {Ray}, {Raymond}, {Raza}, {Razzano}, {Read},
  {Rees}, {Regimbau}, {Rei}, {Reid}, {Reid}, {Reitze}, {Relton}, {Renzini},
  {Rettegno}, {Reza}, {Rezac}, {Ricci}, {Richards}, {Richardson}, {Richardson},
  {Riemenschneider}, {Riles}, {Rinaldi}, {Rink}, {Rizzo}, {Robertson}, {Robie},
  {Robinet}, {Rocchi}, {Rodriguez}, {Rolland}, {Rollins}, {Romanelli},
  {Romano}, {Romel}, {Romero-Rodr{\'\i}guez}, {Romero-Shaw}, {Romie},
  {Ronchini}, {Rosa}, {Rose}, {Rosi{\'n}ska}, {Ross}, {Rowan}, {Rowlinson},
  {Roy}, {Roy}, {Roy}, {Rozza}, {Ruggi}, {Ruiz-Rocha}, {Ryan}, {Sachdev},
  {Sadecki}, {Sadiq}, {Sago}, {Saito}, {Saito}, {Sakai}, {Sakai},
  {Sakellariadou}, {Sakuno}, {Salafia}, {Salconi}, {Saleem}, {Salemi},
  {Samajdar}, {Sanchez}, {Sanchez}, {Sanchez}, {Sanchis-Gual}, {Sanders},
  {Sanuy}, {Saravanan}, {Sarin}, {Sassolas}, {Satari}, {Sathyaprakash}, {Sato},
  {Sato}, {Sauter}, {Savage}, {Sawada}, {Sawant}, {Sawant}, {Sayah},
  {Schaetzl}, {Scheel}, {Scheuer}, {Schiworski}, {Schmidt}, {Schmidt},
  {Schnabel}, {Schneewind}, {Schofield}, {Sch{\"o}nbeck}, {Schulte}, {Schutz},
  {Schwartz}, {Scott}, {Scott}, {Seglar-Arroyo}, {Sekiguchi}, {Sekiguchi},
  {Sellers}, {Sengupta}, {Sentenac}, {Seo}, {Sequino}, {Sergeev}, {Setyawati},
  {Shaffer}, {Shahriar}, {Shams}, {Shao}, {Sharma}, {Sharma}, {Shawhan},
  {Shcheblanov}, {Shibagaki}, {Shikauchi}, {Shimizu}, {Shimoda}, {Shimode},
  {Shinkai}, {Shishido}, {Shoda}, {Shoemaker}, {Shoemaker}, {ShyamSundar},
  {Sieniawska}, {Sigg}, {Singer}, {Singh}, {Singh}, {Singha}, {Sintes},
  {Sipala}, {Skliris}, {Slagmolen}, {Slaven-Blair}, {Smetana}, {Smith},
  {Smith}, {Soldateschi}, {Somala}, {Somiya}, {Son}, {Soni}, {Soni}, {Sordini},
  {Sorrentino}, {Sorrentino}, {Sotani}, {Soulard}, {Souradeep}, {Sowell},
  {Spagnuolo}, {Spencer}, {Spera}, {Srinivasan}, {Srivastava}, {Srivastava},
  {Staats}, {Stachie}, {Steer}, {Steinhoff}, {Steinlechner}, {Steinlechner},
  {Stevenson}, {Stops}, {Stover}, {Strain}, {Strang}, {Stratta}, {Strunk},
  {Sturani}, {Stuver}, {Sudhagar}, {Sudhir}, {Sugimoto}, {Suh}, {Sullivan},
  {Sullivan}, {Summerscales}, {Sun}, {Sun}, {Sunil}, {Sur}, {Suresh}, {Sutton},
  {Suzuki}, {Suzuki}, {Swinkels}, {Szczepa{\'n}czyk}, {Szewczyk}, {Tacca},
  {Tagoshi}, {Tait}, {Takahashi}, {Takahashi}, {Takamori}, {Takano}, {Takeda},
  {Takeda}, {Talbot}, {Talbot}, {Tanaka}, {Tanaka}, {Tanaka}, {Tanaka},
  {Tanaka}, {Tanasijczuk}, {Tanioka}, {Tanner}, {Tao}, {Tao}, {Tapia San
  Mart{\'\i}n}, {Taranto}, {Tasson}, {Telada}, {Tenorio}, {Terhune},
  {Terkowski}, {Thirugnanasambandam}, {Thomas}, {Thomas}, {Thomas}, {Thompson},
  {Thondapu}, {Thorne}, {Thrane}, {Tiwari}, {Tiwari}, {Tiwari}, {Toivonen},
  {Toland}, {Tolley}, {Tomaru}, {Tomigami}, {Tomura}, {Tonelli},
  {Torres-Forn{\'e}}, {Torrie}, {Tosta e Melo}, {T{\"o}yr{\"a}}, {Trapananti},
  {Travasso}, {Traylor}, {Trevor}, {Tringali}, {Tripathee}, {Troiano},
  {Trovato}, {Trozzo}, {Trudeau}, {Tsai}, {Tsai}, {Tsang}, {Tsang}, {Tsao},
  {Tse}, {Tso}, {Tsubono}, {Tsuchida}, {Tsukada}, {Tsuna}, {Tsutsui},
  {Tsuzuki}, {Turbang}, {Turconi}, {Tuyenbayev}, {Ubhi}, {Uchikata},
  {Uchiyama}, {Udall}, {Ueda}, {Uehara}, {Ueno}, {Ueshima}, {Unnikrishnan},
  {Uraguchi}, {Urban}, {Ushiba}, {Utina}, {Vahlbruch}, {Vajente}, {Vajpeyi},
  {Valdes}, {Valentini}, {Valsan}, {van Bakel}, {van Beuzekom}, {van den
  Brand}, {Van Den Broeck}, {Vander-Hyde}, {van der Schaaf}, {van Heijningen},
  {Vanosky}, {van Putten}, {van Remortel}, {Vardaro}, {Vargas}, {Varma},
  {Vas{\'u}th}, {Vecchio}, {Vedovato}, {Veitch}, {Veitch}, {Venneberg},
  {Venugopalan}, {Verkindt}, {Verma}, {Verma}, {Veske}, {Vetrano},
  {Vicer{\'e}}, {Vidyant}, {Viets}, {Vijaykumar}, {Villa-Ortega}, {Vinet},
  {Virtuoso}, {Vitale}, {Vo}, {Vocca}, {von Reis}, {von Wrangel}, {Vorvick},
  {Vyatchanin}, {Wade}, {Wade}, {Wagner}, {Walet}, {Walker}, {Wallace},
  {Wallace}, {Walsh}, {Wang}, {Wang}, {Wang}, {Ward}, {Warner}, {Was},
  {Washimi}, {Washington}, {Watchi}, {Weaver}, {Webster}, {Weinert},
  {Weinstein}, {Weiss}, {Weller}, {Weller}, {Wellmann}, {Wen}, {We{\ss}els},
  {Wette}, {Whelan}, {White}, {Whiting}, {Whittle}, {Wilken}, {Williams},
  {Williams}, {Williams}, {Williamson}, {Willis}, {Willke}, {Wilson},
  {Winkler}, {Wipf}, {Wlodarczyk}, {Woan}, {Woehler}, {Wofford}, {Wong}, {Wu},
  {Wu}, {Wu}, {Wu}, {Wysocki}, {Xiao}, {Xu}, {Yamada}, {Yamamoto}, {Yamamoto},
  {Yamamoto}, {Yamamoto}, {Yamashita}, {Yamazaki}, {Yang}, {Yang}, {Yang},
  {Yang}, {Yang}, {Yap}, {Yeeles}, {Yelikar}, {Ying}, {Yokogawa}, {Yokoyama},
  {Yokozawa}, {Yoo}, {Yoshioka}, {Yu}, {Yu}, {Yuzurihara}, {Zadro{\.z}ny},
  {Zanolin}, {Zeidler}, {Zelenova}, {Zendri}, {Zevin}, {Zhan}, {Zhang},
  {Zhang}, {Zhang}, {Zhang}, {Zhang}, {Zhao}, {Zhao}, {Zhao}, {Zhao}, {Zheng},
  {Zhou}, {Zhou}, {Zhu}, {Zhu}, {Zimmerman}, {Zlochower}, {Zucker}, \&
  {Zweizig}}]{LVK:2022:GWTC-3}
{Abbott}, R., {Abbott}, T.~D., {Acernese}, F., {et~al.} 2021{\natexlab{a}},
  arXiv e-prints, arXiv:2111.03606.
\newblock \doarXiv{2111.03606}

\bibitem[{{Abbott} {et~al.}(2021{\natexlab{b}}){Abbott}, {Abbott}, {Abraham},
  {Acernese}, {Ackley}, {Adams}, {Adams}, {Adhikari}, {Adya}, {Affeldt},
  {Agathos}, {Agatsuma}, {Aggarwal}, {Aguiar}, {Aiello}, {Ain}, {Ajith},
  {Akcay}, {Allen}, {Allocca}, {Altin}, {Amato}, {Anand}, {Ananyeva},
  {Anderson}, {Anderson}, {Angelova}, {Ansoldi}, {Antelis}, {Antier}, {Appert},
  {Arai}, {Araya}, {Areeda}, {Ar{\`e}ne}, {Arnaud}, {Aronson}, {Arun}, {Asali},
  {Ascenzi}, {Ashton}, {Aston}, {Astone}, {Aubin}, {Aufmuth}, {AultONeal},
  {Austin}, {Avendano}, {Babak}, {Badaracco}, {Bader}, {Bae}, {Baer},
  {Bagnasco}, {Baird}, {Ball}, {Ballardin}, {Ballmer}, {Bals}, {Balsamo},
  {Baltus}, {Banagiri}, {Bankar}, {Bankar}, {Barayoga}, {Barbieri}, {Barish},
  {Barker}, {Barneo}, {Barnum}, {Barone}, {Barr}, {Barsotti}, {Barsuglia},
  {Barta}, {Bartlett}, {Bartos}, {Bassiri}, {Basti}, {Bawaj}, {Bayley},
  {Bazzan}, {Becher}, {B{\'e}csy}, {Bedakihale}, {Bejger}, {Belahcene},
  {Beniwal}, {Benjamin}, {Bennett}, {Bentley}, {Bergamin}, {Berger},
  {Bergmann}, {Bernuzzi}, {Berry}, {Bersanetti}, {Bertolini}, {Betzwieser},
  {Bhandare}, {Bhandari}, {Bhattacharjee}, {Bidler}, {Bilenko}, {Billingsley},
  {Birney}, {Birnholtz}, {Biscans}, {Bischi}, {Biscoveanu}, {Bisht}, {Bitossi},
  {Bizouard}, {Blackburn}, {Blackman}, {Blair}, {Blair}, {Blair}, {Blanch},
  {Bobba}, {Bode}, {Boer}, {Boetzel}, {Bogaert}, {Boldrini}, {Bondu},
  {Bonilla}, {Bonnand}, {Booker}, {Boom}, {Bork}, {Boschi}, {Bose},
  {Bossilkov}, {Boudart}, {Bouffanais}, {Bozzi}, {Bradaschia}, {Brady},
  {Bramley}, {Branchesi}, {Brau}, {Breschi}, {Briant}, {Briggs}, {Brighenti},
  {Brillet}, {Brinkmann}, {Brockill}, {Brooks}, {Brooks}, {Brown}, {Brunett},
  {Bruno}, {Bruntz}, {Buikema}, {Bulik}, {Bulten}, {Buonanno}, {Buscicchio},
  {Buskulic}, {Byer}, {Cabero}, {Cadonati}, {Caesar}, {Cagnoli}, {Cahillane},
  {Calder{\'o}n Bustillo}, {Callaghan}, {Callister}, {Calloni}, {Camp},
  {Canepa}, {Cannon}, {Cao}, {Cao}, {Carapella}, {Carbognani}, {Carney},
  {Carpinelli}, {Carullo}, {Carver}, {Casanueva Diaz}, {Casentini}, {Caudill},
  {Cavagli{\`a}}, {Cavalier}, {Cavalieri}, {Cella}, {Cerd{\'a}-Dur{\'a}n},
  {Cesarini}, {Chaibi}, {Chakravarti}, {Chan}, {Chan}, {Chandra}, {Chanial},
  {Chao}, {Charlton}, {Chase}, {Chassande-Mottin}, {Chatterjee},
  {Chattopadhyay}, {Chaturvedi}, {Chatziioannou}, {Chen}, {Chen}, {Chen},
  {Chen}, {Cheng}, {Cheong}, {Chia}, {Chiadini}, {Chierici}, {Chincarini},
  {Chiummo}, {Cho}, {Cho}, {Cho}, {Choate}, {Christensen}, {Chu}, {Chua},
  {Chung}, {Chung}, {Ciani}, {Ciecielag}, {Cie{\'s}lar}, {Cifaldi}, {Ciobanu},
  {Ciolfi}, {Cipriano}, {Cirone}, {Clara}, {Clark}, {Clark}, {Clarke},
  {Clearwater}, {Clesse}, {Cleva}, {Coccia}, {Cohadon}, {Cohen}, {Colleoni},
  {Collette}, {Collins}, {Colpi}, {Constancio}, {Conti}, {Cooper}, {Corban},
  {Corbitt}, {Cordero-Carri{\'o}n}, {Corezzi}, {Corley}, {Cornish}, {Corre},
  {Corsi}, {Cortese}, {Costa}, {Cotesta}, {Coughlin}, {Coughlin}, {Coulon},
  {Countryman}, {Cousins}, {Couvares}, {Covas}, {Coward}, {Cowart}, {Coyne},
  {Coyne}, {Creighton}, {Creighton}, {Croquette}, {Crowder}, {Cudell},
  {Cullen}, {Cumming}, {Cummings}, {Cunningham}, {Cuoco}, {Cury{\l}o},
  {Canton}, {D{\'a}lya}, {Dana}, {DaneshgaranBajastani}, {D'Angelo}, {Danila},
  {Danilishin}, {D'Antonio}, {Danzmann}, {Darsow-Fromm}, {Dasgupta}, {Datrier},
  {Dattilo}, {Dave}, {Davier}, {Davies}, {Davis}, {Daw}, {Dean}, {DeBra},
  {Deenadayalan}, {Degallaix}, {De Laurentis}, {Del{\'e}glise}, {Del Favero},
  {De Lillo}, {De Lillo}, {Del Pozzo}, {DeMarchi}, {De Matteis}, {D'Emilio},
  {Demos}, {Denker}, {Dent}, {Depasse}, {De Pietri}, {De Rosa}, {De Rossi},
  {DeSalvo}, {de Varona}, {Dhurandhar}, {D{\'\i}az}, {Diaz-Ortiz}, {Didio},
  {Dietrich}, {Di Fiore}, {DiFronzo}, {Di Giorgio}, {Di Giovanni}, {Di
  Giovanni}, {Di Girolamo}, {Di Lieto}, {Ding}, {Di Pace}, {Di Palma}, {Di
  Renzo}, {Divakarla}, {Dmitriev}, {Doctor}, {D'Onofrio}, {Donovan}, {Dooley},
  {Doravari}, {Dorrington}, {Downes}, {Drago}, {Driggers}, {Du}, {Ducoin},
  {Dupej}, {Durante}, {D'Urso}, {Duverne}, {Dwyer}, {Easter}, {Eddolls},
  {Edelman}, {Edo}, {Edy}, {Effler}, {Eichholz}, {Eikenberry}, {Eisenmann},
  {Eisenstein}, {Ejlli}, {Errico}, {Essick}, {Estell{\'e}s}, {Estevez},
  {Etienne}, {Etzel}, {Evans}, {Evans}, {Ewing}, {Fafone}, {Fair}, {Fairhurst},
  {Fan}, {Farah}, {Farinon}, {Farr}, {Farr}, {Fauchon-Jones}, {Favata}, {Fays},
  {Fazio}, {Feicht}, {Fejer}, {Feng}, {Fenyvesi}, {Ferguson},
  {Fernandez-Galiana}, {Ferrante}, {Ferreira}, {Fidecaro}, {Figura}, {Fiori},
  {Fiorucci}, {Fishbach}, {Fisher}, {Fishner}, {Fittipaldi}, {Fitz-Axen},
  {Fiumara}, {Flaminio}, {Floden}, {Flynn}, {Fong}, {Font}, {Forsyth},
  {Fournier}, {Frasca}, {Frasconi}, {Frei}, {Freise}, {Frey}, {Frey},
  {Fritschel}, {Frolov}, {Fronz{\'e}}, {Fulda}, {Fyffe}, {Gabbard}, {Gadre},
  {Gaebel}, {Gair}, {Gais}, {Galaudage}, {Gamba}, {Ganapathy}, {Ganguly},
  {Gaonkar}, {Garaventa}, {Garc{\'\i}a-Quir{\'o}s}, {Garufi}, {Gateley},
  {Gaudio}, {Gayathri}, {Gemme}, {Gennai}, {George}, {George}, {George},
  {Gergely}, {Ghonge}, {Ghosh}, {Ghosh}, {Ghosh}, {Giacomazzo}, {Giacoppo},
  {Giaime}, {Giardina}, {Gibson}, {Gier}, {Gill}, {Giri}, {Glanzer}, {Gleckl},
  {Godwin}, {Goetz}, {Goetz}, {Gohlke}, {Goncharov}, {Gonz{\'a}lez},
  {Gopakumar}, {Gossan}, {Gosselin}, {Gouaty}, {Grace}, {Grado}, {Granata},
  {Granata}, {Grant}, {Gras}, {Grassia}, {Gray}, {Gray}, {Greco}, {Green},
  {Green}, {Gretarsson}, {Griggs}, {Grignani}, {Grimaldi}, {Grimes}, {Grimm},
  {Grote}, {Grunewald}, {Gruning}, {Guerrero}, {Guidi}, {Guimaraes},
  {Guix{\'e}}, {Gulati}, {Guo}, {Gupta}, {Gupta}, {Gupta}, {Gustafson},
  {Gustafson}, {Guzman}, {Haegel}, {Halim}, {Hall}, {Hamilton}, {Hammond},
  {Haney}, {Hanke}, {Hanks}, {Hanna}, {Hannam}, {Hannuksela}, {Hannuksela},
  {Hansen}, {Hansen}, {Hanson}, {Harder}, {Hardwick}, {Haris}, {Harms},
  {Harry}, {Harry}, {Hartwig}, {Hasskew}, {Haster}, {Haughian}, {Hayes},
  {Healy}, {Heidmann}, {Heintze}, {Heinze}, {Heinzel}, {Heitmann}, {Hellman},
  {Hello}, {Helmling-Cornell}, {Hemming}, {Hendry}, {Heng}, {Hennes}, {Hennig},
  {Hennig}, {Hernandez Vivanco}, {Heurs}, {Hild}, {Hill}, {Hines}, {Hochheim},
  {Hofgard}, {Hofman}, {Hohmann}, {Holgado}, {Holland}, {Hollows}, {Holmes},
  {Holt}, {Holz}, {Hopkins}, {Horst}, {Hough}, {Howell}, {Hoy}, {Hoyland},
  {Huang}, {H{\"u}bner}, {Huddart}, {Huerta}, {Hughey}, {Hui}, {Husa},
  {Huttner}, {Hutzler}, {Huxford}, {Huynh-Dinh}, {Idzkowski}, {Iess},
  {Imperato}, {Inchauspe}, {Ingram}, {Intini}, {Isi}, {Iyer},
  {JaberianHamedan}, {Jacqmin}, {Jadhav}, {Jadhav}, {James}, {Jani},
  {Janssens}, {Janthalur}, {Jaranowski}, {Jariwala}, {Jaume}, {Jenkins},
  {Jeunon}, {Jiang}, {Johns}, {Johnson-McDaniel}, {Jones}, {Jones}, {Jones},
  {Jones}, {Jones}, {Jonker}, {Ju}, {Junker}, {Kalaghatgi}, {Kalogera},
  {Kamai}, {Kandhasamy}, {Kang}, {Kanner}, {Kapadia}, {Kapasi}, {Karathanasis},
  {Karki}, {Kashyap}, {Kasprzack}, {Kastaun}, {Katsanevas}, {Katsavounidis},
  {Katzman}, {Kawabe}, {K{\'e}f{\'e}lian}, {Keitel}, {Key}, {Khadka},
  {Khalili}, {Khan}, {Khan}, {Khazanov}, {Khetan}, {Khursheed}, {Kijbunchoo},
  {Kim}, {Kim}, {Kim}, {Kim}, {Kim}, {Kim}, {Kimball}, {King}, {Kinley-Hanlon},
  {Kirchhoff}, {Kissel}, {Kleybolte}, {Klimenko}, {Knowles}, {Knyazev}, {Koch},
  {Koehlenbeck}, {Koekoek}, {Koley}, {Kolstein}, {Komori}, {Kondrashov},
  {Kontos}, {Koper}, {Korobko}, {Korth}, {Kovalam}, {Kozak}, {Kr{\"a}mer},
  {Kringel}, {Krishnendu}, {Kr{\'o}lak}, {Kuehn}, {Kumar}, {Kumar}, {Kumar},
  {Kumar}, {Kuns}, {Kwang}, {Lackey}, {Laghi}, {Lalande}, {Lam}, {Lamberts},
  {Landry}, {Lane}, {Lang}, {Lange}, {Lantz}, {Lanza}, {La Rosa},
  {Lartaux-Vollard}, {Lasky}, {Laxen}, {Lazzarini}, {Lazzaro}, {Leaci},
  {Leavey}, {Lecoeuche}, {Lee}, {Lee}, {Lee}, {Lee}, {Lehmann}, {Leon},
  {Leroy}, {Letendre}, {Levin}, {Li}, {Li}, {Li}, {Li}, {Li}, {Linde},
  {Linker}, {Linley}, {Littenberg}, {Liu}, {Liu}, {Llorens-Monteagudo}, {Lo},
  {Lockwood}, {London}, {Longo}, {Lorenzini}, {Loriette}, {Lormand}, {Losurdo},
  {Lough}, {Lousto}, {Lovelace}, {L{\"u}ck}, {Lumaca}, {Lundgren}, {Ma},
  {Macas}, {MacInnis}, {Macleod}, {MacMillan}, {Macquet}, {Maga{\~n}a
  Hernandez}, {Maga{\~n}a-Sandoval}, {Magazz{\`u}}, {Magee}, {Majorana},
  {Maksimovic}, {Maliakal}, {Malik}, {Man}, {Mandic}, {Mangano}, {Mansell},
  {Manske}, {Mantovani}, {Mapelli}, {Marchesoni}, {Marion}, {M{\'a}rka},
  {M{\'a}rka}, {Markakis}, {Markosyan}, {Markowitz}, {Maros}, {Marquina},
  {Marsat}, {Martelli}, {Martin}, {Martin}, {Martinez}, {Martinez}, {Martynov},
  {Masalehdan}, {Mason}, {Massera}, {Masserot}, {Massinger}, {Masso-Reid},
  {Mastrogiovanni}, {Matas}, {Mateu-Lucena}, {Matichard}, {Matiushechkina},
  {Mavalvala}, {Maynard}, {McCann}, {McCarthy}, {McClelland}, {McCormick},
  {McCuller}, {McGuire}, {McIsaac}, {McIver}, {McManus}, {McRae}, {McWilliams},
  {Meacher}, {Meadors}, {Mehmet}, {Mehta}, {Melatos}, {Melchor}, {Mendell},
  {Menendez-Vazquez}, {Mercer}, {Mereni}, {Merfeld}, {Merilh}, {Merritt},
  {Merzougui}, {Meshkov}, {Messenger}, {Messick}, {Metzdorff}, {Meyers},
  {Meylahn}, {Mhaske}, {Miani}, {Miao}, {Michaloliakos}, {Michel}, {Middleton},
  {Milano}, {Miller}, {Millhouse}, {Mills}, {Milotti}, {Milovich-Goff},
  {Minazzoli}, {Minenkov}, {Mir}, {Mishkin}, {Mishra}, {Mistry}, {Mitra},
  {Mitrofanov}, {Mitselmakher}, {Mittleman}, {Mo}, {Mogushi}, {Mohapatra},
  {Mohite}, {Molina}, {Molina-Ruiz}, {Mondin}, {Montani}, {Moore}, {Moraru},
  {Morawski}, {Moreno}, {Morisaki}, {Mours}, {Mow-Lowry}, {Mozzon},
  {Muciaccia}, {Mukherjee}, {Mukherjee}, {Mukherjee}, {Mukherjee}, {Mukund},
  {Mullavey}, {Munch}, {Mu{\~n}iz}, {Murray}, {Nadji}, {Nagar}, {Nardecchia},
  {Naticchioni}, {Nayak}, {Neil}, {Neilson}, {Nelemans}, {Nelson}, {Nery},
  {Neunzert}, {Nitz}, {Ng}, {Ng}, {Nguyen}, {Nguyen}, {Nguyen}, {Nichols},
  {Nissanke}, {Nocera}, {Noh}, {North}, {Nothard}, {Nuttall}, {Oberling},
  {O'Brien}, {O'Dell}, {Oganesyan}, {Ogin}, {Oh}, {Oh}, {Ohme}, {Ohta},
  {Okada}, {Olivetto}, {Oppermann}, {Oram}, {O'Reilly}, {Ormiston}, {Ortega},
  {O'Shaughnessy}, {Ossokine}, {Osthelder}, {Ottaway}, {Overmier}, {Owen},
  {Pace}, {Pagano}, {Page}, {Pagliaroli}, {Pai}, {Pai}, {Palamos}, {Palashov},
  {Palomba}, {Pan}, {Panda}, {Pang}, {Pankow}, {Pannarale}, {Pant}, {Paoletti},
  {Paoli}, {Paolone}, {Parker}, {Pascucci}, {Pasqualetti}, {Passaquieti},
  {Passuello}, {Patel}, {Patricelli}, {Payne}, {Pechsiri}, {Pedraza},
  {Pegoraro}, {Pele}, {Penn}, {Perego}, {Perez}, {P{\'e}rigois}, {Perreca},
  {Perri{\`e}s}, {Petermann}, {Petterson}, {Pfeiffer}, {Pham}, {Phukon},
  {Piccinni}, {Pichot}, {Piendibene}, {Piergiovanni}, {Pierini}, {Pierro},
  {Pillant}, {Pilo}, {Pinard}, {Pinto}, {Piotrzkowski}, {Pirello}, {Pitkin},
  {Placidi}, {Plastino}, {Pluchar}, {Poggiani}, {Polini}, {Pong}, {Ponrathnam},
  {Popolizio}, {Porter}, {Poverman}, {Powell}, {Pracchia}, {Prajapati},
  {Prasai}, {Prasanna}, {Pratten}, {Prestegard}, {Principe}, {Prodi},
  {Prokhorov}, {Prosposito}, {Prudenzi}, {Puecher}, {Punturo}, {Puosi},
  {Puppo}, {P{\"u}rrer}, {Qi}, {Quetschke}, {Quinonez}, {Quitzow-James},
  {Raab}, {Raaijmakers}, {Radkins}, {Radulesco}, {Raffai}, {Rafferty}, {Rail},
  {Raja}, {Rajan}, {Rajbhandari}, {Rakhmanov}, {Ramirez}, {Ramirez},
  {Ramos-Buades}, {Rana}, {Rao}, {Rapagnani}, {Rapol}, {Ratto}, {Raymond},
  {Razzano}, {Read}, {Regimbau}, {Rei}, {Reid}, {Reitze}, {Rettegno}, {Ricci},
  {Richardson}, {Richardson}, {Richardson}, {Ricker}, {Riemenschneider},
  {Riles}, {Rizzo}, {Robertson}, {Robinet}, {Rocchi}, {Rocha}, {Rodriguez},
  {Rodriguez-Soto}, {Rolland}, {Rollins}, {Roma}, {Romanelli}, {Romano},
  {Romel}, {Romero}, {Romero-Shaw}, {Romie}, {Ronchini}, {Rose}, {Rose},
  {Rose}, {Rosell}, {Rosi{\'n}ska}, {Rosofsky}, {Ross}, {Rowan}, {Rowlinson},
  {Roy}, {Roy}, {Ruggi}, {Ryan}, {Sachdev}, {Sadecki}, {Sadiq},
  {Sakellariadou}, {Salafia}, {Salconi}, {Saleem}, {Samajdar}, {Sanchez},
  {Sanchez}, {Sanchez}, {Sanchis-Gual}, {Sanders}, {Sandles}, {Santiago},
  {Santos}, {Saravanan}, {Sarin}, {Sassolas}, {Sathyaprakash}, {Sauter},
  {Savage}, {Savant}, {Sawant}, {Sayah}, {Schaetzl}, {Schale}, {Scheel},
  {Scheuer}, {Schindler-Tyka}, {Schmidt}, {Schnabel}, {Schofield},
  {Sch{\"o}nbeck}, {Schreiber}, {Schulte}, {Schutz}, {Schwarm}, {Schwartz},
  {Scott}, {Scott}, {Seglar-Arroyo}, {Seidel}, {Sellers}, {Sengupta},
  {Sennett}, {Sentenac}, {Sequino}, {Sergeev}, {Setyawati}, {Shaffer},
  {Shahriar}, {Sharifi}, {Sharma}, {Sharma}, {Shawhan}, {Shen}, {Shikauchi},
  {Shink}, {Shoemaker}, {Shoemaker}, {Shukla}, {ShyamSundar}, {Sieniawska},
  {Sigg}, {Singer}, {Singh}, {Singh}, {Singha}, {Singhal}, {Sintes}, {Sipala},
  {Skliris}, {Slagmolen}, {Slaven-Blair}, {Smetana}, {Smith}, {Smith},
  {Somala}, {Son}, {Soni}, {Soni}, {Sorazu}, {Sordini}, {Sorrentino},
  {Sorrentino}, {Soulard}, {Souradeep}, {Sowell}, {Spencer}, {Spera},
  {Srivastava}, {Srivastava}, {Staats}, {Stachie}, {Steer}, {Steinhoff},
  {Steinke}, {Steinlechner}, {Steinlechner}, {Steinmeyer}, {Stevenson},
  {Stolle-McAllister}, {Stops}, {Stover}, {Strain}, {Stratta}, {Strunk},
  {Sturani}, {Stuver}, {S{\"u}dbeck}, {Sudhagar}, {Sudhir}, {Suh},
  {Summerscales}, {Sun}, {Sun}, {Sunil}, {Sur}, {Suresh}, {Sutton}, {Swinkels},
  {Szczepa{\'n}czyk}, {Tacca}, {Tait}, {Talbot}, {Tanasijczuk}, {Tanner},
  {Tao}, {Tapia}, {Tapia San Martin}, {Tasson}, {Taylor}, {Tenorio},
  {Terkowski}, {Thirugnanasambandam}, {Thomas}, {Thomas}, {Thomas}, {Thompson},
  {Thondapu}, {Thorne}, {Thrane}, {Tiwari}, {Tiwari}, {Tiwari}, {Toland},
  {Tolley}, {Tonelli}, {Tornasi}, {Torres-Forn{\'e}}, {Torrie}, {e Melo},
  {T{\"o}yr{\"a}}, {Tran}, {Trapananti}, {Travasso}, {Traylor}, {Tringali},
  {Tripathee}, {Trovato}, {Trudeau}, {Tsai}, {Tsang}, {Tse}, {Tso}, {Tsukada},
  {Tsuna}, {Tsutsui}, {Turconi}, {Ubhi}, {Udall}, {Ueno}, {Ugolini},
  {Unnikrishnan}, {Urban}, {Usman}, {Utina}, {Vahlbruch}, {Vajente}, {Vajpeyi},
  {Valdes}, {Valentini}, {Valsan}, {van Bakel}, {van Beuzekom}, {van den
  Brand}, {Van Den Broeck}, {Vander-Hyde}, {van der Schaaf}, {van Heijningen},
  {Vardaro}, {Vargas}, {Varma}, {Vass}, {Vas{\'u}th}, {Vecchio}, {Vedovato},
  {Veitch}, {Veitch}, {Venkateswara}, {Venneberg}, {Venugopalan}, {Verkindt},
  {Verma}, {Veske}, {Vetrano}, {Vicer{\'e}}, {Viets}, {Vijaykumar},
  {Villa-Ortega}, {Vinet}, {Vitale}, {Vo}, {Vocca}, {Vorvick}, {Vyatchanin},
  {Wade}, {Wade}, {Wade}, {Walet}, {Walker}, {Wallace}, {Wallace}, {Walsh},
  {Wang}, {Wang}, {Wang}, {Wang}, {Ward}, {Warner}, {Was}, {Washington},
  {Watchi}, {Weaver}, {Wei}, {Weinert}, {Weinstein}, {Weiss}, {Wellmann},
  {Wen}, {We{\ss}els}, {Westhouse}, {Wette}, {Whelan}, {White}, {White},
  {Whiting}, {Whittle}, {Wilken}, {Williams}, {Williams}, {Williamson},
  {Willis}, {Willke}, {Wilson}, {Wimmer}, {Winkler}, {Wipf}, {Woan}, {Woehler},
  {Wofford}, {Wong}, {Wrangel}, {Wright}, {Wu}, {Wysocki}, {Xiao}, {Yamamoto},
  {Yang}, {Yang}, {Yang}, {Yap}, {Yeeles}, {Yoon}, {Yu}, {Yu}, {Yuen},
  {Zadro{\.Z}ny}, {Zanolin}, {Zelenova}, {Zendri}, {Zevin}, {Zhang}, {Zhang},
  {Zhang}, {Zhang}, {Zhao}, {Zhao}, {Zheng}, {Zhou}, {Zhou}, {Zhu},
  {Zimmerman}, {Zlochower}, {Zucker}, {Zweizig}, {LIGO Scientific
  Collaboration}, \& {Virgo Collaboration}}]{GWTC-2}
{Abbott}, R., {Abbott}, T.~D., {Abraham}, S., {et~al.} 2021{\natexlab{b}},
  Physical Review X, 11, 021053, \dodoi{10.1103/PhysRevX.11.021053}

\bibitem[{{Abt}(1983)}]{Abt:1983:loguniform}
{Abt}, H.~A. 1983, \araa, 21, 343, \dodoi{10.1146/annurev.aa.21.090183.002015}

\bibitem[{Acernese {et~al.}(2015)}]{AdvancedVirgo}
Acernese, F., {et~al.} 2015, Class. Quant. Grav., 32, 024001,
  \dodoi{10.1088/0264-9381/32/2/024001}

\bibitem[{{Andrew} {et~al.}(2022){Andrew}, {Penoyre}, {Belokurov}, {Evans}, \&
  {Oh}}]{Shion:2022:BH-MS}
{Andrew}, S., {Penoyre}, Z., {Belokurov}, V., {Evans}, N.~W., \& {Oh}, S. 2022,
  \mnras, 516, 3661, \dodoi{10.1093/mnras/stac2532}

\bibitem[{{Andrews} {et~al.}(2022){Andrews}, {Taggart}, \&
  {Foley}}]{Andrews:2022:BH-MS}
{Andrews}, J.~J., {Taggart}, K., \& {Foley}, R. 2022, arXiv e-prints,
  arXiv:2207.00680, \dodoi{10.48550/arXiv.2207.00680}

\bibitem[{Apostolatos {et~al.}(1994)Apostolatos, Cutler, Sussman, \&
  Thorne}]{Apostolatos:1994:Precession}
Apostolatos, T.~A., Cutler, C., Sussman, G.~J., \& Thorne, K.~S. 1994, Phys.
  Rev. D, 49, 6274, \dodoi{10.1103/PhysRevD.49.6274}

\bibitem[{{Barrett} {et~al.}(2018){Barrett}, {Gaebel}, {Neijssel},
  {Vigna-G{\'o}mez}, {Stevenson}, {Berry}, {Farr}, \&
  {Mandel}}]{Barrett:2018:WRWinds}
{Barrett}, J.~W., {Gaebel}, S.~M., {Neijssel}, C.~J., {et~al.} 2018, \mnras,
  477, 4685, \dodoi{10.1093/mnras/sty908}

\bibitem[{{Barthelmy} {et~al.}(2005){Barthelmy}, {Barbier}, {Cummings},
  {Fenimore}, {Gehrels}, {Hullinger}, {Krimm}, {Markwardt}, {Palmer},
  {Parsons}, {Sato}, {Suzuki}, {Takahashi}, {Tashiro}, \&
  {Tueller}}]{Barthelmy2005}
{Barthelmy}, S.~D., {Barbier}, L.~M., {Cummings}, J.~R., {et~al.} 2005, \ssr,
  120, 143, \dodoi{10.1007/s11214-005-5096-3}

\bibitem[{{Bavera} {et~al.}(2020){Bavera}, {Fragos}, {Qin}, {Zapartas},
  {Neijssel}, {Mandel}, {Batta}, {Gaebel}, {Kimball}, \&
  {Stevenson}}]{Bavera:2020:BHSpinsCE}
{Bavera}, S.~S., {Fragos}, T., {Qin}, Y., {et~al.} 2020, \aap, 635, A97,
  \dodoi{10.1051/0004-6361/201936204}

\bibitem[{{Bavera} {et~al.}(2022){Bavera}, {Fragos}, {Zapartas},
  {Ramirez-Ruiz}, {Marchant}, {Kelley}, {Zevin}, {Andrews}, {Coughlin},
  {Dotter}, {Kovlakas}, {Misra}, {Serra-Perez}, {Qin}, {Rocha},
  {Rom{\'a}n-Garza}, {Tran}, \& {Xing}}]{Bavera:2022:LGRB}
{Bavera}, S.~S., {Fragos}, T., {Zapartas}, E., {et~al.} 2022, \aap, 657, L8,
  \dodoi{10.1051/0004-6361/202141979}

\bibitem[{{Belczynski} {et~al.}(2011){Belczynski}, {Bulik}, \&
  {Bailyn}}]{Belczynski:2011:CygX1FutureNSBH}
{Belczynski}, K., {Bulik}, T., \& {Bailyn}, C. 2011, \apjl, 742, L2,
  \dodoi{10.1088/2041-8205/742/1/L2}

\bibitem[{{Belczynski} {et~al.}(2021){Belczynski}, {Done}, \&
  {Lasota}}]{Belczynski:2021:AllApples}
{Belczynski}, K., {Done}, C., \& {Lasota}, J.~P. 2021, arXiv e-prints,
  arXiv:2111.09401.
\newblock \doarXiv{2111.09401}

\bibitem[{{Belczynski} {et~al.}(2008){Belczynski}, {Kalogera}, {Rasio}, {Taam},
  {Zezas}, {Bulik}, {Maccarone}, \& {Ivanova}}]{Belczynski:2008:StarTrack}
{Belczynski}, K., {Kalogera}, V., {Rasio}, F.~A., {et~al.} 2008, \apjs, 174,
  223, \dodoi{10.1086/521026}

\bibitem[{{Belczynski} {et~al.}(2022){Belczynski}, {Romagnolo}, {Olejak},
  {Klencki}, {Chattopadhyay}, {Stevenson}, {Coleman Miller}, {Lasota}, \&
  {Crowther}}]{Belczynski:2022:UncertainMassive}
{Belczynski}, K., {Romagnolo}, A., {Olejak}, A., {et~al.} 2022, \apj, 925, 69,
  \dodoi{10.3847/1538-4357/ac375a}

\bibitem[{Biscoveanu {et~al.}(2021)Biscoveanu, Isi, Vitale, \&
  Varma}]{Biscoveanu:2021:NewSpin}
Biscoveanu, S., Isi, M., Vitale, S., \& Varma, V. 2021, Phys. Rev. Lett., 126,
  171103, \dodoi{10.1103/PhysRevLett.126.171103}

\bibitem[{Bolton(1972)}]{Bolton:1972:CygX1Disco}
Bolton, C.~T. 1972, Nature, 235, 271, \dodoi{10.1038/235271b0}

\bibitem[{{Bondi} \& {Hoyle}(1944)}]{BondiHoyle:1944:accretion}
{Bondi}, H., \& {Hoyle}, F. 1944, \mnras, 104, 273,
  \dodoi{10.1093/mnras/104.5.273}

\bibitem[{{Breivik} {et~al.}(2020){Breivik}, {Coughlin}, {Zevin}, {Rodriguez},
  {Kremer}, {Ye}, {Andrews}, {Kurkowski}, {Digman}, {Larson}, \&
  {Rasio}}]{Breivik:2020:COSMIC}
{Breivik}, K., {Coughlin}, S., {Zevin}, M., {et~al.} 2020, \apj, 898, 71,
  \dodoi{10.3847/1538-4357/ab9d85}

\bibitem[{{Broekgaarden} {et~al.}(2022){Broekgaarden}, {Berger}, {Stevenson},
  {Justham}, {Mandel}, {Chru{\'s}li{\'n}ska}, {van Son}, {Wagg},
  {Vigna-G{\'o}mez}, {de Mink}, {Chattopadhyay}, \&
  {Neijssel}}]{Broekgaarden:2022}
{Broekgaarden}, F.~S., {Berger}, E., {Stevenson}, S., {et~al.} 2022, \mnras,
  516, 5737, \dodoi{10.1093/mnras/stac1677}

\bibitem[{{Brookshaw} \& {Tavani}(1993)}]{BrookshawTavani:1993}
{Brookshaw}, L., \& {Tavani}, M. 1993, \apj, 410, 719, \dodoi{10.1086/172789}

\bibitem[{{Burrows} {et~al.}(2005){Burrows}, {Hill}, {Nousek}, {Kennea},
  {Wells}, {Osborne}, {Abbey}, {Beardmore}, {Mukerjee}, {Short}, {Chincarini},
  {Campana}, {Citterio}, {Moretti}, {Pagani}, {Tagliaferri}, {Giommi},
  {Capalbi}, {Tamburelli}, {Angelini}, {Cusumano}, {Br{\"a}uninger}, {Burkert},
  \& {Hartner}}]{Burrows:2005:Swift}
{Burrows}, D.~N., {Hill}, J.~E., {Nousek}, J.~A., {et~al.} 2005, \ssr, 120,
  165, \dodoi{10.1007/s11214-005-5097-2}

\bibitem[{Chaty(2022)}]{Chaty:2022:HMXBs}
Chaty, S. 2022, in Accreting Binaries, 2514-3433 (IOP Publishing), 6--1 to
  6--58, \dodoi{10.1088/2514-3433/ac595fch6}

\bibitem[{{Chrimes} {et~al.}(2020){Chrimes}, {Stanway}, \&
  {Eldridge}}]{Chrimes:2020:LGRBs}
{Chrimes}, A.~A., {Stanway}, E.~R., \& {Eldridge}, J.~J. 2020, \mnras, 491,
  3479, \dodoi{10.1093/mnras/stz3246}

\bibitem[{{Compas Development Team}(2021)}]{TeamCOMPAS:2021:COMPAS}
{Compas Development Team}. 2021, {COMPAS: Rapid binary population synthesis
  code}, Astrophysics Source Code Library, record ascl:2105.005.
\newblock \doeprint{2105.005}

\bibitem[{Daflon {et~al.}(2001)Daflon, Cunha, Becker, \&
  Smith}]{Daflon:2001:OBstars}
Daflon, S., Cunha, K., Becker, S.~R., \& Smith, V.~V. 2001, The Astrophysical
  Journal, 552, 309, \dodoi{10.1086/320460}

\bibitem[{{Detmers} {et~al.}(2008){Detmers}, {Langer}, {Podsiadlowski}, \&
  {Izzard}}]{Detmers:2008:GRBs}
{Detmers}, R.~G., {Langer}, N., {Podsiadlowski}, P., \& {Izzard}, R.~G. 2008,
  \aap, 484, 831, \dodoi{10.1051/0004-6361:200809371}

\bibitem[{{Duro} {et~al.}(2016){Duro}, {Dauser}, {Grinberg},
  {Mi{\v{s}}kovi{\v{c}}ov{\'a}}, {Rodriguez}, {Tomsick}, {Hanke},
  {Pottschmidt}, {Nowak}, {Kreykenbohm}, {Cadolle Bel}, {Bodaghee}, {Lohfink},
  {Reynolds}, {Kendziorra}, {Kirsch}, {Staubert}, \&
  {Wilms}}]{Duro:2016:CygnusX1Spin}
{Duro}, R., {Dauser}, T., {Grinberg}, V., {et~al.} 2016, \aap, 589, A14,
  \dodoi{10.1051/0004-6361/201424740}

\bibitem[{{Eggleton}(1972)}]{Eggleton:1972:STARS}
{Eggleton}, P.~P. 1972, \mnras, 156, 361, \dodoi{10.1093/mnras/156.3.361}

\bibitem[{{Eggleton} {et~al.}(2011){Eggleton}, {Tout}, {Pols}, {Izzard},
  {Eldridge}, {Lesaffre}, {Stancliffe}, {Church}, \&
  {Lau}}]{Eggleton:2011:STARS}
{Eggleton}, P.~P., {Tout}, C., {Pols}, O., {et~al.} 2011, {STARS: A Stellar
  Evolution Code}, Astrophysics Source Code Library, record ascl:1107.008.
\newblock \doeprint{1107.008}

\bibitem[{{El-Badry} \& {Burdge}(2022)}]{El-BadryBurge:2022:BH-MS}
{El-Badry}, K., \& {Burdge}, K.~B. 2022, \mnras, 511, 24,
  \dodoi{10.1093/mnrasl/slab135}

\bibitem[{{El-Badry} {et~al.}(2022){El-Badry}, {Seeburger}, {Jayasinghe},
  {Rix}, {Almada}, {Conroy}, {Price-Whelan}, \& {Burdge}}]{El-Badry:2022:BH-MS}
{El-Badry}, K., {Seeburger}, R., {Jayasinghe}, T., {et~al.} 2022, \mnras, 512,
  5620, \dodoi{10.1093/mnras/stac815}

\bibitem[{{El-Badry} {et~al.}(2023){El-Badry}, {Rix}, {Quataert}, {Howard},
  {Isaacson}, {Fuller}, {Hawkins}, {Breivik}, {Wong}, {Rodriguez}, {Conroy},
  {Shahaf}, {Mazeh}, {Arenou}, {Burdge}, {Bashi}, {Faigler}, {Weisz},
  {Seeburger}, {Almada Monter}, \& {Wojno}}]{El-Badry:2023A:BH-MS}
{El-Badry}, K., {Rix}, H.-W., {Quataert}, E., {et~al.} 2023, \mnras, 518, 1057,
  \dodoi{10.1093/mnras/stac3140}

\bibitem[{{Eldridge} \& {Stanway}(2009)}]{Eldridge:2009:BPASS}
{Eldridge}, J.~J., \& {Stanway}, E.~R. 2009, \mnras, 400, 1019,
  \dodoi{10.1111/j.1365-2966.2009.15514.x}

\bibitem[{{Fishbach} \& {Kalogera}(2022)}]{Fishbach:2022:ApplesOranges}
{Fishbach}, M., \& {Kalogera}, V. 2022, \apjl, 929, L26,
  \dodoi{10.3847/2041-8213/ac64a5}

\bibitem[{{Fortin} {et~al.}(2023){Fortin}, {Garcia}, {Simaz-Bunzel}, \&
  {Chaty}}]{Fortin:2023:HMXBCatalogue}
{Fortin}, F., {Garcia}, F., {Simaz-Bunzel}, A., \& {Chaty}, S. 2023, arXiv
  e-prints, arXiv:2302.02656, \dodoi{10.48550/arXiv.2302.02656}

\bibitem[{{Fragos} {et~al.}(2022){Fragos}, {Andrews}, {Bavera}, {Berry},
  {Coughlin}, {Dotter}, {Giri}, {Kalogera}, {Katsaggelos}, {Kovlakas},
  {Lalvani}, {Misra}, {Srivastava}, {Qin}, {Rocha}, {Roman-Garza}, {Serra},
  {Stahle}, {Sun}, {Teng}, {Trajcevski}, {Hai Tran}, {Xing}, {Zapartas}, \&
  {Zevin}}]{Fragos:2022:POSYDON}
{Fragos}, T., {Andrews}, J.~J., {Bavera}, S.~S., {et~al.} 2022, arXiv e-prints,
  arXiv:2202.05892.
\newblock \doarXiv{2202.05892}

\bibitem[{{Frost} {et~al.}(2022){Frost}, {Bodensteiner}, {Rivinius}, {Baade},
  {Merand}, {Selman}, {Abdul-Masih}, {Banyard}, {Bordier}, {Dsilva},
  {Hawcroft}, {Mahy}, {Reggiani}, {Shenar}, {Cabezas}, {Hadrava}, {Heida},
  {Klement}, \& {Sana}}]{Frost:2022:HR6819Refute}
{Frost}, A.~J., {Bodensteiner}, J., {Rivinius}, T., {et~al.} 2022, \aap, 659,
  L3, \dodoi{10.1051/0004-6361/202143004}

\bibitem[{{Fryer} {et~al.}(2012){Fryer}, {Belczynski}, {Wiktorowicz},
  {Dominik}, {Kalogera}, \& {Holz}}]{FryerKicks}
{Fryer}, C.~L., {Belczynski}, K., {Wiktorowicz}, G., {et~al.} 2012, \apj, 749,
  91, \dodoi{10.1088/0004-637X/749/1/91}

\bibitem[{{Fuller} \& {Ma}(2019)}]{FullerMa:2019:BHNatalSpin}
{Fuller}, J., \& {Ma}, L. 2019, \apjl, 881, L1,
  \dodoi{10.3847/2041-8213/ab339b}

\bibitem[{{Gallegos-Garcia} {et~al.}(2022){Gallegos-Garcia}, {Fishbach},
  {Kalogera}, {L Berry}, \& {Doctor}}]{GallegosGarcia:2022:HMXBs}
{Gallegos-Garcia}, M., {Fishbach}, M., {Kalogera}, V., {L Berry}, C.~P., \&
  {Doctor}, Z. 2022, \apjl, 938, L19, \dodoi{10.3847/2041-8213/ac96ef}

\bibitem[{{Giesers} {et~al.}(2018){Giesers}, {Dreizler}, {Husser}, {Kamann},
  {Anglada Escud{\'e}}, {Brinchmann}, {Carollo}, {Roth}, {Weilbacher}, \&
  {Wisotzki}}]{Giesers:2018:BH-MS}
{Giesers}, B., {Dreizler}, S., {Husser}, T.-O., {et~al.} 2018, \mnras, 475,
  L15, \dodoi{10.1093/mnrasl/slx203}

\bibitem[{{Giesers} {et~al.}(2019){Giesers}, {Kamann}, {Dreizler}, {Husser},
  {Askar}, {G{\"o}ttgens}, {Brinchmann}, {Latour}, {Weilbacher}, {Wendt}, \&
  {Roth}}]{Giesers:2019:BH-MS}
{Giesers}, B., {Kamann}, S., {Dreizler}, S., {et~al.} 2019, \aap, 632, A3,
  \dodoi{10.1051/0004-6361/201936203}

\bibitem[{{Goldstein} {et~al.}(2016){Goldstein}, {Connaughton}, {Briggs}, \&
  {Burns}}]{Goldstein:2016}
{Goldstein}, A., {Connaughton}, V., {Briggs}, M.~S., \& {Burns}, E. 2016, \apj,
  818, 18, \dodoi{10.3847/0004-637X/818/1/18}

\bibitem[{{Gomez} \& {Grindlay}(2021)}]{GomezGrindlay:2021:HD96670}
{Gomez}, S., \& {Grindlay}, J.~E. 2021, \apj, 913, 48,
  \dodoi{10.3847/1538-4357/abf24c}

\bibitem[{{Gottlieb} {et~al.}(2023){Gottlieb}, {Jacquemin-Ide}, {Lowell},
  {Tchekhovskoy}, \& {Ramirez-Ruiz}}]{Gottlieb:2023:Collapsars}
{Gottlieb}, O., {Jacquemin-Ide}, J., {Lowell}, B., {Tchekhovskoy}, A., \&
  {Ramirez-Ruiz}, E. 2023, arXiv e-prints, arXiv:2302.07271,
  \dodoi{10.48550/arXiv.2302.07271}

\bibitem[{{Gou} {et~al.}(2011){Gou}, {McClintock}, {Reid}, {Orosz}, {Steiner},
  {Narayan}, {Xiang}, {Remillard}, {Arnaud}, \&
  {Davis}}]{Gou:2011:CygnusX1Spin}
{Gou}, L., {McClintock}, J.~E., {Reid}, M.~J., {et~al.} 2011, \apj, 742, 85,
  \dodoi{10.1088/0004-637X/742/2/85}

\bibitem[{{Gou} {et~al.}(2014){Gou}, {McClintock}, {Remillard}, {Steiner},
  {Reid}, {Orosz}, {Narayan}, {Hanke}, \&
  {Garc{\'\i}a}}]{Gou:2014:CygnusX1Spin}
{Gou}, L., {McClintock}, J.~E., {Remillard}, R.~A., {et~al.} 2014, \apj, 790,
  29, \dodoi{10.1088/0004-637X/790/1/29}

\bibitem[{{Green} {et~al.}(2019){Green}, {Schlafly}, {Zucker}, {Speagle}, \&
  {Finkbeiner}}]{Green2019}
{Green}, G.~M., {Schlafly}, E., {Zucker}, C., {Speagle}, J.~S., \&
  {Finkbeiner}, D. 2019, \apj, 887, 93, \dodoi{10.3847/1538-4357/ab5362}

\bibitem[{{Grudzinska} {et~al.}(2015){Grudzinska}, {Belczynski}, {Casares}, {de
  Mink}, {Ziolkowski}, {Negueruela}, {Rib{\'o}}, {Ribas}, {Paredes}, {Herrero},
  \& {Benacquista}}]{Grudzinska:2015:MWC656}
{Grudzinska}, M., {Belczynski}, K., {Casares}, J., {et~al.} 2015, \mnras, 452,
  2773, \dodoi{10.1093/mnras/stv1419}

\bibitem[{{Hamann} \& {Koesterke}(1998)}]{HamannKoesterke:1998:WRWinds}
{Hamann}, W.~R., \& {Koesterke}, L. 1998, \aap, 335, 1003

\bibitem[{{Heger} {et~al.}(2000){Heger}, {Langer}, \&
  {Woosley}}]{Heger:2000:RotatingMassiveStars}
{Heger}, A., {Langer}, N., \& {Woosley}, S.~E. 2000, \apj, 528, 368,
  \dodoi{10.1086/308158}

\bibitem[{{HI4PI Collaboration} {et~al.}(2016){HI4PI Collaboration}, {Ben
  Bekhti}, {Fl{\"o}er}, {Keller}, {Kerp}, {Lenz}, {Winkel}, {Bailin},
  {Calabretta}, {Dedes}, {Ford}, {Gibson}, {Haud}, {Janowiecki}, {Kalberla},
  {Lockman}, {McClure-Griffiths}, {Murphy}, {Nakanishi}, {Pisano}, \&
  {Staveley-Smith}}]{HI4PI2016}
{HI4PI Collaboration}, {Ben Bekhti}, N., {Fl{\"o}er}, L., {et~al.} 2016, \aap,
  594, A116, \dodoi{10.1051/0004-6361/201629178}

\bibitem[{{Hirai} \& {Mandel}(2021)}]{HiraiMandel:2021:AccretionDiskXRays}
{Hirai}, R., \& {Mandel}, I. 2021, \pasa, 38, e056,
  \dodoi{10.1017/pasa.2021.53}

\bibitem[{{Hirai} \& {Mandel}(2022)}]{HiraiMandel:2022:CommonEnvelope}
---. 2022, \apjl, 937, L42, \dodoi{10.3847/2041-8213/ac9519}

\bibitem[{{Hori} {et~al.}(2018){Hori}, {Shidatsu}, {Ueda}, {Kawamuro}, {Morii},
  {Nakahira}, {Isobe}, {Kawai}, {Mihara}, {Matsuoka}, {Morita}, {Nakajima},
  {Negoro}, {Oda}, {Sakamoto}, {Serino}, {Sugizaki}, {Tanimoto}, {Tomida},
  {Tsuboi}, {Tsunemi}, {Ueno}, {Yamaoka}, {Yamada}, {Yoshida}, {Iwakiri},
  {Kawakubo}, {Sugawara}, {Sugita}, {Tachibana}, \& {Yoshii}}]{Hori2018}
{Hori}, T., {Shidatsu}, M., {Ueda}, Y., {et~al.} 2018, \apjs, 235, 7,
  \dodoi{10.3847/1538-4365/aaa89c}

\bibitem[{{Hoyle} \& {Lyttleton}(1939)}]{HoyleLyttleton:939:accretion}
{Hoyle}, F., \& {Lyttleton}, R.~A. 1939, Proceedings of the Cambridge
  Philosophical Society, 35, 405, \dodoi{10.1017/S0305004100021150}

\bibitem[{{Hurley} {et~al.}(2000){Hurley}, {Pols}, \& {Tout}}]{Hurley:2000:BSE}
{Hurley}, J.~R., {Pols}, O.~R., \& {Tout}, C.~A. 2000, \mnras, 315, 543,
  \dodoi{10.1046/j.1365-8711.2000.03426.x}

\bibitem[{{Hurley} {et~al.}(2002){Hurley}, {Tout}, \& {Pols}}]{Hurley:2002:BSE}
{Hurley}, J.~R., {Tout}, C.~A., \& {Pols}, O.~R. 2002, \mnras, 329, 897,
  \dodoi{10.1046/j.1365-8711.2002.05038.x}

\bibitem[{{Illarionov} \&
  {Sunyaev}(1975)}]{IllarionovSunyaev:1975:XRayAccretion}
{Illarionov}, A.~F., \& {Sunyaev}, R.~A. 1975, \aap, 39, 185

\bibitem[{{Jermyn} {et~al.}(2023){Jermyn}, {Bauer}, {Schwab}, {Farmer}, {Ball},
  {Bellinger}, {Dotter}, {Joyce}, {Marchant}, {Mombarg}, {Wolf}, {Sunny Wong},
  {Cinquegrana}, {Farrell}, {Smolec}, {Thoul}, {Cantiello}, {Herwig}, {Toloza},
  {Bildsten}, {Townsend}, \& {Timmes}}]{Jermyn:2023:MESA}
{Jermyn}, A.~S., {Bauer}, E.~B., {Schwab}, J., {et~al.} 2023, \apjs, 265, 15,
  \dodoi{10.3847/1538-4365/acae8d}

\bibitem[{{Karino} {et~al.}(2019){Karino}, {Nakamura}, \&
  {Taani}}]{Karino:2019}
{Karino}, S., {Nakamura}, K., \& {Taani}, A. 2019, \pasj, 71, 58,
  \dodoi{10.1093/pasj/psz034}

\bibitem[{{Kawano} {et~al.}(2017){Kawano}, {Done}, {Yamada}, {Takahashi},
  {Axelsson}, \& {Fukazawa}}]{Kawano:2017:CygnusX1Spin}
{Kawano}, T., {Done}, C., {Yamada}, S., {et~al.} 2017, \pasj, 69, 36,
  \dodoi{10.1093/pasj/psx009}

\bibitem[{{King} \& {Kolb}(1999)}]{KingKolb:1999}
{King}, A.~R., \& {Kolb}, U. 1999, \mnras, 305, 654,
  \dodoi{10.1046/j.1365-8711.1999.02482.x}

\bibitem[{{Kroupa}(2002)}]{Kroupa}
{Kroupa}, P. 2002, Science, 295, 82, \dodoi{10.1126/science.1067524}

\bibitem[{{Lehmer} {et~al.}(2021){Lehmer}, {Eufrasio}, {Basu-Zych}, {Doore},
  {Fragos}, {Garofali}, {Kovlakas}, {Williams}, {Zezas}, \&
  {Santana-Silva}}]{Lehmer:2021:HMXB-L-Z-relation}
{Lehmer}, B.~D., {Eufrasio}, R.~T., {Basu-Zych}, A., {et~al.} 2021, \apj, 907,
  17, \dodoi{10.3847/1538-4357/abcec1}

\bibitem[{{Liotine} {et~al.}(2022){Liotine}, {Zevin}, {Berry}, {Doctor}, \&
  {Kalogera}}]{Liotine:2022:SelectionEffects}
{Liotine}, C., {Zevin}, M., {Berry}, C., {Doctor}, Z., \& {Kalogera}, V. 2022,
  arXiv e-prints, arXiv:2210.01825.
\newblock \doarXiv{2210.01825}

\bibitem[{Long {et~al.}(1981)Long, Dodorico, Charles, \&
  Dopita}]{Long:1981:M33X7Disco}
Long, K., Dodorico, S., Charles, P., \& Dopita, M. 1981, ApJ, 246,
  \dodoi{10.1086/183553}

\bibitem[{{Lower} {et~al.}(2018){Lower}, {Thrane}, {Lasky}, \&
  {Smith}}]{Lower:2018:Eccentricity}
{Lower}, M.~E., {Thrane}, E., {Lasky}, P.~D., \& {Smith}, R. 2018, \prd, 98,
  083028, \dodoi{10.1103/PhysRevD.98.083028}

\bibitem[{{MacLeod} \& {Loeb}(2020)}]{Macleod:2020:MassLoss}
{MacLeod}, M., \& {Loeb}, A. 2020, \apj, 895, 29,
  \dodoi{10.3847/1538-4357/ab89b6}

\bibitem[{{Mahy} {et~al.}(2022){Mahy}, {Sana}, {Shenar}, {Sen}, {Langer},
  {Marchant}, {Abdul-Masih}, {Banyard}, {Bodensteiner}, {Bowman}, {Dsilva},
  {Fabry}, {Hawcroft}, {Janssens}, {Van Reeth}, \&
  {Eldridge}}]{Mahy:2022:HD130298}
{Mahy}, L., {Sana}, H., {Shenar}, T., {et~al.} 2022, \aap, 664, A159,
  \dodoi{10.1051/0004-6361/202243147}

\bibitem[{{Mandel} \& {Fragos}(2020)}]{MandelFragos:2020:GW190412}
{Mandel}, I., \& {Fragos}, T. 2020, \apjl, 895, L28,
  \dodoi{10.3847/2041-8213/ab8e41}

\bibitem[{{Mandel} \& {M{\"u}ller}(2020)}]{MandelMuller:2020:kicks}
{Mandel}, I., \& {M{\"u}ller}, B. 2020, \mnras, 499, 3214,
  \dodoi{10.1093/mnras/staa3043}

\bibitem[{{Mapelli}(2020)}]{Mapelli:2020:Review}
{Mapelli}, M. 2020, Frontiers in Astronomy and Space Sciences, 7, 38,
  \dodoi{10.3389/fspas.2020.00038}

\bibitem[{{Mark} {et~al.}(1969){Mark}, {Price}, {Rodrigues}, {Seward}, \&
  {Swift}}]{Mark:1969:LMCX1Disco}
{Mark}, H., {Price}, R., {Rodrigues}, R., {Seward}, F.~D., \& {Swift}, C.~D.
  1969, \apjl, 155, L143, \dodoi{10.1086/180322}

\bibitem[{{Matsuoka} {et~al.}(2009){Matsuoka}, {Kawasaki}, {Ueno}, {Tomida},
  {Kohama}, {Suzuki}, {Adachi}, {Ishikawa}, {Mihara}, {Sugizaki}, {Isobe},
  {Nakagawa}, {Tsunemi}, {Miyata}, {Kawai}, {Kataoka}, {Morii}, {Yoshida},
  {Negoro}, {Nakajima}, {Ueda}, {Chujo}, {Yamaoka}, {Yamazaki}, {Nakahira},
  {You}, {Ishiwata}, {Miyoshi}, {Eguchi}, {Hiroi}, {Katayama}, \&
  {Ebisawa}}]{MAXI2009}
{Matsuoka}, M., {Kawasaki}, K., {Ueno}, S., {et~al.} 2009, \pasj, 61, 999,
  \dodoi{10.1093/pasj/61.5.999}

\bibitem[{{Miller} \& {Miller}(2015)}]{MillerMiller:2015}
{Miller}, M.~C., \& {Miller}, J.~M. 2015, \physrep, 548, 1,
  \dodoi{10.1016/j.physrep.2014.09.003}

\bibitem[{Miller \& Miller(2015)}]{MillerMiller:2015:Spins}
Miller, M.~C., \& Miller, J.~M. 2015, Physics Reports, 548, 1

\bibitem[{{Miller-Jones} {et~al.}(2021){Miller-Jones}, {Bahramian}, {Orosz},
  {Mandel}, {Gou}, {Maccarone}, {Neijssel}, {Zhao}, {Zi{\'o}{\l}kowski},
  {Reid}, {Uttley}, {Zheng}, {Byun}, {Dodson}, {Grinberg}, {Jung}, {Kim},
  {Marcote}, {Markoff}, {Rioja}, {Rushton}, {Russell}, {Sivakoff}, {Tetarenko},
  {Tudose}, \& {Wilms}}]{MillerJones:2021:CygX1}
{Miller-Jones}, J. C.~A., {Bahramian}, A., {Orosz}, J.~A., {et~al.} 2021,
  Science, 371, 1046, \dodoi{10.1126/science.abb3363}

\bibitem[{{Mineo} {et~al.}(2012){Mineo}, {Gilfanov}, \&
  {Sunyaev}}]{Mineo:2012:XRayBinaryDuration}
{Mineo}, S., {Gilfanov}, M., \& {Sunyaev}, R. 2012, \mnras, 419, 2095,
  \dodoi{10.1111/j.1365-2966.2011.19862.x}

\bibitem[{{Misra} {et~al.}(2022){Misra}, {Kovlakas}, {Fragos}, {Lazzarini},
  {Bavera}, {Lehmer}, {Zezas}, {Zapartas}, {Xing}, {Andrews}, {Dotter},
  {Rocha}, {Srivastava}, \& {Sun}}]{Misra:2022:XrayLuminosityPopsynth}
{Misra}, D., {Kovlakas}, K., {Fragos}, T., {et~al.} 2022, arXiv e-prints,
  arXiv:2209.05505.
\newblock \doarXiv{2209.05505}

\bibitem[{{Mukai}(1993)}]{Mukai1993}
{Mukai}, K. 1993, Legacy, 3, 21

\bibitem[{{Neijssel} {et~al.}(2021){Neijssel}, {Vinciguerra},
  {Vigna-G{\'o}mez}, {Hirai}, {Miller-Jones}, {Bahramian}, {Maccarone}, \&
  {Mandel}}]{Neijssel:2021:WRmassloss}
{Neijssel}, C.~J., {Vinciguerra}, S., {Vigna-G{\'o}mez}, A., {et~al.} 2021,
  \apj, 908, 118, \dodoi{10.3847/1538-4357/abde4a}

\bibitem[{{Neijssel} {et~al.}(2019){Neijssel}, {Vigna-G{\'o}mez}, {Stevenson},
  {Barrett}, {Gaebel}, {Broekgaarden}, {de Mink}, {Sz{\'e}csi}, {Vinciguerra},
  \& {Mandel}}]{Neijssel:2019:MSSFR}
{Neijssel}, C.~J., {Vigna-G{\'o}mez}, A., {Stevenson}, S., {et~al.} 2019,
  \mnras, 490, 3740, \dodoi{10.1093/mnras/stz2840}

\bibitem[{Orosz {et~al.}(2014)Orosz, Steiner, McClintock, Buxton, Bailyn,
  Steeghs, Guberman, \& Torres}]{Orosz:2014:LMCX1}
Orosz, J.~A., Steiner, J.~F., McClintock, J.~E., {et~al.} 2014, The
  Astrophysical Journal, 794, 154, \dodoi{10.1088/0004-637x/794/2/154}

\bibitem[{Orosz {et~al.}(2007)Orosz, McClintock, Narayan, Bailyn, Hartman,
  Macri, Liu, Pietsch, Remillard, Shporer, \& Mazeh}]{Orosz:2007:M33X7}
Orosz, J.~A., McClintock, J.~E., Narayan, R., {et~al.} 2007, Nature, 449, 872,
  \dodoi{10.1038/nature06218}

\bibitem[{Orosz {et~al.}(2009)Orosz, Steeghs, McClintock, Torres, Bochkov, Gou,
  Narayan, Blaschak, Levine, Remillard, Bailyn, Dwyer, \&
  Buxton}]{Orosz:2009:LMCX1}
Orosz, J.~A., Steeghs, D., McClintock, J.~E., {et~al.} 2009, The Astrophysical
  Journal, 697, 573, \dodoi{10.1088/0004-637x/697/1/573}

\bibitem[{{Packet}(1981)}]{Packet:1981:SpinUp}
{Packet}, W. 1981, \aap, 102, 17

\bibitem[{{Paczy{\'n}ski}(1998)}]{Paczynski:1998:GRBs}
{Paczy{\'n}ski}, B. 1998, \apjl, 494, L45, \dodoi{10.1086/311148}

\bibitem[{{Paxton} {et~al.}(2011){Paxton}, {Bildsten}, {Dotter}, {Herwig},
  {Lesaffre}, \& {Timmes}}]{Paxton:2011:MESA}
{Paxton}, B., {Bildsten}, L., {Dotter}, A., {et~al.} 2011, \apjs, 192, 3,
  \dodoi{10.1088/0067-0049/192/1/3}

\bibitem[{{Pietsch} {et~al.}(2006){Pietsch}, {Haberl}, {Sasaki}, {Gaetz},
  {Plucinsky}, {Ghavamian}, {Long}, \& {Pannuti}}]{Pietsch:2006:M33X7}
{Pietsch}, W., {Haberl}, F., {Sasaki}, M., {et~al.} 2006, \apj, 646, 420,
  \dodoi{10.1086/504704}

\bibitem[{{Podsiadlowski} {et~al.}(2003){Podsiadlowski}, {Rappaport}, \&
  {Han}}]{Podsiadlowski:2003:CygnusX1}
{Podsiadlowski}, P., {Rappaport}, S., \& {Han}, Z. 2003, \mnras, 341, 385,
  \dodoi{10.1046/j.1365-8711.2003.06464.x}

\bibitem[{{Podsiadlowski} {et~al.}(2002){Podsiadlowski}, {Rappaport}, \&
  {Pfahl}}]{Podsiadlowski:2002:XRayBinaryEvolution}
{Podsiadlowski}, P., {Rappaport}, S., \& {Pfahl}, E.~D. 2002, \apj, 565, 1107,
  \dodoi{10.1086/324686}

\bibitem[{{Popham} \& {Narayan}(1991)}]{PophamNarayan:1991}
{Popham}, R., \& {Narayan}, R. 1991, \apj, 370, 604, \dodoi{10.1086/169847}

\bibitem[{{Price-Whelan} {et~al.}(2020){Price-Whelan}, {Hogg}, {Rix}, {Beaton},
  {Lewis}, {Nidever}, {Almeida}, {Badenes}, {Barba}, {Beers}, {Carlberg}, {De
  Lee}, {Fern{\'a}ndez-Trincado}, {Frinchaboy}, {Garc{\'\i}a-Hern{\'a}ndez},
  {Green}, {Hasselquist}, {Longa-Pe{\~n}a}, {Majewski}, {Nitschelm}, {Sobeck},
  {Stassun}, {Stringfellow}, \& {Troup}}]{Price-Whelan:2020:BH-MS}
{Price-Whelan}, A.~M., {Hogg}, D.~W., {Rix}, H.-W., {et~al.} 2020, \apj, 895,
  2, \dodoi{10.3847/1538-4357/ab8acc}

\bibitem[{{Qin} {et~al.}(2019){Qin}, {Marchant}, {Fragos}, {Meynet}, \&
  {Kalogera}}]{Qin:2019:HMXBspin}
{Qin}, Y., {Marchant}, P., {Fragos}, T., {Meynet}, G., \& {Kalogera}, V. 2019,
  \apjl, 870, L18, \dodoi{10.3847/2041-8213/aaf97b}

\bibitem[{{Ramachandran} {et~al.}(2022){Ramachandran}, {Oskinova}, {Hamann},
  {Sander}, {Todt}, {Pauli}, {Shenar}, {Torrej{\'o}n}, {Postnov}, {Blondin},
  {Bozzo}, {Hainich}, \& {Massa}}]{Ramachandran:2022:M33X7}
{Ramachandran}, V., {Oskinova}, L.~M., {Hamann}, W.~R., {et~al.} 2022, \aap,
  667, A77, \dodoi{10.1051/0004-6361/202243683}

\bibitem[{{Reynolds}(2021)}]{Reynolds:2020}
{Reynolds}, C.~S. 2021, \araa, 59, 117,
  \dodoi{10.1146/annurev-astro-112420-035022}

\bibitem[{{Rivinius} {et~al.}(2020){Rivinius}, {Baade}, {Hadrava}, {Heida}, \&
  {Klement}}]{Rivinius:2020:HR6819}
{Rivinius}, T., {Baade}, D., {Hadrava}, P., {Heida}, M., \& {Klement}, R. 2020,
  \aap, 637, L3, \dodoi{10.1051/0004-6361/202038020}

\bibitem[{{Rivinius} {et~al.}(2022){Rivinius}, {Klement}, {Chojnowski},
  {Baade}, {Shepard}, \& {Hadrava}}]{Rivinius:2022:MWC656Refute}
{Rivinius}, T., {Klement}, R., {Chojnowski}, S.~D., {et~al.} 2022, arXiv
  e-prints, arXiv:2208.12315, \dodoi{10.48550/arXiv.2208.12315}

\bibitem[{{Romero-Shaw} {et~al.}(2019){Romero-Shaw}, {Lasky}, \&
  {Thrane}}]{Romero-Shaw:2019:SearchingForEccentricity}
{Romero-Shaw}, I.~M., {Lasky}, P.~D., \& {Thrane}, E. 2019, \mnras, 490, 5210,
  \dodoi{10.1093/mnras/stz2996}

\bibitem[{{Sana} {et~al.}(2012){Sana}, {de Mink}, {de Koter}, {Langer},
  {Evans}, {Gieles}, {Gosset}, {Izzard}, {Le Bouquin}, \&
  {Schneider}}]{Sana:2012:MassRatio}
{Sana}, H., {de Mink}, S.~E., {de Koter}, A., {et~al.} 2012, Science, 337, 444,
  \dodoi{10.1126/science.1223344}

\bibitem[{{Sander} {et~al.}(2020){Sander}, {Vink}, \&
  {Hamann}}]{Sander:2020:WRWinds}
{Sander}, A. A.~C., {Vink}, J.~S., \& {Hamann}, W.~R. 2020, \mnras, 491, 4406,
  \dodoi{10.1093/mnras/stz3064}

\bibitem[{{Saracino} {et~al.}(2022){Saracino}, {Kamann}, {Guarcello}, {Usher},
  {Bastian}, {Cabrera-Ziri}, {Gieles}, {Dreizler}, {Da Costa}, {Husser}, \&
  {H{\'e}nault-Brunet}}]{Saracino:2022:BH-MS}
{Saracino}, S., {Kamann}, S., {Guarcello}, M.~G., {et~al.} 2022, \mnras, 511,
  2914, \dodoi{10.1093/mnras/stab3159}

\bibitem[{{Saracino} {et~al.}(2023){Saracino}, {Shenar}, {Kamann}, {Bastian},
  {Gieles}, {Usher}, {Bodensteiner}, {Kochoska}, {Orosz}, \&
  {Sana}}]{Saracino:2023:BH-MS}
{Saracino}, S., {Shenar}, T., {Kamann}, S., {et~al.} 2023, \mnras, 521, 3162,
  \dodoi{10.1093/mnras/stad764}

\bibitem[{{Schr{\o}der} {et~al.}(2021){Schr{\o}der}, {MacLeod}, {Ramirez-Ruiz},
  {Mandel}, {Fragos}, {Loeb}, \&
  {Everson}}]{Schroder:2021:RadiationDrivenWinds}
{Schr{\o}der}, S.~L., {MacLeod}, M., {Ramirez-Ruiz}, E., {et~al.} 2021, arXiv
  e-prints, arXiv:2107.09675.
\newblock \doarXiv{2107.09675}

\bibitem[{{Schulz} {et~al.}(2002){Schulz}, {Cui}, {Canizares}, {Marshall},
  {Lee}, {Miller}, \& {Lewin}}]{Schulz2002}
{Schulz}, N.~S., {Cui}, W., {Canizares}, C.~R., {et~al.} 2002, \apj, 565, 1141,
  \dodoi{10.1086/324482}

\bibitem[{{Seifina} \& {Titarchuk}(2010)}]{Seifina:2010:SS433}
{Seifina}, E., \& {Titarchuk}, L. 2010, \apj, 722, 586,
  \dodoi{10.1088/0004-637X/722/1/586}

\bibitem[{{Sen} {et~al.}(2021){Sen}, {Xu}, {Langer}, {El Mellah},
  {Sch{\"u}rmann}, \& {Quast}}]{Sen:2021:PopSynthXray}
{Sen}, K., {Xu}, X.~T., {Langer}, N., {et~al.} 2021, \aap, 652, A138,
  \dodoi{10.1051/0004-6361/202141214}

\bibitem[{{Shenar} {et~al.}(2022{\natexlab{a}}){Shenar}, {Sana}, {Mahy},
  {El-Badry}, {Marchant}, {Langer}, {Hawcroft}, {Fabry}, {Sen}, {Almeida},
  {Abdul-Masih}, {Bodensteiner}, {Crowther}, {Gieles}, {Gromadzki},
  {H{\'e}nault-Brunet}, {Herrero}, {de Koter}, {Iwanek}, {Koz{\l}owski},
  {Lennon}, {Ma{\'\i}z Apell{\'a}niz}, {Mr{\'o}z}, {Moffat}, {Picco},
  {Pietrukowicz}, {Poleski}, {Rybicki}, {Schneider}, {Skowron}, {Skowron},
  {Soszy{\'n}ski}, {Szyma{\'n}ski}, {Toonen}, {Udalski}, {Ulaczyk}, {Vink}, \&
  {Wrona}}]{Tomer:2022:BH-MS}
{Shenar}, T., {Sana}, H., {Mahy}, L., {et~al.} 2022{\natexlab{a}}, Nature
  Astronomy, 6, 1085, \dodoi{10.1038/s41550-022-01730-y}

\bibitem[{{Shenar} {et~al.}(2022{\natexlab{b}}){Shenar}, {Sana}, {Mahy},
  {Ma{\'\i}z Apell{\'a}niz}, {Crowther}, {Gromadzki}, {Herrero}, {Langer},
  {Marchant}, {Schneider}, {Sen}, {Soszy{\'n}ski}, \&
  {Toonen}}]{Shenar:2022:BH-MS}
---. 2022{\natexlab{b}}, \aap, 665, A148, \dodoi{10.1051/0004-6361/202244245}

\bibitem[{{Shimanskii} {et~al.}(2012){Shimanskii}, {Karitskaya}, {Bochkarev},
  {Galazutdinov}, {Lyuty}, \& {Shimanskaya}}]{Shimanskii:2012:CygnusX1}
{Shimanskii}, V.~V., {Karitskaya}, E.~A., {Bochkarev}, N.~G., {et~al.} 2012,
  Astronomy Reports, 56, 741, \dodoi{10.1134/S106377291210006X}

\bibitem[{{Stanway} \& {Eldridge}(2018)}]{Stanway:2018:BPASS}
{Stanway}, E.~R., \& {Eldridge}, J.~J. 2018, \mnras, 479, 75,
  \dodoi{10.1093/mnras/sty1353}

\bibitem[{{Stevenson} \&
  {Clarke}(2022)}]{StevensonClarke:2022:COMPASConstraints}
{Stevenson}, S., \& {Clarke}, T.~A. 2022, \mnras, 517, 4034,
  \dodoi{10.1093/mnras/stac2936}

\bibitem[{{Tanikawa} {et~al.}(2023){Tanikawa}, {Hattori}, {Kawanaka},
  {Kinugawa}, {Shikauchi}, \& {Tsuna}}]{Tanikawa:2023:BH-MS}
{Tanikawa}, A., {Hattori}, K., {Kawanaka}, N., {et~al.} 2023, \apj, 946, 79,
  \dodoi{10.3847/1538-4357/acbf36}

\bibitem[{{Team COMPAS: Riley} {et~al.}(2022){Team COMPAS: Riley}, {Agrawal},
  {Barrett}, {Boyett}, {Broekgaarden}, {Chattopadhyay}, {Gaebel}, {Gittins},
  {Hirai}, {Howitt}, {Justham}, {Khandelwal}, {Kummer}, {Lau}, {Mandel}, {de
  Mink}, {Neijssel}, {Riley}, {van Son}, {Stevenson}, {Vigna-G{\'o}mez},
  {Vinciguerra}, {Wagg}, {Willcox}, \& {Team Compas}}]{COMPAS:2021}
{Team COMPAS: Riley}, J., {Agrawal}, P., {Barrett}, J.~W., {et~al.} 2022,
  \apjs, 258, 34, \dodoi{10.3847/1538-4365/ac416c}

\bibitem[{{The LIGO Scientific Collaboration} {et~al.}(2021){The LIGO
  Scientific Collaboration}, {the Virgo Collaboration}, {the KAGRA
  Collaboration}, {Abbott}, {Abbott}, {Acernese}, {Ackley}, {Adams},
  {Adhikari}, {Adhikari}, {Adya}, {Affeldt}, {Agarwal}, {Agathos}, {Agatsuma},
  {Aggarwal}, {Aguiar}, {Aiello}, {Ain}, {Ajith}, {Akutsu}, {Albanesi},
  {Allocca}, {Altin}, {Amato}, {Anand}, {Anand}, {Ananyeva}, {Anderson},
  {Anderson}, {Ando}, {Andrade}, {Andres}, {Andri{\'c}}, {Angelova}, {Ansoldi},
  {Antelis}, {Antier}, {Antonini}, {Appert}, {Arai}, {Arai}, {Arai}, {Araki},
  {Araya}, {Araya}, {Areeda}, {Ar{\`e}ne}, {Aritomi}, {Arnaud}, {Aronson},
  {Arun}, {Asada}, {Asali}, {Ashton}, {Aso}, {Assiduo}, {Aston}, {Astone},
  {Aubin}, {Austin}, {Babak}, {Badaracco}, {Bader}, {Badger}, {Bae}, {Bae},
  {Baer}, {Bagnasco}, {Bai}, {Baiotti}, {Baird}, {Bajpai}, {Ball}, {Ballardin},
  {Ballmer}, {Balsamo}, {Baltus}, {Banagiri}, {Bankar}, {Barayoga}, {Barbieri},
  {Barish}, {Barker}, {Barneo}, {Barone}, {Barr}, {Barsotti}, {Barsuglia},
  {Barta}, {Bartlett}, {Barton}, {Bartos}, {Bassiri}, {Basti}, {Bawaj},
  {Bayley}, {Baylor}, {Bazzan}, {B{\'e}csy}, {Bedakihale}, {Bejger},
  {Belahcene}, {Benedetto}, {Beniwal}, {Bennett}, {Bentley}, {BenYaala},
  {Bergamin}, {Berger}, {Bernuzzi}, {Berry}, {Bersanetti}, {Bertolini},
  {Betzwieser}, {Beveridge}, {Bhandare}, {Bhardwaj}, {Bhattacharjee},
  {Bhaumik}, {Bilenko}, {Billingsley}, {Bini}, {Birney}, {Birnholtz},
  {Biscans}, {Bischi}, {Biscoveanu}, {Bisht}, {Biswas}, {Bitossi}, {Bizouard},
  {Blackburn}, {Blair}, {Blair}, {Blair}, {Bobba}, {Bode}, {Boer}, {Bogaert},
  {Boldrini}, {Bonavena}, {Bondu}, {Bonilla}, {Bonnand}, {Booker}, {Boom},
  {Bork}, {Boschi}, {Bose}, {Bose}, {Bossilkov}, {Boudart}, {Bouffanais},
  {Bozzi}, {Bradaschia}, {Brady}, {Bramley}, {Branch}, {Branchesi}, {Brau},
  {Breschi}, {Briant}, {Briggs}, {Brillet}, {Brinkmann}, {Brockill}, {Brooks},
  {Brooks}, {Brown}, {Brunett}, {Bruno}, {Bruntz}, {Bryant}, {Bulik}, {Bulten},
  {Buonanno}, {Buscicchio}, {Buskulic}, {Buy}, {Byer}, {Cadonati}, {Cagnoli},
  {Cahillane}, {Calder{\'o}n Bustillo}, {Callaghan}, {Callister}, {Calloni},
  {Cameron}, {Camp}, {Canepa}, {Canevarolo}, {Cannavacciuolo}, {Cannon}, {Cao},
  {Cao}, {Capocasa}, {Capote}, {Carapella}, {Carbognani}, {Carlin}, {Carney},
  {Carpinelli}, {Carrillo}, {Carullo}, {Carver}, {Casanueva Diaz}, {Casentini},
  {Castaldi}, {Caudill}, {Cavagli{\`a}}, {Cavalier}, {Cavalieri}, {Ceasar},
  {Cella}, {Cerd{\'a}-Dur{\'a}n}, {Cesarini}, {Chaibi}, {Chakravarti},
  {Chalathadka Subrahmanya}, {Champion}, {Chan}, {Chan}, {Chan}, {Chan},
  {Chan}, {Chandra}, {Chanial}, {Chao}, {Charlton}, {Chase},
  {Chassande-Mottin}, {Chatterjee}, {Chatterjee}, {Chatterjee}, {Chaturvedi},
  {Chaty}, {Chatziioannou}, {Chen}, {Chen}, {Chen}, {Chen}, {Chen}, {Chen},
  {Chen}, {Chen}, {Cheng}, {Cheong}, {Cheung}, {Chia}, {Chiadini}, {Chiang},
  {Chiarini}, {Chierici}, {Chincarini}, {Chiofalo}, {Chiummo}, {Cho}, {Cho},
  {Choudhary}, {Choudhary}, {Christensen}, {Chu}, {Chu}, {Chu}, {Chua},
  {Chung}, {Ciani}, {Ciecielag}, {Cie{\'s}lar}, {Cifaldi}, {Ciobanu}, {Ciolfi},
  {Cipriano}, {Cirone}, {Clara}, {Clark}, {Clark}, {Clarke}, {Clearwater},
  {Clesse}, {Cleva}, {Coccia}, {Codazzo}, {Cohadon}, {Cohen}, {Cohen},
  {Colleoni}, {Collette}, {Colombo}, {Colpi}, {Compton}, {Constancio}, {Conti},
  {Cooper}, {Corban}, {Corbitt}, {Cordero-Carri{\'o}n}, {Corezzi}, {Corley},
  {Cornish}, {Corre}, {Corsi}, {Cortese}, {Costa}, {Cotesta}, {Coughlin},
  {Coulon}, {Countryman}, {Cousins}, {Couvares}, {Coward}, {Cowart}, {Coyne},
  {Coyne}, {Creighton}, {Creighton}, {Criswell}, {Croquette}, {Crowder},
  {Cudell}, {Cullen}, {Cumming}, {Cummings}, {Cunningham}, {Cuoco},
  {Cury{\l}o}, {Dabadie}, {Dal Canton}, {Dall'Osso}, {D{\'a}lya}, {Dana},
  {DaneshgaranBajastani}, {D'Angelo}, {Danilishin}, {D'Antonio}, {Danzmann},
  {Darsow-Fromm}, {Dasgupta}, {Datrier}, {Datta}, {Dattilo}, {Dave}, {Davier},
  {Davies}, {Davis}, {Davis}, {Daw}, {Dean}, {DeBra}, {Deenadayalan},
  {Degallaix}, {De Laurentis}, {Del{\'e}glise}, {Del Favero}, {De Lillo}, {De
  Lillo}, {Del Pozzo}, {DeMarchi}, {De Matteis}, {D'Emilio}, {Demos}, {Dent},
  {Depasse}, {De Pietri}, {De Rosa}, {De Rossi}, {DeSalvo}, {De Simone},
  {Dhurandhar}, {D{\'\i}az}, {Diaz-Ortiz}, {Didio}, {Dietrich}, {Di Fiore}, {Di
  Fronzo}, {Di Giorgio}, {Di Giovanni}, {Di Giovanni}, {Di Girolamo}, {Di
  Lieto}, {Ding}, {Di Pace}, {Di Palma}, {Di Renzo}, {Divakarla}, {Dmitriev},
  {Doctor}, {D'Onofrio}, {Donovan}, {Dooley}, {Doravari}, {Dorrington},
  {Drago}, {Driggers}, {Drori}, {Ducoin}, {Dupej}, {Durante}, {D'Urso},
  {Duverne}, {Dwyer}, {Eassa}, {Easter}, {Ebersold}, {Eckhardt}, {Eddolls},
  {Edelman}, {Edo}, {Edy}, {Effler}, {Eguchi}, {Eichholz}, {Eikenberry},
  {Eisenmann}, {Eisenstein}, {Ejlli}, {Engelby}, {Enomoto}, {Errico}, {Essick},
  {Estell{\'e}s}, {Estevez}, {Etienne}, {Etzel}, {Evans}, {Evans}, {Ewing},
  {Fafone}, {Fair}, {Fairhurst}, {Farah}, {Farinon}, {Farr}, {Farr}, {Farrow},
  {Fauchon-Jones}, {Favaro}, {Favata}, {Fays}, {Fazio}, {Feicht}, {Fejer},
  {Fenyvesi}, {Ferguson}, {Fernandez-Galiana}, {Ferrante}, {Ferreira},
  {Fidecaro}, {Figura}, {Fiori}, {Fishbach}, {Fisher}, {Fittipaldi}, {Fiumara},
  {Flaminio}, {Floden}, {Fong}, {Font}, {Fornal}, {Forsyth}, {Franke},
  {Frasca}, {Frasconi}, {Frederick}, {Freed}, {Frei}, {Freise}, {Frey},
  {Fritschel}, {Frolov}, {Fronz{\'e}}, {Fujii}, {Fujikawa}, {Fukunaga},
  {Fukushima}, {Fulda}, {Fyffe}, {Gabbard}, {Gadre}, {Gair}, {Gais},
  {Galaudage}, {Gamba}, {Ganapathy}, {Ganguly}, {Gao}, {Gaonkar}, {Garaventa},
  {Garc{\'\i}a-N{\'u}{\~n}ez}, {Garc{\'\i}a-Quir{\'o}s}, {Garufi}, {Gateley},
  {Gaudio}, {Gayathri}, {Ge}, {Gemme}, {Gennai}, {George}, {Gerberding},
  {Gergely}, {Gewecke}, {Ghonge}, {Ghosh}, {Ghosh}, {Ghosh}, {Ghosh},
  {Giacomazzo}, {Giacoppo}, {Giaime}, {Giardina}, {Gibson}, {Gier}, {Giesler},
  {Giri}, {Gissi}, {Glanzer}, {Gleckl}, {Godwin}, {Goetz}, {Goetz}, {Gohlke},
  {Golomb}, {Goncharov}, {Gonz{\'a}lez}, {Gopakumar}, {Gosselin}, {Gouaty},
  {Gould}, {Grace}, {Grado}, {Granata}, {Granata}, {Grant}, {Gras}, {Grassia},
  {Gray}, {Gray}, {Greco}, {Green}, {Green}, {Gretarsson}, {Gretarsson},
  {Griffith}, {Griffiths}, {Griggs}, {Grignani}, {Grimaldi}, {Grimm}, {Grote},
  {Grunewald}, {Gruning}, {Guerra}, {Guidi}, {Guimaraes}, {Guix{\'e}},
  {Gulati}, {Guo}, {Guo}, {Gupta}, {Gupta}, {Gupta}, {Gustafson}, {Gustafson},
  {Guzman}, {Ha}, {Haegel}, {Hagiwara}, {Haino}, {Halim}, {Hall}, {Hamilton},
  {Hammond}, {Han}, {Haney}, {Hanks}, {Hanna}, {Hannam}, {Hannuksela},
  {Hansen}, {Hansen}, {Hanson}, {Harder}, {Hardwick}, {Haris}, {Harms},
  {Harry}, {Harry}, {Hartwig}, {Hasegawa}, {Haskell}, {Hasskew}, {Haster},
  {Hattori}, {Haughian}, {Hayakawa}, {Hayama}, {Hayes}, {Healy}, {Heidmann},
  {Heidt}, {Heintze}, {Heinze}, {Heinzel}, {Heitmann}, {Hellman}, {Hello},
  {Helmling-Cornell}, {Hemming}, {Hendry}, {Heng}, {Hennes}, {Hennig},
  {Hennig}, {Hernandez}, {Hernandez Vivanco}, {Heurs}, {Hild}, {Hill},
  {Himemoto}, {Hines}, {Hiranuma}, {Hirata}, {Hirose}, {Hochheim}, {Hofman},
  {Hohmann}, {Holcomb}, {Holland}, {Hollows}, {Holmes}, {Holt}, {Holz}, {Hong},
  {Hopkins}, {Hough}, {Hourihane}, {Howell}, {Hoy}, {Hoyland}, {Hreibi},
  {Hsieh}, {Hsu}, {Huang}, {Huang}, {Huang}, {Huang}, {Huang}, {Huang},
  {H{\"u}bner}, {Huddart}, {Hughey}, {Hui}, {Hui}, {Husa}, {Huttner},
  {Huxford}, {Huynh-Dinh}, {Ide}, {Idzkowski}, {Iess}, {Ikenoue}, {Imam},
  {Inayoshi}, {Ingram}, {Inoue}, {Ioka}, {Isi}, {Isleif}, {Ito}, {Itoh},
  {Iyer}, {Izumi}, {JaberianHamedan}, {Jacqmin}, {Jadhav}, {Jadhav}, {James},
  {Jan}, {Jani}, {Janquart}, {Janssens}, {Janthalur}, {Jaranowski}, {Jariwala},
  {Jaume}, {Jenkins}, {Jenner}, {Jeon}, {Jeunon}, {Jia}, {Jin}, {Johns},
  {Jones}, {Jones}, {Jones}, {Jones}, {Jones}, {Jonker}, {Ju}, {Jung}, {Jung},
  {Junker}, {Juste}, {Kaihotsu}, {Kajita}, {Kakizaki}, {Kalaghatgi},
  {Kalogera}, {Kamai}, {Kamiizumi}, {Kanda}, {Kandhasamy}, {Kang}, {Kanner},
  {Kao}, {Kapadia}, {Kapasi}, {Karat}, {Karathanasis}, {Karki}, {Kashyap},
  {Kasprzack}, {Kastaun}, {Katsanevas}, {Katsavounidis}, {Katzman}, {Kaur},
  {Kawabe}, {Kawaguchi}, {Kawai}, {Kawasaki}, {K{\'e}f{\'e}lian}, {Keitel},
  {Key}, {Khadka}, {Khalili}, {Khan}, {Khazanov}, {Khetan}, {Khursheed},
  {Kijbunchoo}, {Kim}, {Kim}, {Kim}, {Kim}, {Kim}, {Kim}, {Kimball}, {Kimura},
  {Kinley-Hanlon}, {Kirchhoff}, {Kissel}, {Kita}, {Kitazawa}, {Kleybolte},
  {Klimenko}, {Knee}, {Knowles}, {Knyazev}, {Koch}, {Koekoek}, {Kojima},
  {Kokeyama}, {Koley}, {Kolitsidou}, {Kolstein}, {Komori}, {Kondrashov},
  {Kong}, {Kontos}, {Koper}, {Korobko}, {Kotake}, {Kovalam}, {Kozak},
  {Kozakai}, {Kozu}, {Kringel}, {Krishnendu}, {Kr{\'o}lak}, {Kuehn}, {Kuei},
  {Kuijer}, {Kumar}, {Kumar}, {Kumar}, {Kumar}, {Kume}, {Kuns}, {Kuo}, {Kuo},
  {Kuromiya}, {Kuroyanagi}, {Kusayanagi}, {Kuwahara}, {Kwak}, {Lagabbe},
  {Laghi}, {Lalande}, {Lam}, {Lamberts}, {Landry}, {Landry}, {Lane}, {Lang},
  {Lange}, {Lantz}, {La Rosa}, {Lartaux-Vollard}, {Lasky}, {Laxen},
  {Lazzarini}, {Lazzaro}, {Leaci}, {Leavey}, {Lecoeuche}, {Lee}, {Lee}, {Lee},
  {Lee}, {Lee}, {Lee}, {Lehmann}, {Lema{\^\i}tre}, {Leonardi}, {Leroy},
  {Letendre}, {Levesque}, {Levin}, {Leviton}, {Leyde}, {Li}, {Li}, {Li}, {Li},
  {Li}, {Li}, {Lin}, {Lin}, {Lin}, {Lin}, {Lin}, {Linde}, {Linker}, {Linley},
  {Littenberg}, {Liu}, {Liu}, {Liu}, {Liu}, {Llamas}, {Llorens-Monteagudo},
  {Lo}, {Lockwood}, {London}, {Longo}, {Lopez}, {Lopez Portilla}, {Lorenzini},
  {Loriette}, {Lormand}, {Losurdo}, {Lott}, {Lough}, {Lousto}, {Lovelace},
  {Lucaccioni}, {L{\"u}ck}, {Lumaca}, {Lundgren}, {Luo}, {Lynam}, {Macas},
  {MacInnis}, {Macleod}, {MacMillan}, {Macquet}, {Maga{\~n}a Hernandez},
  {Magazz{\`u}}, {Magee}, {Maggiore}, {Magnozzi}, {Mahesh}, {Majorana},
  {Makarem}, {Maksimovic}, {Maliakal}, {Malik}, {Man}, {Mandic}, {Mangano},
  {Mango}, {Mansell}, {Manske}, {Mantovani}, {Mapelli}, {Marchesoni},
  {Marchio}, {Marion}, {Mark}, {M{\'a}rka}, {M{\'a}rka}, {Markakis},
  {Markosyan}, {Markowitz}, {Maros}, {Marquina}, {Marsat}, {Martelli},
  {Martin}, {Martin}, {Martinez}, {Martinez}, {Martinez}, {Martinovic},
  {Martynov}, {Marx}, {Masalehdan}, {Mason}, {Massera}, {Masserot},
  {Massinger}, {Masso-Reid}, {Mastrogiovanni}, {Matas}, {Mateu-Lucena},
  {Matichard}, {Matiushechkina}, {Mavalvala}, {McCann}, {McCarthy},
  {McClelland}, {McClincy}, {McCormick}, {McCuller}, {McGhee}, {McGuire},
  {McIsaac}, {McIver}, {McRae}, {McWilliams}, {Meacher}, {Mehmet}, {Mehta},
  {Meijer}, {Melatos}, {Melchor}, {Mendell}, {Menendez-Vazquez}, {Menoni},
  {Mercer}, {Mereni}, {Merfeld}, {Merilh}, {Merritt}, {Merzougui}, {Meshkov},
  {Messenger}, {Messick}, {Meyers}, {Meylahn}, {Mhaske}, {Miani}, {Miao},
  {Michaloliakos}, {Michel}, {Michimura}, {Middleton}, {Milano}, {Miller},
  {Miller}, {Miller}, {Miller}, {Millhouse}, {Mills}, {Milotti}, {Minazzoli},
  {Minenkov}, {Mio}, {Mir}, {Miravet-Ten{\'e}s}, {Mishra}, {Mishra}, {Mistry},
  {Mitra}, {Mitrofanov}, {Mitselmakher}, {Mittleman}, {Miyakawa}, {Miyamoto},
  {Miyazaki}, {Miyo}, {Miyoki}, {Mo}, {Moguel}, {Mogushi}, {Mohapatra},
  {Mohite}, {Molina}, {Molina-Ruiz}, {Mondin}, {Montani}, {Moore}, {Moraru},
  {Morawski}, {More}, {Moreno}, {Moreno}, {Mori}, {Morisaki}, {Moriwaki},
  {Mours}, {Mow-Lowry}, {Mozzon}, {Muciaccia}, {Mukherjee}, {Mukherjee},
  {Mukherjee}, {Mukherjee}, {Mukherjee}, {Mukund}, {Mullavey}, {Munch},
  {Mu{\~n}iz}, {Murray}, {Musenich}, {Muusse}, {Nadji}, {Nagano}, {Nagano},
  {Nagar}, {Nakamura}, {Nakano}, {Nakano}, {Nakashima}, {Nakayama}, {Napolano},
  {Nardecchia}, {Narikawa}, {Naticchioni}, {Nayak}, {Nayak}, {Negishi}, {Neil},
  {Neilson}, {Nelemans}, {Nelson}, {Nery}, {Neubauer}, {Neunzert}, {Ng}, {Ng},
  {Nguyen}, {Nguyen}, {Nguyen}, {Nguyen Quynh}, {Ni}, {Nichols}, {Nishizawa},
  {Nissanke}, {Nitoglia}, {Nocera}, {Norman}, {North}, {Nozaki}, {Nuttall},
  {Oberling}, {O'Brien}, {Obuchi}, {O'Dell}, {Oelker}, {Ogaki}, {Oganesyan},
  {Oh}, {Oh}, {Oh}, {Ohashi}, {Ohishi}, {Ohkawa}, {Ohme}, {Ohta}, {Okada},
  {Okutani}, {Okutomi}, {Olivetto}, {Oohara}, {Ooi}, {Oram}, {O'Reilly},
  {Ormiston}, {Ormsby}, {Ortega}, {O'Shaughnessy}, {O'Shea}, {Oshino},
  {Ossokine}, {Osthelder}, {Otabe}, {Ottaway}, {Overmier}, {Pace}, {Pagano},
  {Page}, {Pagliaroli}, {Pai}, {Pai}, {Palamos}, {Palashov}, {Palomba}, {Pan},
  {Pan}, {Panda}, {Pang}, {Pang}, {Pankow}, {Pannarale}, {Pant}, {Panther},
  {Paoletti}, {Paoli}, {Paolone}, {Parisi}, {Park}, {Park}, {Parker},
  {Pascucci}, {Pasqualetti}, {Passaquieti}, {Passuello}, {Patel}, {Pathak},
  {Patricelli}, {Patron}, {Paul}, {Payne}, {Pedraza}, {Pegoraro}, {Pele},
  {Pe{\~n}a Arellano}, {Penn}, {Perego}, {Pereira}, {Pereira}, {Perez},
  {P{\'e}rigois}, {Perkins}, {Perreca}, {Perri{\`e}s}, {Petermann},
  {Petterson}, {Pfeiffer}, {Pham}, {Phukon}, {Piccinni}, {Pichot},
  {Piendibene}, {Piergiovanni}, {Pierini}, {Pierro}, {Pillant}, {Pillas},
  {Pilo}, {Pinard}, {Pinto}, {Pinto}, {Piotrzkowski}, {Pirello}, {Pitkin},
  {Placidi}, {Planas}, {Plastino}, {Pluchar}, {Poggiani}, {Polini}, {Pong},
  {Ponrathnam}, {Popolizio}, {Porter}, {Poulton}, {Powell}, {Pracchia},
  {Pradier}, {Prajapati}, {Prasai}, {Prasanna}, {Pratten}, {Principe}, {Prodi},
  {Prokhorov}, {Prosposito}, {Prudenzi}, {Puecher}, {Punturo}, {Puosi},
  {Puppo}, {P{\"u}rrer}, {Qi}, {Quetschke}, {Quitzow-James}, {Raab},
  {Raaijmakers}, {Radkins}, {Radulesco}, {Raffai}, {Rail}, {Raja}, {Rajan},
  {Ramirez}, {Ramirez}, {Ramos-Buades}, {Rana}, {Rapagnani}, {Rapol}, {Ray},
  {Raymond}, {Raza}, {Razzano}, {Read}, {Rees}, {Regimbau}, {Rei}, {Reid},
  {Reid}, {Reitze}, {Relton}, {Renzini}, {Rettegno}, {Rezac}, {Ricci},
  {Richards}, {Richardson}, {Richardson}, {Riemenschneider}, {Riles},
  {Rinaldi}, {Rink}, {Rizzo}, {Robertson}, {Robie}, {Robinet}, {Rocchi},
  {Rodriguez}, {Rolland}, {Rollins}, {Romanelli}, {Romano}, {Romel},
  {Romero-Rodr{\'\i}guez}, {Romero-Shaw}, {Romie}, {Ronchini}, {Rosa}, {Rose},
  {Rosi{\'n}ska}, {Ross}, {Rowan}, {Rowlinson}, {Roy}, {Roy}, {Roy}, {Rozza},
  {Ruggi}, {Ryan}, {Sachdev}, {Sadecki}, {Sadiq}, {Sago}, {Saito}, {Saito},
  {Sakai}, {Sakai}, {Sakellariadou}, {Sakuno}, {Salafia}, {Salconi}, {Saleem},
  {Salemi}, {Samajdar}, {Sanchez}, {Sanchez}, {Sanchez}, {Sanchis-Gual},
  {Sanders}, {Sanuy}, {Saravanan}, {Sarin}, {Sassolas}, {Satari},
  {Sathyaprakash}, {Sato}, {Sato}, {Sauter}, {Savage}, {Sawada}, {Sawant},
  {Sawant}, {Sayah}, {Schaetzl}, {Scheel}, {Scheuer}, {Schiworski}, {Schmidt},
  {Schmidt}, {Schnabel}, {Schneewind}, {Schofield}, {Sch{\"o}nbeck}, {Schulte},
  {Schutz}, {Schwartz}, {Scott}, {Scott}, {Seglar-Arroyo}, {Sekiguchi},
  {Sekiguchi}, {Sellers}, {Sengupta}, {Sentenac}, {Seo}, {Sequino}, {Sergeev},
  {Setyawati}, {Shaffer}, {Shahriar}, {Shams}, {Shao}, {Sharma}, {Sharma},
  {Shawhan}, {Shcheblanov}, {Shibagaki}, {Shikauchi}, {Shimizu}, {Shimoda},
  {Shimode}, {Shinkai}, {Shishido}, {Shoda}, {Shoemaker}, {Shoemaker},
  {ShyamSundar}, {Sieniawska}, {Sigg}, {Singer}, {Singh}, {Singh}, {Singha},
  {Sintes}, {Sipala}, {Skliris}, {Slagmolen}, {Slaven-Blair}, {Smetana},
  {Smith}, {Smith}, {Soldateschi}, {Somala}, {Somiya}, {Son}, {Soni}, {Soni},
  {Sordini}, {Sorrentino}, {Sorrentino}, {Sotani}, {Soulard}, {Souradeep},
  {Sowell}, {Spagnuolo}, {Spencer}, {Spera}, {Srinivasan}, {Srivastava},
  {Srivastava}, {Staats}, {Stachie}, {Steer}, {Steinlechner}, {Steinlechner},
  {Stops}, {Stover}, {Strain}, {Strang}, {Stratta}, {Strunk}, {Sturani},
  {Stuver}, {Sudhagar}, {Sudhir}, {Sugimoto}, {Suh}, {Summerscales}, {Sun},
  {Sun}, {Sunil}, {Sur}, {Suresh}, {Sutton}, {Suzuki}, {Suzuki}, {Swinkels},
  {Szczepa{\'n}czyk}, {Szewczyk}, {Tacca}, {Tagoshi}, {Tait}, {Takahashi},
  {Takahashi}, {Takamori}, {Takano}, {Takeda}, {Takeda}, {Talbot}, {Talbot},
  {Tanaka}, {Tanaka}, {Tanaka}, {Tanaka}, {Tanaka}, {Tanasijczuk}, {Tanioka},
  {Tanner}, {Tao}, {Tao}, {Tapia San Mart{\'\i}n}, {Taranto}, {Tasson},
  {Telada}, {Tenorio}, {Terhune}, {Terkowski}, {Thirugnanasambandam}, {Thomas},
  {Thomas}, {Thompson}, {Thondapu}, {Thorne}, {Thrane}, {Tiwari}, {Tiwari},
  {Tiwari}, {Toivonen}, {Toland}, {Tolley}, {Tomaru}, {Tomigami}, {Tomura},
  {Tonelli}, {Torres-Forn{\'e}}, {Torrie}, {Tosta e Melo}, {T{\"o}yr{\"a}},
  {Trapananti}, {Travasso}, {Traylor}, {Trevor}, {Tringali}, {Tripathee},
  {Troiano}, {Trovato}, {Trozzo}, {Trudeau}, {Tsai}, {Tsai}, {Tsang}, {Tsang},
  {Tsao}, {Tse}, {Tso}, {Tsubono}, {Tsuchida}, {Tsukada}, {Tsuna}, {Tsutsui},
  {Tsuzuki}, {Turbang}, {Turconi}, {Tuyenbayev}, {Ubhi}, {Uchikata},
  {Uchiyama}, {Udall}, {Ueda}, {Uehara}, {Ueno}, {Ueshima}, {Unnikrishnan},
  {Uraguchi}, {Urban}, {Ushiba}, {Utina}, {Vahlbruch}, {Vajente}, {Vajpeyi},
  {Valdes}, {Valentini}, {Valsan}, {van Bakel}, {van Beuzekom}, {van den
  Brand}, {Van Den Broeck}, {Vander-Hyde}, {van der Schaaf}, {van Heijningen},
  {Vanosky}, {van Putten}, {van Remortel}, {Vardaro}, {Vargas}, {Varma},
  {Vas{\'u}th}, {Vecchio}, {Vedovato}, {Veitch}, {Veitch}, {Venneberg},
  {Venugopalan}, {Verkindt}, {Verma}, {Verma}, {Veske}, {Vetrano},
  {Vicer{\'e}}, {Vidyant}, {Viets}, {Vijaykumar}, {Villa-Ortega}, {Vinet},
  {Virtuoso}, {Vitale}, {Vo}, {Vocca}, {von Reis}, {von Wrangel}, {Vorvick},
  {Vyatchanin}, {Wade}, {Wade}, {Wagner}, {Walet}, {Walker}, {Wallace},
  {Wallace}, {Walsh}, {Wang}, {Wang}, {Wang}, {Ward}, {Warner}, {Was},
  {Washimi}, {Washington}, {Watchi}, {Weaver}, {Webster}, {Weinert},
  {Weinstein}, {Weiss}, {Weller}, {Wellmann}, {Wen}, {We{\ss}els}, {Wette},
  {Whelan}, {White}, {Whiting}, {Whittle}, {Wilken}, {Williams}, {Williams},
  {Williamson}, {Willis}, {Willke}, {Wilson}, {Winkler}, {Wipf}, {Wlodarczyk},
  {Woan}, {Woehler}, {Wofford}, {Wong}, {Wu}, {Wu}, {Wu}, {Wu}, {Wysocki},
  {Xiao}, {Xu}, {Yamada}, {Yamamoto}, {Yamamoto}, {Yamamoto}, {Yamamoto},
  {Yamashita}, {Yamazaki}, {Yang}, {Yang}, {Yang}, {Yang}, {Yang}, {Yap},
  {Yeeles}, {Yelikar}, {Ying}, {Yokogawa}, {Yokoyama}, {Yokozawa}, {Yoo},
  {Yoshioka}, {Yu}, {Yu}, {Yuzurihara}, {Zadro{\.z}ny}, {Zanolin}, {Zeidler},
  {Zelenova}, {Zendri}, {Zevin}, {Zhan}, {Zhang}, {Zhang}, {Zhang}, {Zhang},
  {Zhang}, {Zhao}, {Zhao}, {Zhao}, {Zhao}, {Zhou}, {Zhou}, {Zhu}, {Zhu},
  {Zimmerman}, {Zlochower}, {Zucker}, \& {Zweizig}}]{LVK:2021:GWTC3-pop}
{The LIGO Scientific Collaboration}, {the Virgo Collaboration}, {the KAGRA
  Collaboration}, {et~al.} 2021, arXiv e-prints, arXiv:2111.03634,
  \dodoi{10.48550/arXiv.2111.03634}

\bibitem[{{Thompson} {et~al.}(2019){Thompson}, {Kochanek}, {Stanek}, {Badenes},
  {Post}, {Jayasinghe}, {Latham}, {Bieryla}, {Esquerdo}, {Berlind}, {Calkins},
  {Tayar}, {Lindegren}, {Johnson}, {Holoien}, {Auchettl}, \&
  {Covey}}]{Thompson:2019:BH-MS}
{Thompson}, T.~A., {Kochanek}, C.~S., {Stanek}, K.~Z., {et~al.} 2019, Science,
  366, 637, \dodoi{10.1126/science.aau4005}

\bibitem[{{Thompson} {et~al.}(2020){Thompson}, {Kochanek}, {Stanek}, {Badenes},
  {Jayasinghe}, {Tayar}, {Johnson}, {Holoien}, {Auchettl}, \&
  {Covey}}]{Thompson:2020:BH-MS}
---. 2020, Science, 368, eaba4356, \dodoi{10.1126/science.aba4356}

\bibitem[{{Tong} {et~al.}(2022){Tong}, {Galaudage}, \&
  {Thrane}}]{Tong:2022:PopSpinGWTC3}
{Tong}, H., {Galaudage}, S., \& {Thrane}, E. 2022, arXiv e-prints,
  arXiv:2209.02206, \dodoi{10.48550/arXiv.2209.02206}

\bibitem[{{Valsecchi} {et~al.}(2010){Valsecchi}, {Glebbeek}, {Farr}, {Fragos},
  {Willems}, {Orosz}, {Liu}, \& {Kalogera}}]{Valsecchi:2010}
{Valsecchi}, F., {Glebbeek}, E., {Farr}, W.~M., {et~al.} 2010, \nat, 468, 77,
  \dodoi{10.1038/nature09463}

\bibitem[{{van den Heuvel} \& {Yoon}(2007)}]{vandenHeuvel:2007:LGRBs}
{van den Heuvel}, E.~P.~J., \& {Yoon}, S.~C. 2007, \apss, 311, 177,
  \dodoi{10.1007/s10509-007-9583-8}

\bibitem[{{van Son} {et~al.}(2023){van Son}, {de Mink}, {Chru{\'s}li{\'n}ska},
  {Conroy}, {Pakmor}, \& {Hernquist}}]{vanSon:2022}
{van Son}, L.~A.~C., {de Mink}, S.~E., {Chru{\'s}li{\'n}ska}, M., {et~al.}
  2023, \apj, 948, 105, \dodoi{10.3847/1538-4357/acbf51}

\bibitem[{{Vinciguerra} {et~al.}(2020){Vinciguerra}, {Neijssel},
  {Vigna-G{\'o}mez}, {Mandel}, {Podsiadlowski}, {Maccarone}, {Nicholl},
  {Kingdon}, {Perry}, \& {Salemi}}]{Vinciguerra:2020:MTEfficiency}
{Vinciguerra}, S., {Neijssel}, C.~J., {Vigna-G{\'o}mez}, A., {et~al.} 2020,
  \mnras, 498, 4705, \dodoi{10.1093/mnras/staa2177}

\bibitem[{{Vink}(2017)}]{Vink:2017:WRWinds}
{Vink}, J.~S. 2017, \aap, 607, L8, \dodoi{10.1051/0004-6361/201731902}

\bibitem[{{Vink} \& {de Koter}(2005)}]{VinkdeKoter:2005:m}
{Vink}, J.~S., \& {de Koter}, A. 2005, \aap, 442, 587,
  \dodoi{10.1051/0004-6361:20052862}

\bibitem[{Webster \& Murdin(1972)}]{WebsterMurdin:1972:CygX1Disco}
Webster, B.~L., \& Murdin, P. 1972, Nature, 235, 37, \dodoi{10.1038/235037a0}

\bibitem[{{Willcox} {et~al.}(in prep.){Willcox}, {MacLeod}, {Mandel}, \&
  {Hirai}}]{Willcox:inprep:fGamma}
{Willcox}, R., {MacLeod}, M., {Mandel}, I., \& {Hirai}, R. in prep., {The
  Importance of Angular Momentum Loss on the Outcomes of Binary Mass Transfer}

\bibitem[{{Wojdowski} {et~al.}(1998){Wojdowski}, {Clark}, {Levine}, {Woo}, \&
  {Zhang}}]{Wojdowski:1998:LMCX1Disco}
{Wojdowski}, P., {Clark}, G.~W., {Levine}, A.~M., {Woo}, J.~W., \& {Zhang},
  S.~N. 1998, \apj, 502, 253, \dodoi{10.1086/305893}

\bibitem[{{Woosley}(1993)}]{Woosley:1993:GRBs}
{Woosley}, S.~E. 1993, \apj, 405, 273, \dodoi{10.1086/172359}

\bibitem[{{Zdziarski} {et~al.}(2018){Zdziarski}, {Malyshev}, {Dubus}, {Pooley},
  {Johnson}, {Frankowski}, {De Marco}, {Chernyakova}, \&
  {Rao}}]{Zdziarski:2018:CygnusX3}
{Zdziarski}, A.~A., {Malyshev}, D., {Dubus}, G., {et~al.} 2018, \mnras, 479,
  4399, \dodoi{10.1093/mnras/sty1618}

\bibitem[{{Zhao} {et~al.}(2021){Zhao}, {Gou}, {Dong}, {Zheng}, {Steiner},
  {Miller-Jones}, {Bahramian}, {Orosz}, \& {Feng}}]{Zhao:2021:CygnusX1Spin}
{Zhao}, X., {Gou}, L., {Dong}, Y., {et~al.} 2021, \apj, 908, 117,
  \dodoi{10.3847/1538-4357/abbcd6}

\end{thebibliography}
\end{document}